\documentclass{ar-1col}

\usepackage[comma]{natbib}
\usepackage{url}
\usepackage{amssymb}
\usepackage{hyperref}

\newcommand{\be}{\begin{equation}}
\newcommand{\ee}{\end{equation}}
\newcommand\ion[2]{#1$\;${\scshape{#2}}}

\begin{document}

\markboth{Chomiuk, Metzger \& Shen}{Shocking Novae}

\title{New Insights into Classical Novae}

\author{Laura Chomiuk$^1$, Brian D.~Metzger$^2$, Ken J.~Shen$^3$
\affil{$^1$Center for Data Intensive and Time Domain Astronomy, 
Department of Physics and Astronomy, Michigan State University, East Lansing, MI 48824, USA; email: chomiukl@msu.edu}
\affil{$^2$Department of Physics, Columbia University, New York, NY 10027, USA; email: bdm2129@columbia.edu}
\affil{$^3$Department of Astronomy and Theoretical Astrophysics Center, University of California, Berkeley, CA 94720, USA; email: kenshen@astro.berkeley.edu}}

\begin{abstract}
\vspace{-0.3cm}
We survey our understanding of classical novae---non-terminal, thermonuclear eruptions on the surfaces of white dwarfs in binary systems. The recent and unexpected discovery of GeV gamma-rays from Galactic novae has highlighted the complexity of novae and their value as laboratories for studying shocks and particle acceleration. We review half a century of nova literature through this new lens, and conclude: 
\begin{itemize}
\parbox[0cm]{4in}{
\item The basics of the thermonuclear runaway theory of novae are confirmed by observations. The white dwarf sustains surface nuclear burning for some time after runaway, and until recently, it was commonly believed that radiation from this nuclear burning solely determines the nova's bolometric luminosity.
\vspace{0.15cm}}
\vspace{0.15cm}
\parbox[0cm]{4in}{
\item The processes by which novae eject material from the binary system remain poorly understood.  Mass loss from novae is complex (sometimes fluctuating in rate, velocity, and morphology) and often prolonged in time over weeks, months, or years. }
\vspace{0.15cm}
\parbox[0cm]{4in}{
\item The complexity of the mass ejection leads to gamma-ray producing shocks internal to the nova ejecta. When gamma-rays are detected (around optical maximum), the shocks are deeply embedded and the surrounding gas is very dense. 
}
\parbox[0cm]{4in}{
\item 
Observations of correlated optical and gamma-ray light curves confirm that the shocks are radiative and contribute significantly to the bolometric luminosity of novae.  Novae are therefore the closest and most common ``interaction-powered" transients.
}
\end{itemize}
\end{abstract}

\maketitle

\tableofcontents

\section{INTRODUCTION}

Classical and recurrent novae (derived from the Latin {\it stella nova}, or {\it new star}) are luminous eruptions that take place in binary star systems in which a white dwarf (WD) accretes matter from a non-degenerate stellar companion \citep{Gallagher&Starrfield78}. 
As an accreted layer accumulates on the WD surface, the density and temperature at its base rise, leading to an increase in the nuclear burning rate. 
Under circumstances that depend sensitively on the WD mass and accretion rate, the layer undergoes unstable (``runaway") nuclear burning once it reaches a critical mass \citep[e.g.,][]{Starrfield+72, Prialnik&Kovetz95,Townsley&Bildsten04}.
The resulting energy release causes the accreted envelope to expand enormously, 
ultimately leading to its ejection,
often along with heavier elements dredged up from deeper layers of the WD.  Novae, with an estimated frequency of 
$\sim$20--70 eruptions per year in our Galaxy \citep[e.g.,][]{Darnley+06, Shafter17}, are the second most common type of thermonuclear eruptions after Type I X-ray bursts from neutron stars \citep{Galloway&Keek17}.
However, even with the advent of synoptic time-domain surveys, the discovery rate of Galactic novae\footnote{See \url{https://asd.gsfc.nasa.gov/Koji.Mukai/novae/novae.html} for a compilation of recent Galactic novae.} remains modest ($\sim 5-15$ yr$^{-1}$), probably due to gaps in optical monitoring and dust obscuration in the Galactic plane.

Among the brightest transients in the night sky, novae were sometimes called ``guest stars"
(see \citealt{Duerbeck08} for a historical perspective and \citealt{Hoffmann+20} for recent work).  Given their storied role in the history of Astronomy, it is striking---despite substantial observational and theoretical progress---that our understanding of these 
common transients remains incomplete.  Nothing highlights this state of affairs better than the nearly universal\footnote{Though evidence for relativistic particle acceleration was already present from radio synchrotron emission (e.g., \citealt{Hjellming+86, Taylor+87, Rupen+01}; \S\ref{sec:radiononthermal})
and through other, indirect inference \citep{Tatischeff&Hernanz07}.}
surprise that accompanied the discovery of GeV gamma-ray emission from novae by the Large Area Telescope (LAT) on NASA's {\it Fermi Gamma-Ray Space Telescope} \citep{Abdo+10, Ackermann+14}.  
Evidence has long existed for complex mass-loss patterns and internal shocks within nova outflows, but the energetic importance of these shocks was not fully appreciated until recently.

\begin{marginnote}[]
\entry{Large Area Telescope (LAT)}{Instrument on the {\it Fermi Gamma-Ray Space Telescope} that monitors $\sim$60\% of the sky simultaneously at photon energies $\sim$0.1--300 GeV \citep{Atwood+09}.}
\end{marginnote}

In the standard paradigm, novae are almost exclusively driven by thermal emission from the hot WD.
After the thermonuclear runaway, a shell of gas is expelled from the WD, expanding into the surrounding environment at hundreds to thousands of kilometers per second.  The thermal and kinetic energies of the ejecta, whether released in a short-lived episode or as a longer, continuous wind, are powered by radiation from nuclear burning on the surface of the WD.
Novae are best known as optical transients with light curves that rise rapidly to maximum and decay over timescales of days to months \citep{Payne-Gaposchkin57}.  However, as mass loss subsides and the ejecta dilute, they become increasingly transparent to radiation at shorter wavelengths and the spectral energy distribution of the nova shifts to the ultraviolet (UV; \citealt{Gallagher&Code74}).
Eventually, the photosphere recedes far enough inwards that the WD can be observed as a luminous supersoft X-ray source, sustained by residual nuclear burning for weeks to years \citep{Kahabka&vandenHeuvel97}.
The expanding ejecta are photo-ionized by the hot central WD, and produce thermal radio emission on timescales of years \citep{Seaquist&Bode08}.  Many novae form dust in their ejecta, revealed by sudden rises in their infrared (IR) emission and sometimes by drops in their optical emission due to extinction along the line of sight \citep{Gehrz88}.

\begin{marginnote}
\entry{Supersoft X-ray Sources}{Thermal X-ray sources powered by surface nuclear burning on WDs with luminosities $L_{\rm X} \approx 10^{36}-10^{38}$ erg/s and effective temperatures $T_{\rm eff} \approx 10^5-10^6$ K.}
\end{marginnote}

Although this basic picture remains largely intact, multi-wavelength observations over the past decade have increasingly revealed a non-thermal, shock-powered side to novae, which is providing new insights into old mysteries about these events. The shocks occur either as multiple phases of ejecta collide during the eruption, or as the ejecta crash into a pre-existing medium surrounding the binary.
The continuum gamma-ray ($\gtrsim 100$ MeV) emission observed by {\it Fermi}-LAT is clear evidence of the acceleration of relativistic particles by shocks \citep{Martin&Dubus13}.  Shocks have also long been implied by X-ray observations of hot ($\sim 10^7-10^8$ K) presumably shock-heated gas in novae
\citep{OBrien+94}, observed weeks to months after eruption \citep{Mukai+08}. 
These X-rays might be absorbed earlier in the eruption, and reprocessed into the ultraviolet/optical/infrared (UVOIR) bands, thus contributing to the optical light curve and its variability \citep{Metzger+14, Li+17, Aydi+20}.  Relativistic electrons accelerated at shocks also generate synchrotron radiation, sometimes seen as a distinct early component of radio emission 
\citep{Taylor+87, Weston+16a, Finzell+18}.  Shock interaction may also play a crucial role in shaping the complex large-scale morphology of nova ejecta and may---via compression and hydrodynamical instabilities---generate the large densities and inhomogeneities needed for molecule and dust formation \citep{Evans&Rawlings08, Derdzinski+17}.

As rich and complex phenomena, novae are worthy of study in their own right.  However, they also serve as bright, nearby probes of a number of physical processes, such as binary mass transfer, explosive nuclear burning, and radiative shocks, that are relevant to many other astrophysical systems.  The conditions under which an accreting WD can gain mass, despite mass loss in novae, have implications for the viability of the single degenerate channel for thermonuclear supernovae (SNe; e.g., Type Ia or Type Iax; \citealt{Nomoto82,Foley+13}).  
The  morphologies of nova ejecta---often described as a bipolar outflow with an equatorial ring---resemble the outflows from other dynamical binary mass transfer events, such as planetary nebulae \citep{DeMarco09} or those expected from the common envelope phase of stellar mergers \citep{Ivanova+13}.
A better understanding of novae could provide new insights into a wide variety of other---rarer and more distant---transient events, especially those that may be shock-powered, such as Type IIn (interacting) SNe, tidal disruption events, and stellar mergers.

\begin{figure}[!b]
\includegraphics[width=5in, clip]{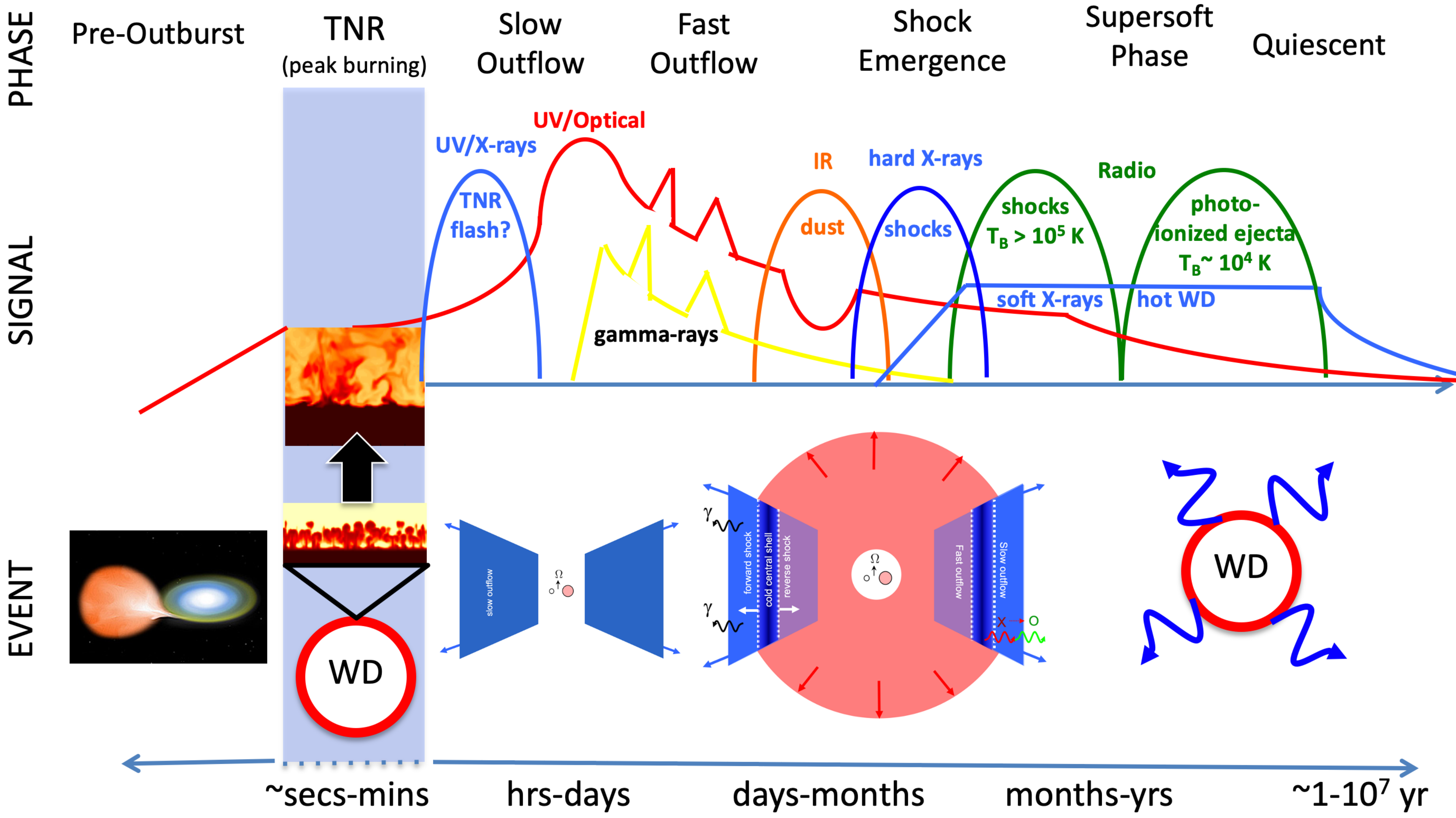}
\vspace{-0.2in}
\caption{Schematic timeline of the physical processes and electromagnetic signals from novae.  The figure includes modified images of convection/mixing during the thermonuclear runaway from \citet[][reproduced with permission $\copyright$ESO]{Casanova+16} and internal shocks from \citet{Metzger+15}.}
\label{fig:schematic}
\end{figure}

Although we touch upon most aspects of the nova phenomenon in this work, a comprehensive review is not possible. 
Instead, motivated by areas of rapid observational development, we shape the discussion around two broad and connected themes: mass loss during novae ($\S\ref{sec:massloss}$) and the newly appreciated role of shocks in shaping their electromagnetic emission ($\S\ref{sec:shocksobs}, \S\ref{sec:shockstheory}$). In \S \ref{sec:implications}, we discuss the implications for other WD phenomena and astrophysical transients, and suggest strategies for answering critical open questions surrounding nova mass loss and shocks.
Throughout the article we point the reader to reviews that provide greater depth on specific topics.  
Figure \ref{fig:schematic} shows a schematic timeline of physical processes and electromagnetic emission in novae that will be useful to refer to as we proceed.

\begin{marginnote}[]
\entry{Classical Nova}{Thermonuclear eruption from a WD accreting hydrogen-rich material from a Roche lobe-overflowing main sequence or moderately evolved companion.}
\entry{Cataclysmic Variable}{A mass-transferring binary system containing a WD and main sequence secondary that is overflowing its Roche lobe.}
\end{marginnote}

Before diving in, we introduce a few common definitions.  
With one exception (the helium nova V445~Pup; see {\bf Box 1}), the donor stars in nova-hosting binaries are hydrogen-rich. 
A ``classical nova" is one in which the secondary is typically a main-sequence star overflowing its Roche lobe;
the host binaries are ``cataclysmic variables" (CVs) with short-period orbits, $P \approx 1.4-10$ hours (\citealt{Warner95,Diaz&Bruch97}; although moderately evolved donors are not uncommon, e.g., \citealt{Darnley+12}).
CV mass transfer is thought to be relatively conservative, with at most a few percent of the transferred mass lost in outflows; searches for circumstellar material around CVs point to low density surroundings \citep[e.g.,][]{Froning05, Hoard+14}.
By contrast, an ``embedded" nova is one in which the secondary is an evolved giant star, typically on a long orbit with period $P \gtrsim 100$ days \citep{Mikolajewska10}. They are described as ``embedded" because interaction with the companion wind often shapes their observational signatures.  Novae with red giant companions likely make up 20--40\% of observed events \citep{Williams+16}. 
``Symbiotic" novae are the subset of embedded novae with eruptions that evolve very slowly, over decades or even centuries \citep{Kenyon&Truran83}.  

\begin{marginnote}[]
\entry{Embedded Nova}{A nova where the WD is fed material from a giant companion star, usually via wind accretion.}
\entry{Symbiotic Novae}{The subset of embedded novae that evolve slowly, over decades or even centuries.}
\entry{Recurrent Nova}{A nova observed to undergo more than one thermonuclear eruption in recorded history.}
\end{marginnote}

The binary remains intact after a nova eruption, so all novae are expected to recur, with periods ranging from $\sim 1$ yr to $\gtrsim 10^{7}$ yr as required to accrete and accumulate a critical-mass envelope \citep[e.g.,][]{Yaron+05}.
``Recurrent" novae are the subset of systems observed to undergo more than one eruption in recorded history, but are otherwise driven by the same physical processes as other novae.
Hereafter, we use ``novae" as a general term for all thermonuclear novae (classical and recurrent), unless additional specificity is warranted.

\begin{textbox}[b]\section{BOX 1: Helium Novae}
The vast majority of classical and recurrent novae are the result of runaway hydrogen burning.  However, qualitatively similar phenomena can occur on WDs due to unstable helium burning \citep{Taam80,Shen&Bildsten09}.  The best ``helium nova" candidate is V445 Puppis \citep{Ashok&Banerjee03}.  Early phases of its 2000 eruption showed notably hydrogen-deficient spectra containing prominent He, C, and Fe features \citep{Iijima&Nakanishi08}.  Near-IR imaging of the ejecta showed a tightly collimated bipolar outflow and an equatorial dust disk (Figure \ref{fig:opticalimages}; \citealt{Woudt+09}), similar to the ejecta geometry frequently observed in ordinary hydrogen novae (\S\ref{sec:imaging}), but with higher ejecta speeds 
($\sim$8500 km/s).  Controversy initially surrounded the interpretation of this event, because the non-detection of a post-nova central object was taken as evidence for a destructive explosion \citep{Goranskij+10}.  
However, the dust around the eruption site has finally begun to clear, and a photometric period has recently been detected at $P \approx 1.8-3.7$\,d, consistent with the orbital period of a He star transferring mass to a WD companion (D.~Steeghs, private communication).  
 \end{textbox}

\section{MASS LOSS IN NOVAE}
\label{sec:massloss}

Mass loss is key to driving the observational appearance of nova eruptions.  
The mass-loss rate and outflow velocity as functions of time---whether impulsive or in the form of a sustained outflow---control how the light curve and spectral energy distribution evolve.  Stochastic or secular variability in these properties can generate internal shocks within the ejecta \citep{OBrien&Lloyd94,Mukai&Ishida01, Chomiuk+14}. 
Mass loss can carry angular momentum away from the binary, with implications for the long-term evolution of CVs and thermonuclear SN progenitors (\S\ref{sec:longterm}).  Despite decades of theoretical and observational work, the mechanisms giving rise to nova outflows, including the role of the binary companion, remain a topic of debate.  

In addressing the issue of mass loss, we first describe the root cause of the eruption---the thermonuclear runaway (\S\ref{sec:TNR}) and the subsequent steady burning phase (\S\ref{sec:SSS}), and observational tests of the basic theory.  We then discuss several proposed mass-loss mechanisms in \S\ref{sec:massejection}.  In $\S\ref{sec:observations}$ we describe observations that probe mass loss in novae across the electromagnetic spectrum.

\subsection{Nuclear Burning in Novae}

The outcome of hydrogen-rich mass transfer depends on several properties of the accreting WD including its mass ($M_{\rm WD}$), accretion rate ($\dot{M}$), core temperature, and the composition of the accreted gas (e.g., \citealt{Fujimoto82a,Starrfield+00,Townsley&Bildsten04,Yaron+05,Nomoto+07,Shen&Bildsten09b,Chen+19}).
When significant mixing occurs between the WD and the accreted layer, the outcome will also depend on the composition of the WD being carbon/oxygen (CO) or oxygen/neon (ONe).  CO WDs are the expected remnants of stars with zero-age main sequence masses $M_{\rm ZAMS} \lesssim (7-8)\, M_{\odot}$, while ONe WDs come from stars with $(7-8)\, M_{\odot} \lesssim M_{\rm ZAMS} \lesssim (9-10)\, M_{\odot}$;  these ranges are metallicity-dependent and theoretically uncertain \citep{Doherty+15}. 

\begin{marginnote}
\entry{$\dot{M}$}{The accretion rate onto the WD, driven by mass transfer from a binary companion.}
\end{marginnote}

\begin{figure}[t]
\includegraphics[width=0.8\textwidth]{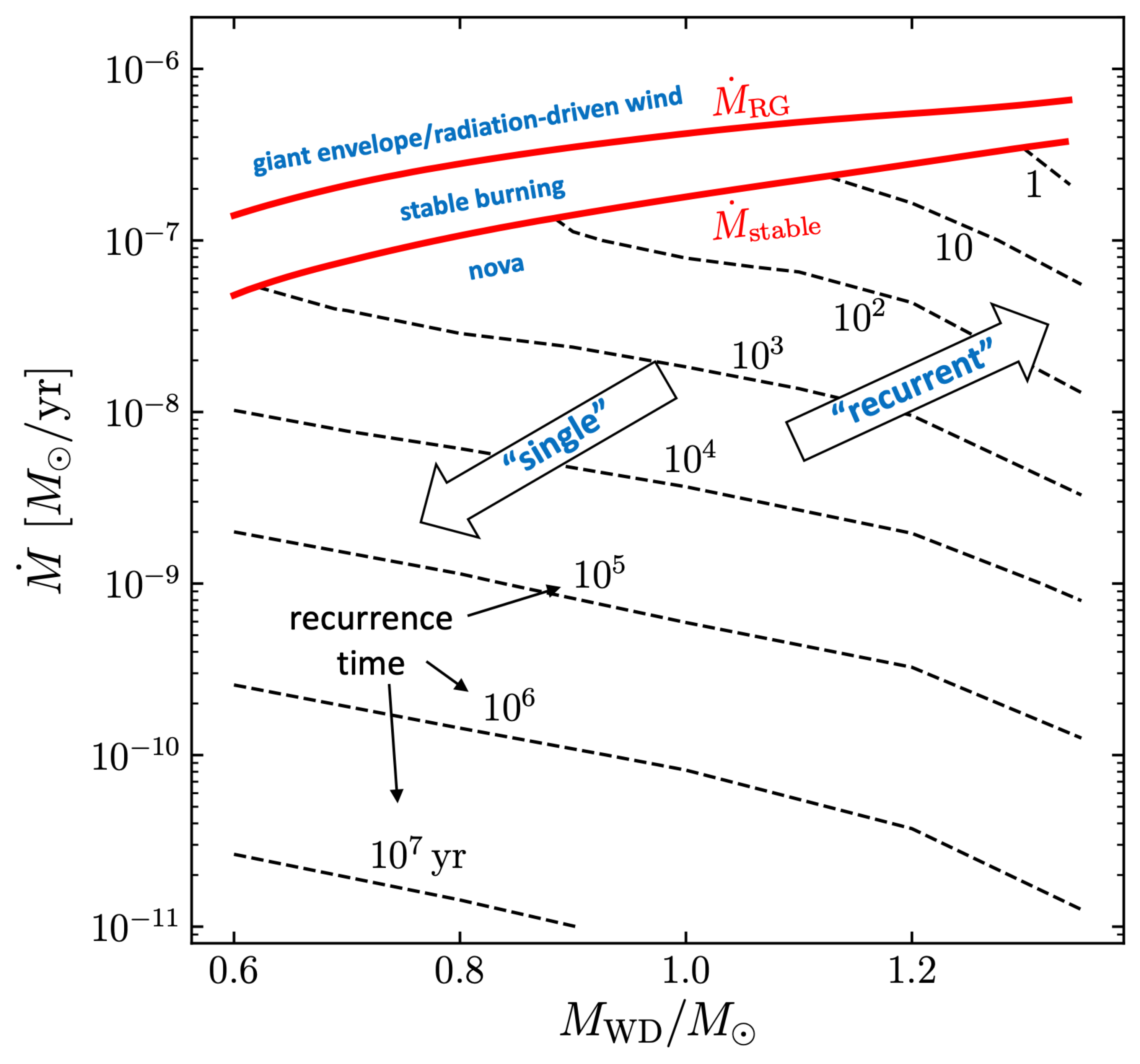}
\vspace{-0.15in}\caption{Depending on the WD mass and accretion rate ($\dot{M}$), a wide variety of phenomena are theoretically expected.  The lower red line is $\dot{M}_{\rm stable}$, the lowest accretion rate leading to stable burning for a given WD mass. 
At higher accretion rates, $\dot{M} > \dot{M}_{\rm RG}$, hydrogen will still burn stably, but more slowly than the matter is accreted. The matter will pile up to form a red giant-like structure or it must be carried away in a wind. Below $\dot{M}_{\rm stable}$, burning is unstable, resulting in nova eruptions.
Dashed black lines are lines of constant nova recurrence time (labelled in units of years). Nova ejecta masses can be estimated as the product of recurrence time and $\dot{M}$.  The stability lines are taken from \citet[][reproduced with permission $\copyright$AAS]{Wolf+13}.  Recurrence timescales are calculated using the stellar evolution code \texttt{MESA} (v12115; \citealt{Paxton+11}) in a similar procedure as described in \citet{Wolf+13}, but with a central core temperature of $10^7 \, {\rm K}$ and covering a larger range in accretion rates.}
\label{fig:wolf13a}
\end{figure}

A nova eruption is due to a ``thermonuclear runaway" (TNR), the unstable burning of hydrogen on the WD surface \citep{Gallagher&Starrfield78}.
A TNR is the outcome of mass accretion at low rates, $\dot{M} \lesssim \dot{M}_{\rm stable}$, where the threshold value for steady, thermally stable burning, $\dot{M}_{\rm stable} \approx 4 \times (10^{-8}-10^{-7})
M_{\odot}$/yr, is an increasing function of $M_{\rm WD}$ (Figure \ref{fig:wolf13a}). 
For accretion rates just above $\dot{M}_{\rm stable}$, hydrogen burns stably at the rate it is accreted, powering persistent soft X-ray emission (\citealt{Fujimoto82b, Nomoto82, Nomoto+07, Shen&Bildsten07, Wolf+13}).
For still higher accretion rates ($\dot{M} \gtrsim 3\,\dot{M}_{\rm stable}$), the fusion rate cannot match the accretion rate, and the accreted matter piles up into an extended red-giant-like structure, or it is lost in a radiation-driven wind \citep{Hachisu+96}.  The CV hosts of classical novae typically have mass transfer rates $\dot{M} \approx 10^{-10}-10^{-8}\, M_{\odot}$/yr \citep{Patterson84}, which place them in the unstable regime.

\subsubsection{The Thermonuclear Runaway}
\label{sec:TNR}
Novae are challenging to model because myriad hydrodynamical effects such as convection, mixing, and instabilities combine with a variety of nuclear processes.  However, the basic mechanism of the TNR can be understood through a comparison of several physical timescales:
\begin{itemize}
\item{The accretion/recurrence timescale, $\tau_{\rm rec} = M_{\rm acc}/\dot{M}$, depends on the accretion rate ($\dot{M} \approx 10^{-11}-10^{-7}\, M_{\odot}$/yr) and the critical accreted mass for TNR ($M_{\rm acc} \approx 10^{-7}-10^{-3}\, M_{\odot}$). 
It is typically $\sim 10^{4}-10^{7}$ years, but it can be as short as 1 year.}
\item{The nuclear timescale, $\tau_{\rm nuc} = c_P T/\epsilon_{\rm nuc}$, where $T$ is temperature, $c_P$ is the specific heat, and $\epsilon_{\rm nuc}$ is the nuclear energy generation rate from hydrogen burning in the burning layer at the base of the accreted envelope.
During peak burning in the TNR, $\tau_{\rm nuc}$ can become as small as seconds if there is significant dredge-up from the underlying WD, which provides catalytic seed nuclei for CNO-cycle reactions.}
\item{The convective overturn timescale, $\tau_{\rm eddy} = H_P / v_{\rm conv}$, is the timescale over which eddies can rise over a pressure scale-height, $H_P$, and redistribute the heat generated in the thin burning layer at the base of the accreted material.  Convective velocities, $v_{\rm conv}$, can reach hundreds of km/s at the peak of burning, implying a convective timescale of $\sim$seconds, somewhat faster than the shortest nuclear timescales.}
\item{The hydrodynamic timescale, $\tau_{\rm hd} = H_P/c_s$, required for hydrostatic readjustment of the pressure scale-height given the sound speed, $c_s$. At the peak of burning, $\tau_{\rm hd}$ is a few tenths of a second and is the shortest relevant timescale.}
\end{itemize} 

\begin{marginnote}[]
\entry{$M_{\rm acc}$}{The mass transferred to the WD between novae.}
\entry{$\tau_{\rm rec}$}{Recurrence time between novae, i.e., time required to build up, through accretion, the critical $M_{\rm acc}$ to initiate the TNR.}
\end{marginnote}

During the accretion phase, $\tau_{\rm rec} \ll \tau_{\rm nuc}$ (for most novae, models predict proton-proton and $^3$He reactions occur over much of $\tau_{\rm rec}$, but at a low rate; \citealt{Kovetz&Prialnik85, Townsley&Bildsten04, Shen&Bildsten09b}). 
The envelope mass increases, along with the density and temperature at its base.  After enough mass has been accreted, energy release from nuclear burning becomes too rapid to be transported by radiation and electron conduction, and the TNR phase begins with the onset of convection \citep{Fujimoto82b}.  The TNR is a product of the thin shell instability, brought on by the relatively small value of the specific thermal energy compared to the specific gravitational binding energy (i.e., the envelope cannot initially expand enough to quench the TNR). While electron degenerate conditions can aid the onset of the TNR due to the density's lack of response to an increase in temperature, novae are ignited under ideal gas conditions as well.  This occurs at higher accretion rates $\gtrsim 3 \times 10^{-9} \, M_\odot \, /{\rm yr}$.  An increase in the temperature under such conditions does lead to a stabilizing decrease in the density, but since the relevant nuclear burning rates scale much more strongly with temperature than density, the TNR can still occur.

The nuclear burning becomes very vigorous during the convective burning phase associated with the TNR, but the timescale ordering $\tau_{\rm hd} < \tau_{\rm eddy} < \tau_{\rm nuc}$ is always preserved. No deflagration or detonation occurs, because convection is always able to efficiently redistribute the heat generated in the burning layer throughout the envelope. 
Eventually, the specific thermal energy in the envelope approaches the specific gravitational binding energy of the WD and the envelope expands, quenching runaway nuclear burning and causing a transition to a prolonged phase of steady burning.

Two types of nuclear timescales come into play during the TNR: those related to $\beta$ decays (yielding timescale $\tau_{\beta}$), 
and those related to proton capture reactions ($\tau_{(p,\gamma)}$).  
As the density and temperature in the accreted envelope approach critical levels, nuclear reactions transition to the CNO cycle, which first operates in equilibrium ($\tau_{\beta} < \tau_{(p,\gamma)}$).  However, as the temperature at the base of the accreted envelope reaches $\gtrsim 10^{8}\,$K, the timescales reverse ($\tau_{\beta} > \tau_{(p,\gamma)}$), and the CNO cycle is limited by the $\beta$-decay timescales (the ``hot CNO cycle").  
Convection transports the $\beta$-unstable nuclei to the outer cooler regions of the nova envelope, where they are preserved from destruction and available to decay later on (e.g.,~\citealt{Starrfield+16}).
The resulting energy release drives expansion of the envelope and may contribute to mass ejection (\S\ref{sec:impulsive}).

The critical mass of the hydrogen layer needed to trigger the runaway spans a wide range, $M_{\rm acc} \approx 10^{-7}-10^{-3}\,M_{\odot}$, and is a decreasing function of both $M_{\rm WD}$ and $\dot{M}$ (e.g.,~\citealt{Yaron+05,Wolf+13}).  For a fixed $M_{\rm WD}$, a higher value of $\dot{M}$ implies higher temperatures at a given $M_{\rm acc}$, so the ignition mass is smaller.  For a fixed $\dot{M}$, a higher $M_{\rm WD}$ means higher densities and temperatures for a given value of $M_{\rm acc}$, so the ignition mass is smaller.

The nova recurrence time, $\tau_{\rm rec} = M_{\rm acc}/\dot{M}$, is the time to build up the critical layer through accretion, and it also decreases with both $M_{\rm WD}$ and $\dot{M}$, as shown with dashed lines in Figure~\ref{fig:wolf13a}.  
Recurrence times can vary from $\sim$1 yr for rapidly accreting massive WDs approaching the Chandrasekhar limit ($M_{\rm Ch} \approx 1.4\,M_{\odot}$), to $\tau_{\rm rec} \gtrsim 10^{7}$ years for slowly accreting, low-mass WDs ($M_{\rm WD} \lesssim 0.8\,M_{\odot}$).  Because shorter recurrence times result in more frequent novae---and thus a higher detection rate in surveys---the observed nova sample is biased to higher WD masses and higher accretion rates \citep{Truran&Livio86, Ritter+91,Iben+92_mdot}.

Observations of novae generally affirm these theoretical predictions. Recurrent novae ($\tau_{\rm rec} \lesssim 100$ yr; \citealt{Schaefer10}) tend to have small ejecta masses and occur on high-mass WDs \citep[e.g.,][]{Diaz+10, Osborne+11, Orio+13, Page+15}---although there are confounding exceptions (e.g., T~Pyx; \citealt{Uthas+10, Nelson+14, Patterson+17}; see {\bf Box 7}). Observations of the steady-burning phase following the TNR (\S\ref{sec:SSS}) imply that ejecta mass scales inversely with WD mass, as theoretically predicted \citep{Wolf+13, Henze+14}.  However, other observational estimates of ejecta mass consistently yield values more than an order of magnitude greater than theoretically predicted, and show only weak correlations with other nova properties \citep{Roy+12, Tarasova19}. Much of this discrepancy may be addressed by more accurately and consistently correcting for the effects of clumping and aspherical geometries in the ejecta, which would drive observational estimates downward \citep{Ribeiro+14, Wendeln+17}.  

Spectroscopic studies of nova ejecta often show overabundances in elements such as carbon, oxygen, and neon \citep[e.g.,][]{Ferland&Shields78, Williams+85, Gehrz+98, Schwarz+01, Downen+13}. These metallicity enhancements cannot be the result of nuclear burning because the temperatures achieved during the TNR ($\sim$ few $\times 10^8$ K) are not high enough to synthesize such heavy elements. Instead, they indicate that mixing must take place between the accreted hydrogen envelope and the underlying CO or ONe WD \citep{Starrfield+78,Prialnik+78}.  Enrichment of heavy elements into the burning region is a,lso needed to generate outflows of sufficient mass and kinetic energy to be consistent with observations \citep{Starrfield+98,Jose&Hernanz98}.
The luminosity from nuclear burning achieved during the peak of the TNR is ultimately limited by $\beta$-decay timescales (and hence is temperature insensitive), but it does scale with the abundance of CNO nuclei in the burning region. 

Several mechanisms have been proposed to generate mixing at the interface between the accreted material and the underlying WD, which can operate gradually prior to the TNR or rapidly during the TNR.  Pre-eruption diffusion can generate moderate enrichment \citep{Prialnik&Kovetz84}, but may not have sufficient time to operate in high-$\dot{M}$ systems with short recurrence times ($\dot{M} \gtrsim 10^{-9}\, M_{\odot}$/yr; \citealt{Livio&Truran87}). 
Instabilities that feed off differential rotation in the accreted layer, such as the baroclinic instability, could also play a role in long-term gradual mixing (e.g., \citealt{Kippenhahn&Thomas78, Fujimoto93}). On the other hand, Kelvin-Helmholtz instabilities driven by turbulent convection could cause rapid mixing at the onset of the TNR, akin to the process of ``convective overshoot" in stellar evolution  \citep{Glasner+97}. Recent multi-dimensional hydrodynamical simulations have demonstrated that this process can reproduce observed levels of metal enrichment in nova envelopes (e.g., \citealt{Casanova+10,Jose+20}). 
However, debate remains whether such mixing will always be capable of mixing material through the helium-rich buffer layer generated after each nova by stable H-burning ($\S\ref{sec:SSS}$; e.g.,~\citealt{Iben+92_mixing, Starrfield+98, Denissenkov+13}). 
The amount of mixing, and how it depends on parameters like $\dot{M}$ and $M_{\rm WD}$, remains one of the largest uncertainties in modelling nova TNRs \citep[e.g.,][]{Starrfield+20}. 

Novae are not generally believed to be major contributors to Galactic-scale nucleosynthesis.  
While classical novae occur $\gtrsim 10^{3}$ times more frequently than SNe, they typically eject $\lesssim 10^{-5}$ less mass per event ($M_{\rm ej} \lesssim 10^{-4}\, M_{\odot}$ in novae versus $M_{\rm ej}  \gtrsim 1-10\, M_{\odot}$ in SNe).  Nevertheless, novae likely are major contributors of
isotopes produced by $\beta$-decay ``bottlenecks" in the CNO process,
such as $^{13}$C, $^{15}$N, and $^{17}$O \citep[e.g.,][]{Starrfield+72, Jose&Hernanz98}.  They may also be significant contributors 
of radioactive nuclei such as $^{22}$Na and $^{26}$Al \citep{Jose+97, Hernanz12}.  Recent spectroscopic observations have shown that novae can be prolific producers of $^{7}$Be and its decay product $^{7}$Li (e.g., \citealt{Tajitsu+15, Izzo+15}; see \citealt{DellaValle&Izzo20} for a review), confirming early theoretical predictions \citep{Starrfield+78b}.  
\citet{Jose&Hernanz07} and \citet{Starrfield+16} provide focused reviews of nucleosynthesis in novae.

\begin{textbox}[h]\section{BOX 2: Early UV/X-ray Flash from the TNR}
In addition to the X-ray emission from the later stages of novae ($\S\ref{sec:SSS}, \S\ref{sec:Xrays}$), a short-lived phase of UV/X-ray emission is predicted soon after the TNR, as the effective temperature $T_{\rm eff}$ rises prior to the envelope expansion \citep{Hillman+14}.  The luminosity of this early UV/X-ray flash is close to the Eddington limit and can last from hours to days for typical parameters.  Pre-maximum UV emission has been seen in some novae (e.g., \citealt{Cao+12}), but the early X-ray flash has yet to be unambiguously detected (\citealt{Morii+16,Kato+16}).
This is unsurprising given its relatively short duration and occurrence prior to the visual outburst that generally triggers nova discoveries.  Extant and future wide-field UV or X-ray monitors, such as MAXI \citep{Negoro+16} and
the {\it Einstein Probe} \citep{Yuan+18}, have a chance to detect this emission, which would confirm an important prediction of nova theory, and provide better constraints on the exact time of the TNR, the WD mass, and the amount of pre-existing circumbinary material (e.g., via X-ray absorption).
 \end{textbox}
 
\subsubsection{Supersoft X-rays from Sustained Nuclear Burning}
\label{sec:SSS}

Following the TNR, much of the accreted envelope expands and is eventually expelled from the binary (\S\ref{sec:massejection}). Once enough mass has been removed, the remaining envelope finds a much more compact hydrostatic solution and burns steadily until it reaches the minimum envelope mass for steady burning \citep{Schwarz+11, Wolf+13}.  During this steady-burning period, which lasts for days to years, the WD maintains a luminosity determined by the core mass--luminosity relation of 
\citet[][$L \approx 10^{37}-10^{38}$ erg/s]{Paczynski70},
which approaches the Eddington luminosity as $M_{\rm WD}$ approaches the Chandrasekhar mass $M_{\rm Ch} \simeq 1.4\, M_{\odot}$ (see \citealt{Gehrz+98} for a detailed discussion).  The WD's effective temperature:
\begin{equation}
T_{\rm eff} \approx \left(\frac{L}{4\pi \sigma R_{\rm WD}^{2}}\right)^{1/4} \approx 6 \times 10^5\ {\rm K}\ \left(\frac{L}{10^{38}\ {\rm erg/s}}\right)^{1/4} \left(\frac{R_{\rm WD}}{10^{9}\ {\rm cm}}\right)^{-1/2},
\end{equation}
ranges from $(0.3-1.5) \times 10^6$ K. $T_{\rm eff}$ is primarily dependent on $M_{\rm WD}$ \citep{Wolf+13}, because of the WD mass--radius relationship \citep{Nauenberg72}:
\begin{equation} \label{eq:wd_rm}
R_{\rm WD} \approx 8.9 \times 10^8\ {\rm cm} \left( \frac{M_{\rm WD}}{M_{\odot}} \right)^{-1/3}\ \left[1-\left (\frac{M_{\rm WD}}{M_{\rm Ch}}\right)^{4/3}\right]^{1/2}.
\end{equation}
The nuclear-burning WD's spectrum peaks in the extreme UV and very soft X-ray bands, and is therefore called a ``supersoft" source.

 \begin{marginnote}[]
\entry{Eddington Luminosity}{The critical luminosity, $L_{\rm edd} = 4\pi GMc/\kappa$, above which outwards radiation pressure exceeds gravity for material of mass $M$ and opacity $\kappa$.}
\end{marginnote}

This supersoft X-ray emission is not immediately visible after the TNR, because the dense ejecta absorb the WD's emission.  As the expanding ejecta drop in column density, the photosphere recedes through the ejecta until the supersoft emission is finally revealed (typically on timescales of weeks to months, primarily depending on the ejecta mass and expansion velocity; \citealt{Schwarz+11, Henze+14}).
This supersoft phase lasts for a time $t_{\rm off}$ that primarily depends on $M_{\rm WD}$ \citep{Starrfield+74, Sala&Hernanz05, Wolf+13}, after which the hydrogen layer is too low in mass to support a steadily burning solution, and the WD cools. 

This basic picture has been confirmed by X-ray and UV observations. Early space-based UV observations showed that novae remain UV bright months after eruption, even after their optical emission had substantially faded \citep{Gallagher&Code74}. A decade later, similar results were found for soft X-rays (\citealt{Ogelman+87}; see also \citealt{Shore+94}). 
High-resolution X-ray grating spectra obtained with {\it Chandra} and {\it XMM-Newton} confirmed that the emission arises near the WD surface \citep[e.g.,][]{Nelson+08,Rauch+10,Ness+11,Orio12}.  In some cases the soft X-ray flux is continuum emission directly from the WD atmosphere, while in other cases it is in the form of strong emission lines, likely powered by photospheric emission that is obscured from the line of sight (e.g., \citealt{Ness+13} and references therein).  
High-cadence and long-term monitoring of novae by the {\it Neil Gehrels Swift Observatory}  \citep{Gehrels+04} has substantially improved our understanding of X-ray emission from novae in recent years (see \citealt{Osborne15} and \citealt{Page+20} for reviews). 

A measurement of the temperature of the supersoft source is the most direct strategy for estimating the WD mass during a nova eruption \citep{Wolf+13}, although caution should be exercised in fitting models to the X-ray spectrum \citep{Krautter+96}. 
The supersoft ``turn-off time", $t_{\rm off}$, is sometimes taken as a proxy for WD mass, and broad agreement exists between theoretical predictions and the observed $T_{\rm eff}-t_{\rm off}$ relation \citep{Henze+11, Schwarz+11, Wolf+13}. However, there are significant discrepancies for lower-mass WDs \citep{Henze+14}, perhaps because $t_{\rm off}$ depends on the uncertain mechanism of envelope removal (\S\ref{sec:massejection}; \citealt{Wolf+13}).  Constraints on the WD mass, spin period, and properties of the burning layer could, in principle, also be derived from the short-period ($30-70 \,$s) supersoft oscillations that have been observed for several novae (e.g., \citealt{Ness+15}); however, attempts to match these oscillations to theoretical models have so far been unsuccessful \citep{Wolf+18}.

\subsection{Mechanisms of Mass Ejection}
\label{sec:massejection}

The observable signatures of nova eruptions---and possibly even their overall radiated energy (\S\ref{sec:shockstheory})---are intimately tied to how mass is ejected from the binary.
An important clue to the mass loss mechanism(s) comes from the velocities of nova ejecta, which are observed to span a wide range $v_{\rm ej} \approx 200-7000$ km/s \citep[e.g.,][]{Munari+11, Naito+12, Darnley+16}, across different events and even within an individual nova eruption ($\S$\ref{sec:spectra}).  Most stellar outflows reach velocities comparable to the local escape speed, $v_{\rm ej} \approx v_{\rm esc} = (2GM/R_{\rm w})^{1/2}$ (e.g., 
\citealt{Castor+75}), where $R_{\rm w}$ is the outflow launching radius and $M$ is the central mass (in novae, typically dominated by the WD).  Turning this around, the ejecta velocities in novae imply outflow-launching radii of:
\be
R_{\rm w} \approx 2.7\times 10^{10}\,{\rm cm} \left(\frac{M}{M_{\odot}}\right) \left(\frac{v_{\rm ej}}{1000\, {\rm km/s}}\right)^{-2}.
\label{eq:Rw}
\ee 
For comparison, the radius of a cold WD is $R_{\rm WD} \lesssim 10^{9}$ cm (Eq.\ \ref{eq:wd_rm}), while the semi-major axis of a binary of period $P$ is:
\be
a_{\rm bin} \approx 3.5\times 10^{10}\,{\rm cm}\left(\frac{M}{M_{\odot}}\right)^{1/3}\left(\frac{P}{\rm hr}\right)^{2/3} .
\label{eq:abin}
\ee
This simple analysis implies that different physical mechanisms may drive slow ($\sim$few hundred km/s) and fast ($\sim$few thousand km/s) outflows in novae.  Indeed, several distinct mechanisms driving nova mass loss have long been proposed, described in \S\ref{sec:impulsive}--\S\ref{sec:RLOF}. 

\begin{marginnote}[]
\entry{$v_{\rm ej}$}{The ejecta velocity of material expelled in a nova eruption.}
\end{marginnote}

\subsubsection{Impulsive Ejection at the TNR}
\label{sec:impulsive}
Observations of nova ejecta are often interpreted as a single, impulsive mass ejection coincident with the TNR (e.g., \citealt{Seaquist&Bode08, Mason+18}).
Indeed, some hydrodynamical models of nova eruptions find that a portion of the envelope is ejected in the minutes to hours following the TNR, powered by radioactive heating from $\beta$-unstable nuclei  \citep{Starrfield+08}. 
Early models found that this material is expelled as a shock-driven shell \citep{Sparks69, Prialnik86},
but later this picture was revised to a wind whose mass-loss rate declines by an order of magnitude over just a few hours \citep{Prialnik&Kovetz92}.

The mass and velocity of this initial ejection are primarily determined by the amount of mixing between the accreted envelope and the underlying WD \citep{Starrfield+78, Starrfield+98}.
Even when significant prompt ejection occurs in nova models, only a fraction of the envelope is expelled. 
If this were the only form of mass loss from novae, the sustained-burning supersoft X-ray
phase would last for longer than observed ($\sim$centuries; \citealt{Starrfield+78}). Combined with the $\sim$Eddington luminosities of novae after eruption, this implies a significant role for prolonged winds driven by radiation pressure (\S\ref{sec:wind}).

Rotation of the WD could in principle play a role in shaping the geometry of the prompt ejecta (e.g., \citealt{Porter+98, Scott00}).  If the WD is rotating rapidly, the pressure of the burning layer---and hence the peak temperature achieved during the TNR---will depend on latitude, being significantly higher at the rotational poles than the equator.  As the rate of energy production is extremely sensitive to temperature, material ejected from the earliest phases of the eruption could show departures from spherical symmetry.  Rotation could also play a role in shaping any centrifugally driven mass loss, if the expanding nova envelope---instead of simply conserving angular momentum---maintains co-rotation with the rapidly spinning WD surface via efficient magnetic coupling \citep{Zhao&Fuller20}.

\subsubsection{Prolonged Optically Thick Winds}
\label{sec:wind}

The highest velocities seen in novae, $v_{\rm ej} > 1000$ km/s, imply small launching radii $R_{\rm w} \lesssim a_{\rm bin}$.  A compelling physical model for such fast outflows is an optically thick wind (i.e., where acceleration occurs deep below the photosphere) driven by radiation pressure  from the near-constant luminosity of the nuclear-burning WD \citep{Friedjung66, Bath&Shaviv76}.
The Eddington luminosity at a particular location in the nova envelope depends on temperature- and density-dependent mean opacities (as estimated e.g., in the OPAL tables; \citealt{Iglesias&Rogers96}), and the most important wind driver is the iron opacity bump,
which occurs at a temperature of $T \approx 1.6\times 10^{5}$\,K in the nova envelope \citep{Kato&Hachisu94}.  

As the mass of the hydrogen-rich envelope gradually decreases due to both nuclear burning and wind mass loss, the wind mass-loss rate also decreases \citep{Kato&Hachisu94}.
This makes the photosphere recede to smaller radii and higher temperatures, and the wind velocity increases as the wind-launching radius moves inward. Once the photospheric temperature exceeds $1.6\times 10^{5}$ K, there are no more significant sources of opacity, and the wind ceases \citep{Kato&Hachisu94}. The duration of the optically thick wind phase can be days to months, with lower $M_{\rm WD}$ and larger ignition masses leading to longer wind phases.

Essentially all hydrodynamical models of nova eruptions to date are one-dimensional.  Even putting aside the potential for large-scale asphericity in nova ejecta, this simplification may gloss over important physics in nova winds.   Multi-dimensional (magneto)hydrodynamical models of massive star envelopes with radiative transport reveal that, near the Fe opacity bump, the interplay between convection and radiative transport is complex.  Multi-dimensional effects and magnetic fields lead to clumpy envelope structures that qualitatively differ from those predicted by one-dimensional models (e.g., \citealt{Jiang+17}).  If similar physical processes occur in nova outflows, they could drive the inferred clumpiness of nova ejecta (\S\ref{sec:spectra}, \S\ref{sec:imaging}).

Some novae reach peak luminosities that exceed the electron-scattering Eddington luminosity by up to an order of magnitude ($\sim 10^{39}$ erg/s; e.g., \citealt{Duerbeck81, Schwarz+98, Schwarz+01, Aydi+18a, Skopal19}).\footnote{\citet{Kato&Hachisu05, Kato&Hachisu07} propose that super-Eddington luminosities can be attained by including free-free emission in the wind.  However, this violates energy conservation unless the wind is heated from below by some mechanism at a rate which itself exceeds the Eddington luminosity.}  
To achieve photon luminosities in excess of the Eddington limit requires an additional energy source in the hydrostatic WD atmosphere that is itself super-Eddington \citep{Quataert+16}.  A substantial contribution to the luminosity of novae may come from internal shocks within the ejecta ($\S\ref{sec:shockoptical}$), but even in this case super-Eddington {\it kinetic} luminosities would be required of the outflow.  Alternatively, apparently super-Eddington luminosities can be achieved with nuclear burning, if the Eddington luminosity is increased due to a decrease in the effective opacity ($\kappa$; $L_{\rm edd} \propto \kappa^{-1}$), which might be achieved if the WD atmosphere is clumpy or ``porous" (e.g., \citealt{Shaviv01}).

\subsubsection{Common Envelope-Like Mass-Loss}
\label{sec:RLOF}

The orbital motion of the binary is another potential source of energy \citep[e.g.,][]{MacDonald80}.  Around light curve maximum, the nova photosphere reaches radii $\approx 10^{12}-10^{13}$ cm, engulfing the binary system for novae with main sequence companions ($P \lesssim$ 12 hr). The proposed picture is broadly similar to a ``common envelope" phase, in which a compact secondary star enters the dilute envelope of a giant companion following a phase of unstable binary mass transfer \citep{Paczynski76, Ivanova+13}.  The influence of the binary should be particularly apparent for the lowest ejecta velocities, $v_{\rm ej} \lesssim 1000$ km/s, for which the inferred launching radii reside near or exterior to the companion orbit ($R_{\rm w} \gtrsim a_{\rm bin}$; Eq.~\ref{eq:Rw} and \ref{eq:abin}) .  

Early generations of opacity tables implied that radiation-driven winds were insufficient to eject the envelope, and works like \citet{Macdonald+85} and \citet{Shankar+91} found that frictional drag in this common-envelope phase could power mass ejection in slow novae.
However, other models that also included radiation-driven winds found that the envelope density near the binary orbit was too low for frictional heating to play a significant role \citep{Kato&Hachisu91b, Kato&Hachisu91a}.  With the inclusion of the Fe opacity bump, radiation-driven winds seemed sufficient to remove the envelope \citep{Kato&Hachisu94}. Studies of the common-envelope phase in novae fell out of fashion, although \citet{Kato&Hachisu11} find that it may play a role in the slowest novae.  However, we note that even in the common envelope community, the efficiency and timescale of envelope removal remain uncertain \citep{Ivanova+13}, and modern common envelope simulations fail to unbind stellar envelopes without contentious additional sources of energy like recombination \citep{Ivanova18}, jets \citep{Soker17}, or pulsation-driven shocks \citep{Clayton+17}. 

In principle, nature has provided a relatively clean test of the role of frictional drag in nova ejections, as embedded novae driven by wind accretion and with large orbital separations should not undergo a common envelope phase. Indeed, embedded novae seem to be polarized to two extremes: very fast novae that evolve over just a week or so, and symbiotic novae that are among the slowest evolving transients known ($\sim$ decades to centuries; \citealt{Mikolajewska10}). The lack of embedded novae that evolve on intermediate timescales ($\sim$ months) may be an indication that nova envelope removal is less efficient in long-period systems, due to the lack of frictional drag \citep{Kenyon&Truran83}. 

Even if frictional drag from the binary is not a significant contributor to the expulsion of the envelope, binary orbital motions may still shape the ejecta, concentrating them in the equatorial plane \citep{Hutchings72, Livio+90}.
As the WD envelope expands to encompass the binary orbit following the TNR, it will be spun-up and preferentially focused towards the binary orbital plane by centrifugal forces and the companion star's gravity.  
Mass loss may preferentially occur through the minimum in the gravitational potential near the outer $L_2$ Lagrange point, located in the orbital plane on the far side of the secondary companion.  Mass that leaks through this ``nozzle" emerges from the binary in a sprinkler-like spiral pattern and is accelerated outwards by a combination of pressure gradients and non-axisymmetric torques from the binary motion \citep{Shu+79,Pejcha+16a}.  
As discussed in \S\ref{sec:imaging}, novae are commonly observed to have bipolar outflows and equatorial rings/disks, 
which suggest that the binary orbit plays an important role in shaping, if not driving, outflows during at least some phases of the eruption.  Spatially resolved images suggest that the remnants of slower novae are more elongated than those of higher-velocity novae \citep[e.g.,][]{Slavin+95, OBrien&Bode08}, as expected if orbital motion shapes the ejecta.

\subsection{Multi-Wavelength Observations of Mass Loss}
\label{sec:observations}

\subsubsection{Optical Light Curves}
\label{sec:lightcurves}
Novae are generally discovered as optical transients, and a large fraction of their total radiative output emerges in the optical.  The luminosity increases by $8-15$ magnitude from pre-outburst to peak in the $V$ band, typically over just a few days \citep{Payne-Gaposchkin57, Warner95}.  Peak absolute magnitudes are generally in the range $M_{V} = -5$ to $-10$ mag \citep{Shafter17, Ozdonmez+18}, implying luminosities close to---or in excess of---the WD Eddington luminosity $L_{\rm edd} \approx 10^{38}$ erg/s.  

\begin{figure}[t]
\includegraphics[width=4in]{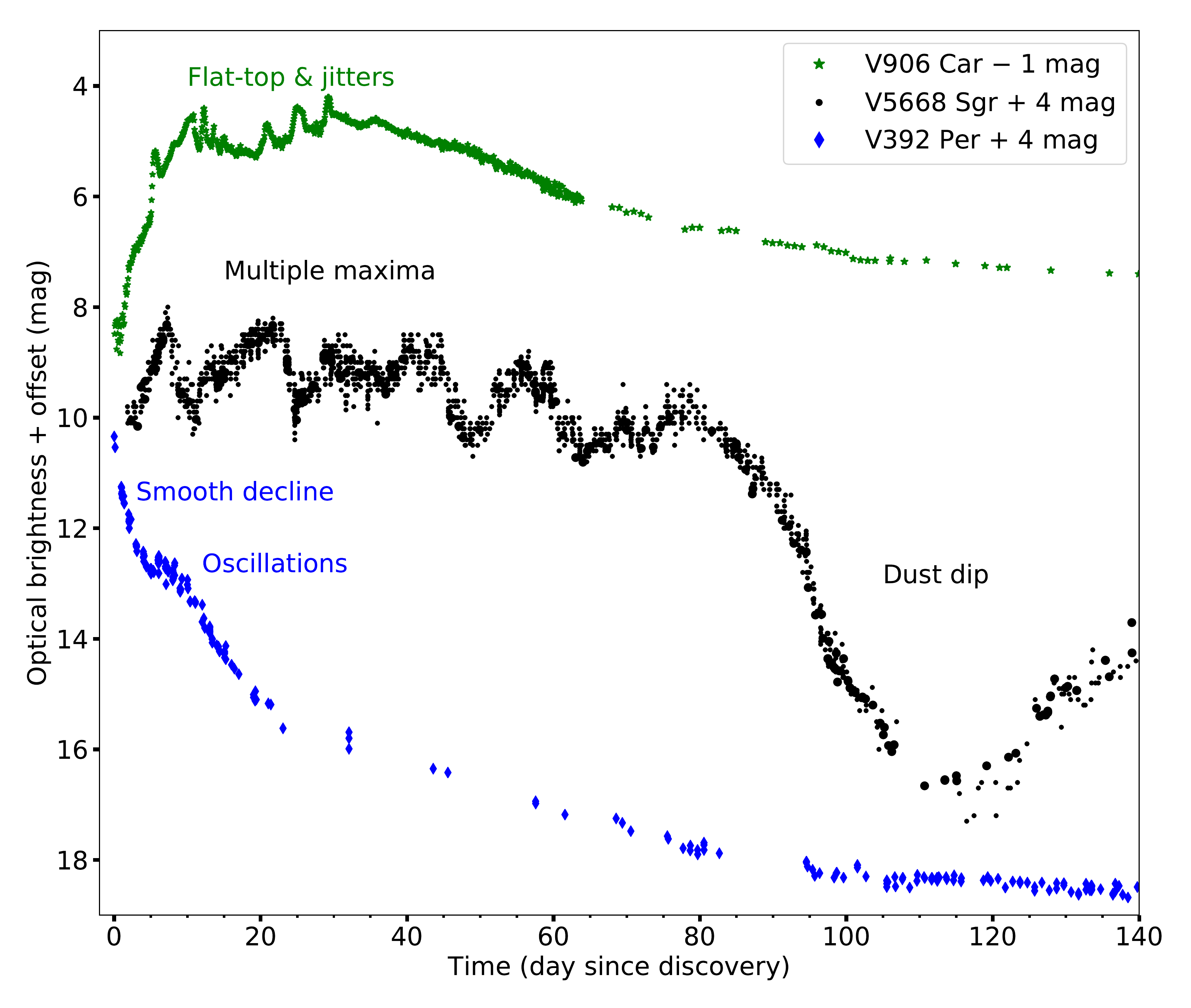}
\vspace{-0.3in}\caption{A selection of optical light curves of three novae (all detected by \emph{Fermi}-LAT), illustrating the broad range of nova light curve morphologies. Common features are marked---several of which remain unexplained. V906~Car data are in the $R$ band, from the \emph{BRITE} satellite and the Stonybrook/SMARTS atlas of novae \citep{Aydi+20, Walter+12}. The V5668~Sgr light curve is composed of $V$ (larger circle symbols) and Visual band (smaller circles) data from the AAVSO \citep{Kafka20}. The V392~Per data are in the $V$ band, as reported by the AAVSO. Figure created by E.\ Aydi.}
\label{fig:rogues}
\end{figure}

To first order, a simple nova model of expanding ejecta and constant bolometric luminosity powered by sustained nuclear burning on the WD (\S\ref{sec:SSS}) predicts a smooth decline like that seen over most of V392~Per's evolution in Figure \ref{fig:rogues}.  The optical emission declines because the peak of the spectral energy distribution shifts blueward, into the UV and X-ray, as the ejecta thin and the photosphere recedes.  At late times, the ejecta become optically thin and the optical luminosity is dominated by emission lines.  While this has been the standard picture for decades \citep[e.g.,][]{Gallagher&Starrfield76, Shore+94}, recent work has demonstrated that reprocessed emission from internal shocks may also contribute or even dominate the optical luminosity in some cases (see \S\ref{sec:shockoptical} for more discussion).

Nova light curves are often labeled according to their speed class, quantified as the time  to decay by 2 or 3 visual magnitudes from peak brightness (denoted as $t_2$ or $t_3$, respectively; \citealt{Payne-Gaposchkin57}).  Speed classes range from ``very fast" ($t_2 <$ 10 days) to ``very slow" ($t_2 \gtrsim 150$ days).  Roughly speaking, the speed class is related to the timescale for envelope expansion and removal; a nova fades faster for lower ejecta masses or higher expansion velocities.
Theory predicts that $M_{\rm ej}$ is a decreasing function of WD mass (\S\ref{sec:TNR}), so faster novae are typically thought to arise from more massive WDs.
However, the relationship between speed class and WD mass has relatively little direct empirical proof and should be applied with caution. Also note that $t_2$ is far from a perfect metric, given the complex evolution of many nova light curves. For example, the optical brightness of V5668~Sgr drops two magnitudes below its maximum several times during its evolution (Figure \ref{fig:rogues}).

\citet{Strope+10} present a catalog of 93 well-observed nova light curves, mainly from the American Association of Variable Star Observers database (AAVSO; \citealt{Kafka20}), that they use to classify light curves based on both their speed class and light curve shape.
While many novae (38\% of their sample) show smoothly declining light curves as might be expected from a simple nova model, they note that the rest of the sample show curious features in their light curves like flat tops, oscillations, and jitters.  Figure \ref{fig:rogues} shows three optical light curves illustrating these morphologies.  With the exception of ``dust dips" ($\S\ref{sec:dust}$), most of these features currently lack compelling theoretical explanations. Increasingly, the diversity and complexity of nova light curve structure---and the present lack of explanation for them---is being highlighted by the higher cadence and precision light curves from space-based facilities like  the \emph{Solar Mass Ejection Imager} \citep{Hounsell+10, Hounsell+16}, \emph{STEREO HI-1B} \citep{Eyres+17,Thompson17}, \emph{BRITE} \citep{Aydi+20}, and, in the future, \emph{TESS}.

For example, many novae exhibit multiple distinct optical maxima, also called ``jitters" or ``flares" (e.g., \citealt{Bianchini+92, Strope+10, Walter+12, Aydi+19, Aydi+20}).  During optical flares, the radius of the photosphere often appears to temporarily expand \citep[e.g.,][]{Tanaka+11, Munari+15, Aydi+19}---behavior which is accompanied by the appearance of new absorption line systems, suggestive of distinct mass ejection events (see \S\ref{sec:spectra}; although \citealt{Williams16} tries to explain them with ballistic, expanding clumps).  
There are several mechanisms that could in principle give rise to such variability in the outflow properties.  One is a change in the nature of the mass-loss mechanism, like a global transition from a slower, quasi-hydrostatic common envelope phase to a faster wind \citep{Kato&Hachisu11}.  Alternatively, some studies have found that contraction of the envelope may accelerate nuclear burning at the base of the envelope, 
triggering envelope re-expansion and a second mass-loss episode \citep[e.g.,][]{Prialnik&Livio95, Hillman+14}, although this has not been found by other work \citep[e.g.,][]{Townsley&Bildsten04,Denissenkov+13}.  As we shall discuss in \S\ref{sec:gammaray}, there is evidence from gamma-ray observations that optical flares in at least some cases are powered by internal shocks (see \citealt{Sanyal74} for an early discussion of this scenario). 

\begin{textbox}[h]\section{BOX 3: Maximum Magnitude--Rate of Decline Relationship?}
An empirical relationship has long been claimed between the absolute optical magnitude of novae at maximum, $M_V$, and light curve decline time, such that more luminous novae have shorter $t_2$ or $t_3$ \citep[e.g.,][]{McLaughlin45, deVaucouleurs78, Capaccioli+89, DellaValle&Livio95}.  Although physical interpretations of this ``maximum magnitude-rate of decline" (MMRD) relation have been proposed \citep{Livio92}, they rely on model-dependent predictions for how the peak nova luminosity scales with the WD mass.   The MMRD relationship is frequently used to determine distances to Galactic novae, if the visual extinction is known.  However, recent observations show a significant population of outliers from this relationship that have called the existence of the MMRD relation into question (particularly fast and faint novae; \citealt{Kasliwal+11, Shara+17_m87}).  There are also theoretical reasons to believe nova properties should not conform to a one-parameter family \citep[e.g.,][]{Prialnik&Kovetz95}.  Given this uncertainty in the MMRD, substantial and hard-to-quantify uncertainties exist in the distances to most Galactic novae, although there have been recent promising advances using three-dimensional dust maps \citep{Ozdonmez+16} and \emph{Gaia} astrometry \citep{Schaefer18}.
 \end{textbox}

\subsubsection{Optical Spectra}
\label{sec:spectra}
Novae have been studied using optical spectroscopy for more than a century \citep{McLaughlin43, Payne-Gaposchkin57}.  Over the course of an eruption, nova spectra transition from being photosphere-dominated to emission-line dominated as the ejecta thin, and the ionization states of the lines generally increase.
The widths, shapes, and evolution of spectral features hold information about ejecta dynamics and morphology. Here we focus on spectroscopic constraints on mass ejection in novae, but note that the relative strengths of emission lines also constrain the abundances of the ejecta, and can reveal whether the nova occurred on a C/O or O/Ne WD (see \citealt{Gehrz+98} for a review of abundance measurements).
The optical spectra of novae are strongly affected by physical processes that occur in the UV, where the bulk of the WD luminosity is emitted and the ejecta opacity from atomic transitions is highest.  
\citet{Shore08} provides a description of the UV spectral evolution that accompanies what is more commonly observed in the optical.

\begin{figure}[!t]
\includegraphics[width=5in]{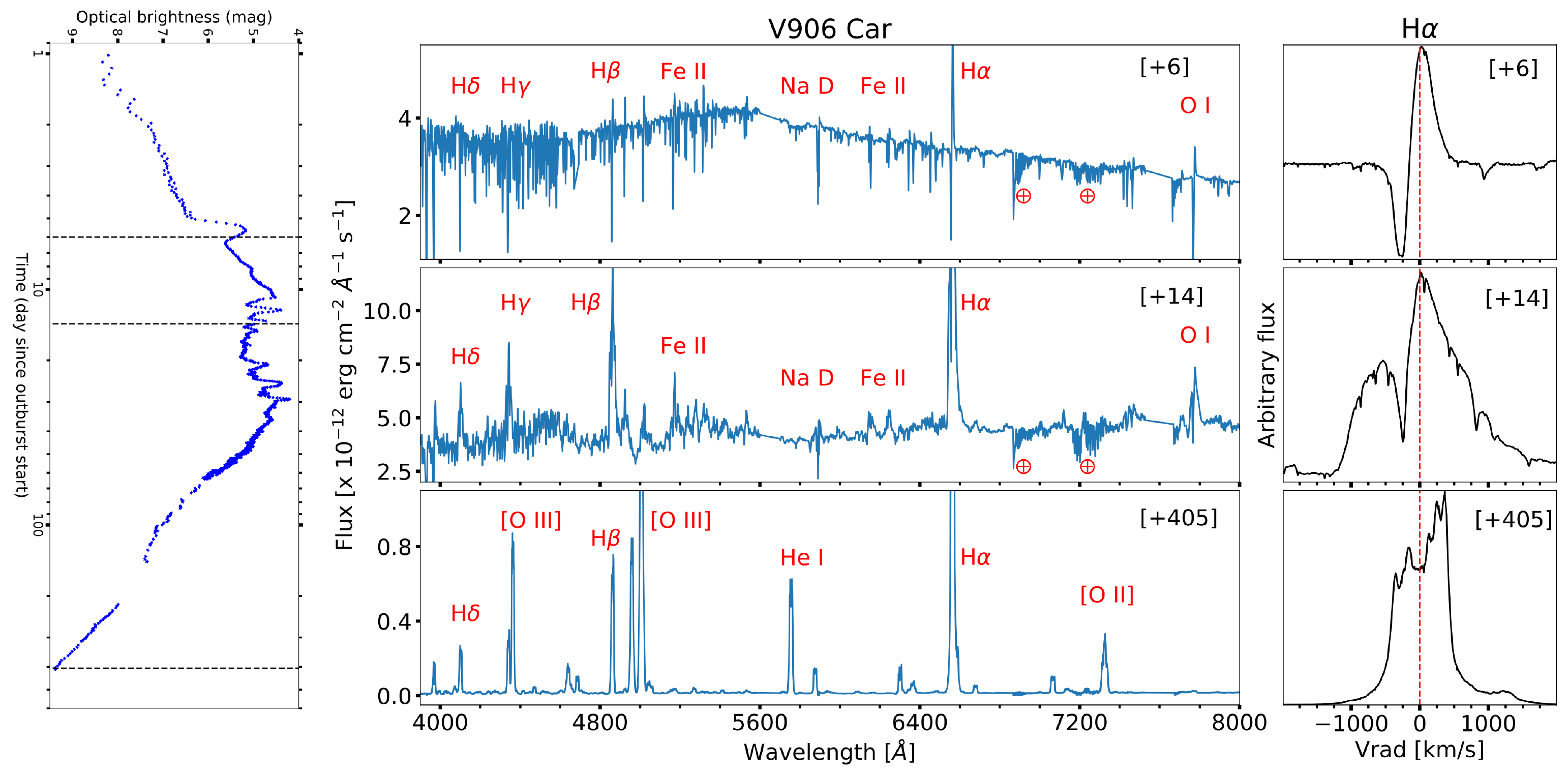}
\vspace{-0.2in}\caption{The optical evolution of nova V906~Car (2018), showing the optical ($R$-band) light curve in the left panel, optical spectra in the middle panel, and zoom-ins around the H$\alpha$ line in the right panel. The top spectrum is from 5 days before light curve maximum (6 days after the start of eruption) and shows a photospheric spectrum with relatively narrow P~Cygni profiles. The middle spectrum is from 3 days after light curve maximum, and shows the strengthening and broadening of emission lines. The absorption component from the previous spectrum is still superimposed on the emission line. The bottom spectrum is from more than a year after maximum, and shows a nebular spectrum dominated by high-excitation and forbidden emission lines.   Data from \citet{Aydi+20}; figure created by E.\ Aydi.}
\label{fig:v906spec}
\end{figure}

Novae caught early are generally first seen during a photospheric or  ``fireball" phase, in which the nova envelope expands and the optical light curve rises, as the photosphere radius grows and the effective temperature decreases \citep{Gehrz88, Hauschildt+94b}.  
This expansion is accompanied by the onset of mass loss, which is variously modelled as an impulsive ``shell'' ejection (e.g., \citealt{Shore14,Mason+18}), a continuous ``wind" \citep{Kato&Hachisu94}, or a combination of the two (e.g.,~\citealt{Friedjung87}).  
As the nova transitions through optical maximum, most spectral lines show P Cygni profiles (e.g, the top spectrum in Figure \ref{fig:v906spec}). With time, the emission component strengthens and the absorption weakens, consistent with a receding photosphere.
Around maximum, the ejecta are dense enough to rapidly recombine, and the Str$\ddot{\rm o}$mgren sphere powered by the nuclear-burning WD is ionization bounded, such that the outer ejecta are neutral \citep{Beck+90, Williams90}.  The dominant form of opacity is line blanketing by nearly neutral metals 
(sometimes called the ``iron curtain'', since most of the opacity is from Fe lines; \citealt{Shore&Aufdenberg93, Shore+94}).
Modelling of spectra before and around maximum finds that the density profiles of nova ejecta appear to require a sharp outer profile ($\rho \propto r^{-n}$ with $n \approx 15$ for the outermost ejecta; \citealt{Hauschildt+94b}) and a much shallower profile for the inner ejecta ($n \approx 3$; \citealt{Hauschildt+92}). This density profile implies that the optical photospheric radius can vary by over a factor of $\sim$100 with wavelength and thereby sample a range of temperatures and ionization states \citep{Hauschildt+95}; it is therefore often called a ``pseudo-photosphere".

As the ejecta expand and densities drop, the ionization front moves outwards (in both mass and physical coordinates), until eventually the ejecta become fully ionized \citep{Williams90}.  Spectral features transition from low ionization states around maximum (e.g., \ion{H}{i}, \ion{O}{i}, \ion{Na}{i}, \ion{Fe}{ii}; top and middle spectra of Figure \ref{fig:v906spec}) to highly ionized species. As the density drops, the features also migrate from permitted transitions to forbidden transitions (bottom spectrum of Figure \ref{fig:v906spec}). 
Eventually, the nova spectrum relaxes to a ``nebular" phase characterized by forbidden lines from ionization states expected in a $\sim 10^4$ K gas (e.g., [\ion{N}{ii}], [\ion{O}{iii}]). The ejecta can remain ionized long after the supersoft source turns off, ``frozen in" by the now low densities and long recombination times.

The evolution of the spectral line profiles is less well explained. 
Prior to optical peak, the emission lines show P Cygni profiles characterized by low velocities, $200-1000$ km/s (top H$\alpha$ profile in Figure \ref{fig:v906spec}).  Shortly after the nova reaches maximum light, a broad emission component emerges with roughly double the width of the initial P Cygni profile, while the absorption component from the pre-existing P Cygni profile remains superimposed on top of the broader emission (middle H$\alpha$ profile in Figure \ref{fig:v906spec}).
This general spectral evolution has long been recognized \citep[e.g.,][]{McLaughlin43, Gallagher&Starrfield78}, and has recently been revisited by \citet{Aydi+20b}, who show that it is nearly universal across diverse novae. 
The field has long grappled with interpreting the coexistence of broad emission and lower-velocity absorption, along with the abrupt transition of the spectral profiles over a matter of $\sim$days. The longest-standing and most cohesive explanation is to have two physically distinct---slow and fast---ejecta components (\S\ref{sec:imaging}; \citealt{McLaughlin47, Friedjung87, Friedjung11, Friedjung&Duerbeck93}).
Additional components can subsequently appear in the line profiles, usually at higher velocities (e.g., \citealt{Hutchings70a}). Their appearance is frequently associated with new ``flares" in the optical light curve (\citealt{Csak+05, Tanaka+11}).

Another curiosity is that spectral components are often seen to ``accelerate" to higher velocities (e.g., ~\citealt{Hutchings70b, Shore+11}). This phenomenon may reflect driving of clumps or shells by radiation pressure (e.g., \citealt{Williams+08}),
or it can potentially be explained by inner ejecta expanding faster than the outer ejecta (so as the photosphere recedes inwards in mass, it samples faster material; \citealt{Friedjung11}).  If the line-forming regions are thin shells compressed by internal shocks, momentum added to the shell by the WD outflow leads to acceleration (\citealt{Steinberg&Metzger20}; \S\ref{sec:shockstheory}). 

Spectroscopy also reveals that the morphologies of nova ejecta are complex. The ejecta are clumpy, as optical line ratios yield estimates of volume filling factors $\lesssim$0.1 in the line-emitting gas (e.g., \citealt{Ederoclite+06, Shore+13}). Structures in the spectral lines often persist at a particular velocity for months during eruption, and have been interpreted as originating in discrete clumps within the ejecta \citep{Williams13, Mason+18}. The coexistence of a wide range of ionization species indicates a striking range of densities and temperatures within the ejecta, and is again often explained as self-shielding clumps \citep{Williams94, Saizar&Ferland94}. A variety of evidence at other wavelengths also supports nova ejecta being highly clumpy.  This includes clearly resolved inhomogeneities in images of nova ejecta (\S\ref{sec:imaging}) and large-amplitude variability in the supersoft X-rays  due to time-variable absorption (likely by clumps in the ejecta; \citealt{Osborne+11, Page+20}). Finally, emission line profiles, especially in the nebular stage, often imply bulk asphericity in the ejecta. Double- or triple-peaked line profiles are common and usually interpreted as bipolar ejecta, sometimes surrounded by an equatorial torus or disk \citep{Hutchings72, Shore+13, Ribeiro+13}.

\begin{textbox}[h]\section{BOX 4: Transient Heavy-Element Absorption (THEA) Features}
Narrow absorption features from heavy elements (e.g., \ion{Fe}{ii}, \ion{Ti}{ii}, \ion{Cr}{ii}; FWHM $\approx$ 30--300 km/s) are seen in high-resolution spectra of some novae around and after maximum light \citep{Williams+08, Williams&Mason10}.  The physical origin of these ``THEA" features is uncertain.  \cite{Williams&Mason10} proposed that the absorbing material originates from a massive pre-existing circumbinary disk, inspired by the model of \cite{Taam&Spruit01}.
However, IR searches rule out such disks around most CVs, finding $\lesssim 10^{-10}$ M$_{\odot}$ in circumbinary material (e.g., \citealt{Hoard+14}). \citet{Williams12} proposed that the THEA lines originate in material irradiated or ablated from the secondary star during the nova; however, simulations find that mass loss from the secondary is likely not sufficient to produce the THEA lines \citep{Figueira+18}. 

The possibility remains that the THEA lines are associated with the early slow ejecta (e.g., the narrow  P~Cygni component in the top spectrum in Figure \ref{fig:v906spec}; \citealt{Aydi+20b}).
Some novae show hints that the THEA features are not present at the start of the eruption \citep{Williams12}, and the velocity and timing of their appearance matches the ``pre-maximum" and/or ``principal" components long recognized in nova spectra \citep{McLaughlin43, Williams13}.
Future high-resolution spectra from early in nova eruptions will test if the THEA lines originate in nova ejecta or have a more exotic source. 
\end{textbox}

\subsubsection{Dust and Molecule Formation}
\label{sec:dust}
Many novae show dramatic drops in their optical light curves on timescales of $20-100$ days after the eruption (e.g., V5668~Sgr in Figure~\ref{fig:rogues}).  These features arise from obscuration of the optical photosphere by dust forming within the expanding ejecta. 

These ``dust dips" are accompanied by a simultaneous increase in the mid-IR emission, originating from the outer ejecta coincident with the newly formed dust layer
(e.g., \citealt{Hyland&Neugebauer70,Gehrz+80}).
Dust formation is rapid, with grains growing to large sizes ($\sim$1 $\mu$m) compared to the dust in the interstellar medium over $\sim$month timescales \citep[e.g.,][]{Helton+10, Gehrz+18}. 
It is sometimes preceded by the formation of molecules like carbon monoxide (CO), which can be detected in near-IR spectra of novae within weeks of eruption \citep{Banerjee&Ashok12}. 
The molecular and dust phases of nova ejecta, and the complex chemistry therein, are reviewed by \cite{Gehrz88} and \cite{Evans&Rawlings08}.

The timescale for dust formation can be understood from basic considerations.  The equilibrium temperature in an ejected shell is set by the radiation of the WD of luminosity $L$ and given by $T_{\rm eq} = (L/4\pi \sigma R^{2})^{1/4}$, where $R = v_{\rm ej}t$ is the ejecta radius at time $t$ after the start of expansion and $v_{\rm ej}$ is the velocity. 
Dust formation should take place when $T_{\rm eq}$ reaches the condensation temperature for solids, $T_{\rm s} \approx 1200$ K, at time:
\be
t_{\rm dust} \equiv \left(\frac{L}{4\pi \sigma v_{\rm ej}^{2} T_{\rm s}^{4}}\right)^{1/2} \approx 76\,{\rm d}\left(\frac{L}{10^{38}\ {\rm erg/s}}\right)^{1/2}\left(\frac{v_{\rm ej}}{10^{3}\ {\rm km/s}}\right)^{-1}\left(\frac{T_{\rm s}}{1200\ {\rm K}}\right)^{-1/2}
\label{eq:tdust}
\ee
This estimate is in reasonable accord with the observed timescale for dust formation.  A more detailed calculation accounts for the fact that the dust grains do not re-radiate the absorbed light of the nova with perfect efficiency, and thus do not share the effective temperature of the radiation (e.g., \citealt{Evans+17}).

However, the nova ejecta are irradiated by the UV/X-ray luminous WD, leading to hostile conditions for the formation of molecules and dust.  The chemistry leading to dust formation requires carbon atoms to be neutral (e.g., \citealt{Rawlings&Williams89}), and shielded from the WD's harsh radiation (e.g.,~\citealt{Bath&Harkness89}).
This is generally only possible if the gas density is $\gtrsim 10^9-10^{10}$ cm$^{-3}$ (e.g., \citealt{Gehrz&Ney87}), higher than one would predict if the thickness of the ejecta shell were comparable to its radius ($R_{\rm dust} \approx v_{\rm ej}t_{\rm dust}$) at the dust formation epoch. 
Clumps in the ejecta (as observed with optical spectroscopy and imaging; \S\ref{sec:spectra}, \S\ref{sec:imaging}) may be dense enough to serve as sites of dust formation \citep{Woodward+92}.
As we discuss in $\S\ref{sec:multiD}$, shocks within the nova ejecta also offer a mechanism to produce regions of dense, self-shielded gas \citep{Derdzinski+17}.  

Although optical dust dips are seen in only $\sim$20\% of novae \citep{Strope+10}, many more show signs of dust formation in the form of IR excess \citep{Gehrz+98}. A population study of how many novae show dust signatures is currently lacking, but is increasingly feasible in an era where multi-band optical/IR light curves are common.  While some of the variation in dust signatures may be due to differences in the amount of dust formed  \citep{Gehrz88}, inclination effects from aspherically distributed dust formation probably play an important role.  In the well-resolved helium nova V445~Pup, some dust emission is seen in a bipolar outflow, but the primary evidence for dust is seen in absorption, as an equatorial disk that blocks the light of the binary (bottom right panel of Figure \ref{fig:opticalimages}; \citealt{Woudt+09}).  
In the slow nova V1280 Sco, emission from warm dust is again structured in two distinct lobes, but it is unclear if an absorbing equatorial dust disk is located between them, or if all the dust is located in the bipolar outflow \citep{Chesneau+12}. 
The sites of dust formation in novae remain poorly understood, and additional high-resolution IR and millimeter observations are needed to understand the link between dust formation and bulk ejecta properties.

\subsubsection{Thermal Radio Emission}
\label{sec:radiothermal}

Nova ejecta emit free-free thermal radiation at radio wavelengths, often modelled as an isothermal expanding \ion{H}{ii} region \citep{Seaquist&Bode08}. Temperatures of $\sim 10^{4}$ K are maintained throughout the ejecta by the ionizing radiation from the central WD (\S\ref{sec:SSS}),
despite the loss of thermal energy due to adiabatic expansion (\citealt{Cunningham+15}).   

Nova radio light curves evolve over months to years, brightening as the radio photosphere expands with the ejecta, and then fading as the ejecta become optically thin at radio wavelengths and continue decreasing in density \citep{Seaquist&Palimaka77, Hjellming+79}. Free-free opacity decreases with increasing frequency, so emission at higher frequencies emerges from deeper in the ejecta and evolves more rapidly. The spectral energy distribution of the free-free emission has been observed over $>$2 decades in frequency, from 1 GHz to 600 GHz \citep{Ivison+93, Hjellming96}, although there is a poorly explored possibility that dust emission can contribute at $>$100 GHz (e.g., \citealt{Nielbock&Schmidtobreick03}). 

The radial profile of the ejecta density $\rho(r,t)$ and the total mass of ionized gas can be measured by  monitoring of the radio spectral evolution. 
The radio data are generally well fit with a density profile, $\rho \propto r^{-2}$ or $r^{-3}$ (e.g., \citealt{Seaquist+80, Weston+16a}), consistent with models of UV/optical spectra (e.g., \citealt{Hauschildt+92}; \S\ref{sec:spectra}). The derived ejecta profiles show a cutoff at an inner radius, and can be explained as either a single homologous ejection or as a prolonged wind \citep{Hjellming+79, Kwok83}. In addition, by tracking the expanding photosphere, radio light curves can trace when the ejecta began expanding, sometimes revealing surprising  weeks-long delays between the rise of the optical light curve and the ejection of the radio-emitting material \citep{Nelson+14, Linford+17} 
 
Ejecta masses obtained by integrating the inferred density profile are typically in the range $M_{\rm ej} \approx 10^{-5}-10^{-3}$ M$_{\odot}$ (\citealt{Roy+12,Wendeln+17}).
Because the free-free luminosity traces the emission measure of the ionized gas ($n_e^2\, dl$, where $n_e$ is the number density of electrons and $dl$ is the path length through the ejecta), ejecta mass estimates require assumptions about the filling factor, which can be $<$1 due to ejecta clumping or aspherical geometry. 
Analysis of optical spectra imply filling factors less than unity (\S\ref{sec:spectra}), implying that many ejecta mass estimates in the literature are in fact upper limits, because they assume it is unity.  Spherical symmetry is generally assumed when modelling the nova's radio light curve, but morpho-kinematic codes like {\tt SHAPE} now allow this assumption to be relaxed \citep{Steffen+11,Ribeiro+14}.
In the future, radio images from high-resolution interferometers (\S\ref{sec:imaging}) can inform the choice of ejecta morphology, and be analyzed along with radio light curves and optical spectroscopy to provide better measurements of the distribution and total mass of ejecta.

\subsubsection{High-resolution Imaging}
\label{sec:imaging}

As they expand, the nova ejecta grow to an angular diameter:
\be
\theta = 0.115^{\prime\prime}\ \left(\frac{d}{1\ {\rm kpc}}\right)^{-1} \left(\frac{v_{\rm ej}}{1000\ {\rm km/s}}\right) \left(\frac{t}{100\ {\rm d}}\right)
\ee
where $t$ is the time since the start of expansion and $d$ is the source distance. This means that high-resolution facilities like the \emph{Hubble Space Telescope} (\emph{HST}), radio/mm interferometers, and near-IR adaptive-optics imagers can resolve nova ejecta starting several months to years after eruption, while ground-based optical imaging is able to resolve novae years to decades after eruption.

\begin{figure}[h]
\includegraphics[width=4.5in]{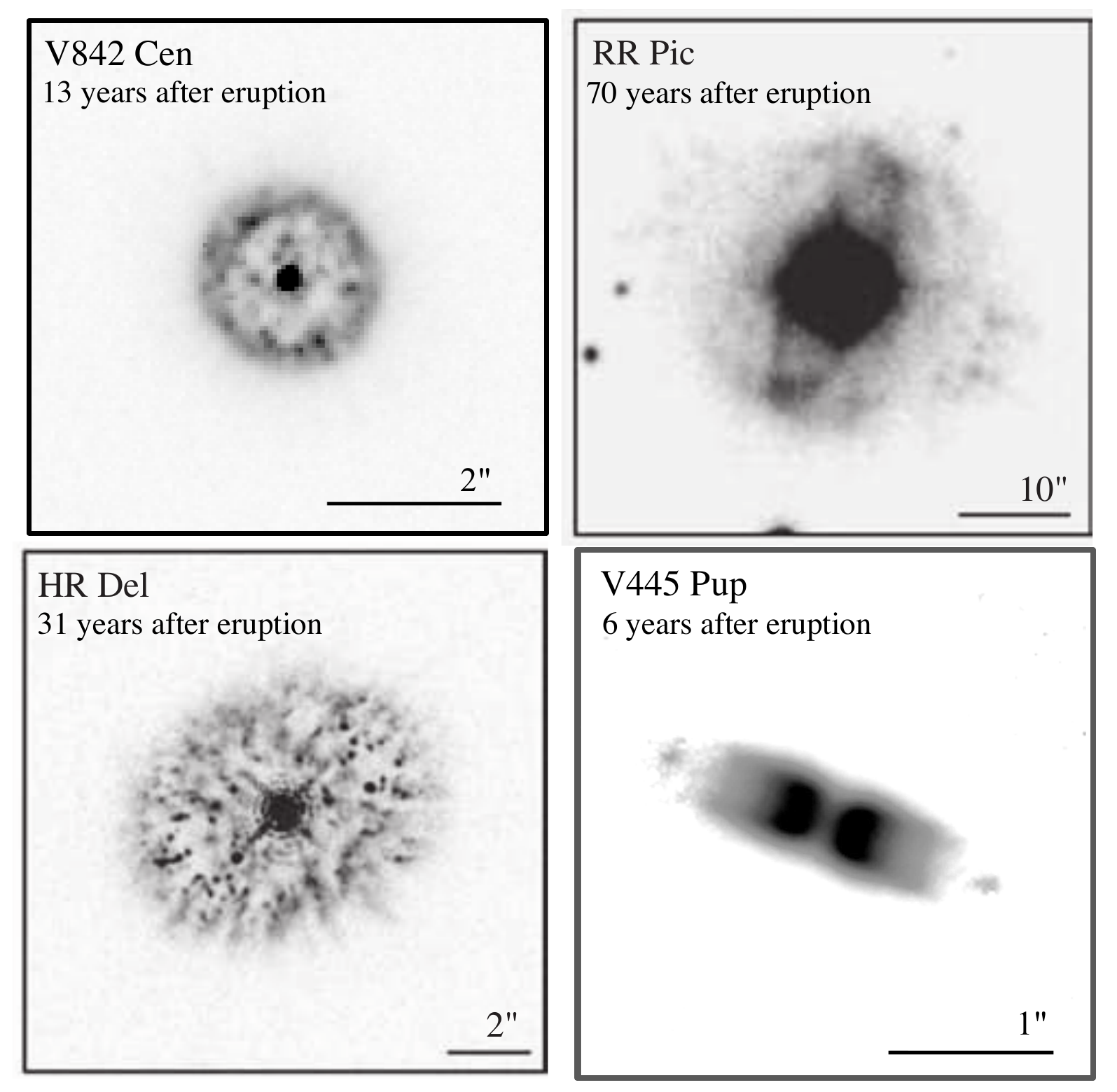}
\vspace{-0.15in}\caption{Optical/IR images of novae, highlighting  the diversity of structures observed in their ejecta. The time elapsed between eruption and imaging is noted in the top-left corner of each panel. The {\bf top left} image shows V842~Cen (1986), observed with \emph{HST} and consistent with a clumpy near-spherical ejection (image obtained from the Hubble Legacy Archive). The {\bf bottom left} panel shows HR~Del (1967) imaged with \emph{HST}, revealing ``cometary tails" extending from clumps, prolate ellipticity, and on closer analysis, bipolar morphology \citep{Harman_OBrien03}. The {\bf top right} image shows RR~Pic (1925) observed with the Anglo-Australian Telescope, and exhibits a clear equatorial ring and bipolar outflows \citep{Gill&OBrien98}. The {\bf bottom right} panel shows an image of the helium nova V445~Pup (2000) obtained using adaptive optics on the Very Large Telescope, showing a collimated bipolar outflow with distinct high-velocity knots at the outer extent of each lobe (panel adapted with permission from \citealt{Woudt+09}; $\copyright$AAS). The V842~Cen, HR~Del, and RR~Pic images trace the H$\alpha$+[\ion{N}{ii}] emission line flux, while the V445~Pup image was obtained in the near-IR $K$ band and traces both warm dust and plasma.}
\label{fig:opticalimages}
\end{figure}

The largest samples of nova images have been collected at optical wavelengths, usually using an H$\alpha$+[\ion{N}{ii}] narrow-band filter, at relatively late times ($\sim$5--50 yr; e.g., \citealt{Cohen85, Slavin+95, Gill&OBrien98, Downes&Duerbeck00}).
Most resolved nova ejecta are consistent with clumpy, elliptical shells (e.g., V842~Cen in Figure \ref{fig:opticalimages}; see \citealt{OBrien&Bode08} for more examples). A minority show more complex structures, like bipolar overdensities and equatorial rings (e.g., RR~Pic in Figure \ref{fig:opticalimages}).  When imaged at enough resolution, the emission usually resolves into clumps, and sometimes ``cometary tail" features extend radially outward from the clumps (e.g., HR~Del in Figure \ref{fig:opticalimages}), which could be the result of a fast wind moving past a slower inhomogeneous shell \citep{Lloyd+95}.

\begin{figure}[t]
\includegraphics[width=5in]{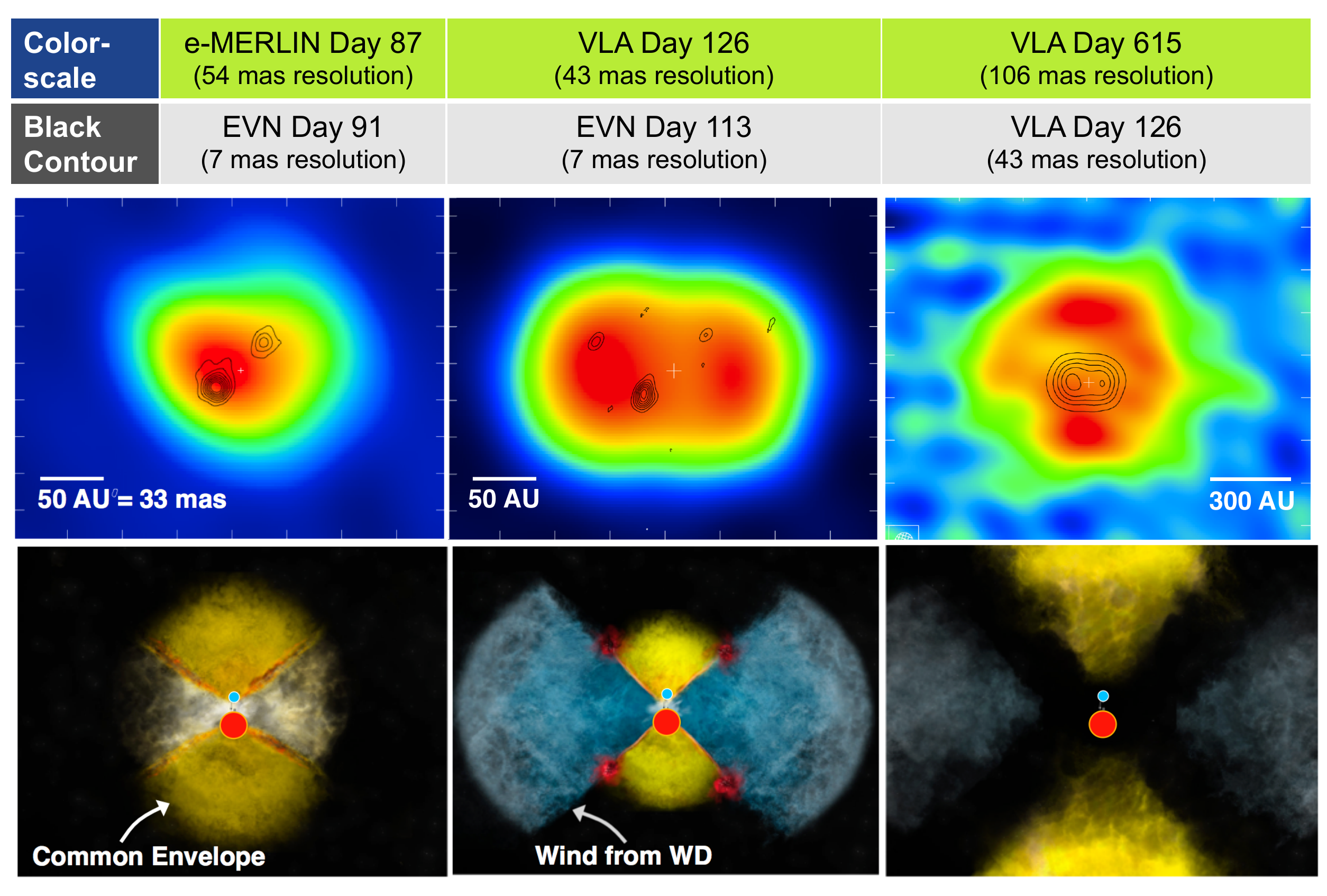}
\vspace{-0.15in}\caption{Images of nova V959~Mon show an ejecta geometry consistent with a slowly expanding equatorial torus, a faster bipolar outflow, and shocks where the two ejections collide.
The top row shows radio observations (labelled at top), and the bottom row is a series of artist's representations. {\bf Left panels:} the thermal nova ejecta are barely resolved at early times (colorscale in top row), while synchrotron knots are superimposed on the ejecta (black contours in top row; likely tracing shocks and particle acceleration; \S\ref{sec:radiononthermal}). {\bf Middle panels}: The thermal ejecta are concentrated in two lobes (colorscale), and  the synchrotron knots (black contours) surround the lobes. This is consistent with a fast wind funneled along the binary's poles, crashing into dense equatorial material and producing shocks at the interface (red regions in artist's representation). {\bf Right panels:} The polar outflow has faded, and the brightest, densest material is now oriented perpendicularly, corresponding to the slower equatorial material (colorscale; the black contours here represent thermal emission from day 126, for comparison). This can be explained if the WD stops powering a wind at late times, and the ejecta drift off into space. Radio images adapted from \citet{Chomiuk+14}; Artist's impressions created by B.\ Saxton/NRAO/AUI/NSF and reproduced with permission.}
\label{fig:Chomiuk14}
\end{figure}

High-resolution imaging at earlier times is more informative, because it can trace how the ejecta evolve over time and resolve structures that may be too low density to detect at late times, thereby constraining the mass ejection processes (\S\ref{sec:massejection})  and the origin of shocks (\S\ref{sec:shockstheory}).
Early images of nova ejecta are consistent with an elliptical expanding shell, and ellipticity that changes with time, either from rounder to more prolate \citep{Paresce+95, Schaefer+14} or from more prolate/bipolar to rounder \citep{Taylor+88, Pavelin+93, Hjellming96}. This evolution is suggestive of at least two distinct flows with different morphologies \citep{Taylor+88, Chochol+97}, but the clearest evidence comes from V959~Mon, where the binary has an edge-on inclination convenient for imaging \citep{Page+13}.
\citet{Chomiuk+14} present multi-epoch radio images of V959 Mon, and capture an orthogonal ``flip" in the major axis of the ejecta, as shown in Figure \ref{fig:Chomiuk14}. They interpret the bi-lobed ejecta imaged on day 126 as a fast bipolar flow (elongated along the horizontal axis in the central panel of Figure \ref{fig:Chomiuk14}), which diffuses and fades by day 615 when the emission is dominated by a slower, equatorially concentrated flow (elongated along the  vertical axis in the right panel of Figure \ref{fig:Chomiuk14}).
\emph{HST} imaging confirms an edge-on equatorial torus and faster, bipolar flows in V959~Mon \citep{Sokoloski+16}.
The helium nova V445~Pup showed a similar morphology (bottom right panel of Figure \ref{fig:opticalimages}), and, in a clear indication of non-impulsive mass ejection, \citet{Woudt+09} showed that the high-velocity knots visible at the extremities of the bipolar outflow were not ejected until 345 days after the start of eruption.
Novae that are near enough for well-resolved, early imaging campaigns are rare, and every future example should be exploited as a test of whether the bipolar outflow/equatorial disk configuration can explain the morphologies of all novae.

\section{SHOCK SIGNATURES ACROSS THE ELECTROMAGNETIC SPECTRUM}
\label{sec:shocksobs}

Many of the observational probes of nova ejecta discussed in \S\ref{sec:observations}, especially optical spectroscopy and high-resolution imaging, support the presence of distinct outflow components
that span a range of velocities, which may be launched with substantial delays following TNR and which tend to accelerate as the eruption proceeds.
Interactions of these components with each other---or with pre-existing material surrounding the binary---will inevitably give rise to shocks.
Since the typical sound speeds in the $\sim10^{4}$ K nova ejecta are $c_{\rm s} \approx 10$ km/s, while the ejecta velocities are $\gtrsim 100-1000$ km/s, the resulting shocks are highly supersonic.  These shocks heat the plasma to millions of degrees and accelerate a small fraction of particles to relativistic speeds, generating thermal and non-thermal emission, respectively.  

\begin{figure}[!t]
\includegraphics[width=3.5in]{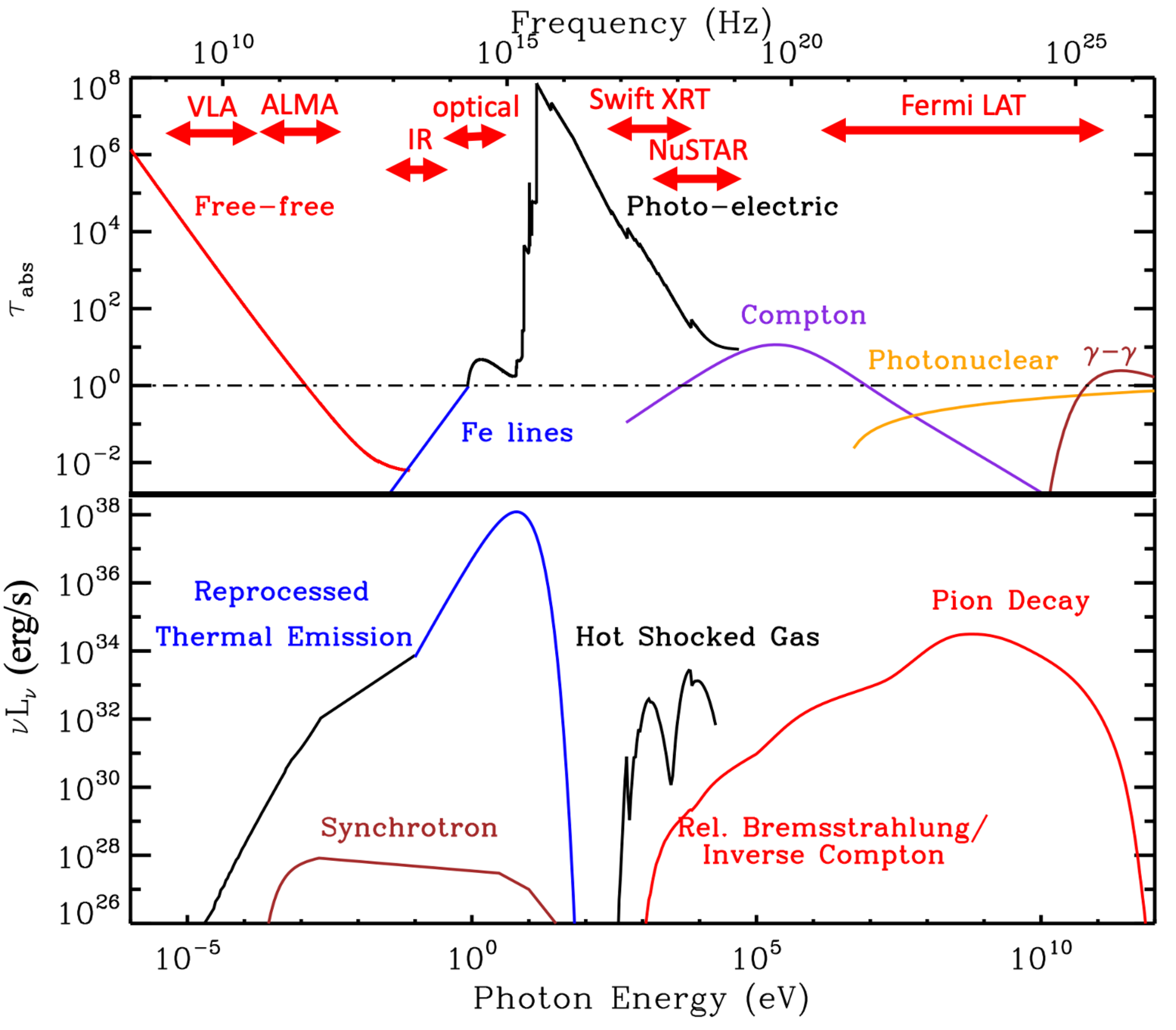}
\vspace{-0.2in}\caption{The processes governing opacity and emission from internal shocks in nova ejecta, 
from radio to gamma-ray frequencies.
{\bf Top panel:} Absorptive optical depth to the shocked region, measured outwards through the nova ejecta, as a function of photon energy (frequency); $\tau_{\rm abs}$ = 1 is marked as a horizontal dashed line. This is for a relatively early epoch in a nova eruption, when the gas column ahead of the shocks is $\sim 10^{25}$ cm$^{-2}$.  We have assumed solar composition gas with a temperature of $10^{4}$ K upstream of the shock, and a nova thermal luminosity of $1.5\times 10^{38}$ erg/s that primarily emerges at optical wavelengths.
The free-free and photo-electric opacities were calculated from CLOUDY \citep{Ferland+13}; the effective optical depth due to inelastic Compton scattering was calculated using a Monte Carlo method (I.~Vurm, private communication); the optical depth due to photonuclear pair creation with the ejecta and $\gamma-\gamma$ pair creation on the nova optical light were calculated using standard expressions (e.g.,~\citealt{Zdziarski&Svensson89}).  {\bf Bottom panel:} The spectral energy distribution of shock-powered emission in novae, accounting for absorption by the ejecta.  This is based on one-dimensional internal shock simulations from \citet{Steinberg&Metzger20}, combined with non-thermal emission from \citet[][see Figure \ref{fig:Indrek}]{Vurm&Metzger18}.  A hadronic model is assumed for the gamma-ray emission, while the radio emission results from electron acceleration at the shock with an assumed efficiency of $\epsilon_e \approx$ 0.01.  A possible extension of the non-thermal gamma-rays to $\gtrsim$ TeV energies accessible to Cherenkov telescopes is not shown (\S\ref{sec:Emax}). Panel adapted from \citet{Steinberg&Metzger20}.  }
\label{fig:taueff}
\end{figure}

This section describes the observational evidence for shocks in novae.  To aid our discussion, Figure \ref{fig:taueff} shows the optical depth $\tau_{\rm abs}$ through the ejecta as a function of photon frequency, along with the shock-powered spectral energy distribution, for conditions characteristic of a week after eruption when the shocks are still relatively deeply embedded.  Places in the spectrum where $\tau_{\rm abs}$ reaches a minimum, such as the optical/IR and GeV gamma-ray bands (\S\ref{sec:gammaray}), will reveal shock-powered emission first.  However, as the shocks propagate to larger radii where $\tau_{\rm abs}$ is smaller, shock-powered X-ray (\S\ref{sec:Xrays}) and radio (\S\ref{sec:radiononthermal}) emission become observable.

\subsection{GeV Gamma-Rays}
\label{sec:gammaray}

\begin{figure}[!t]
\includegraphics[width=4.5in]{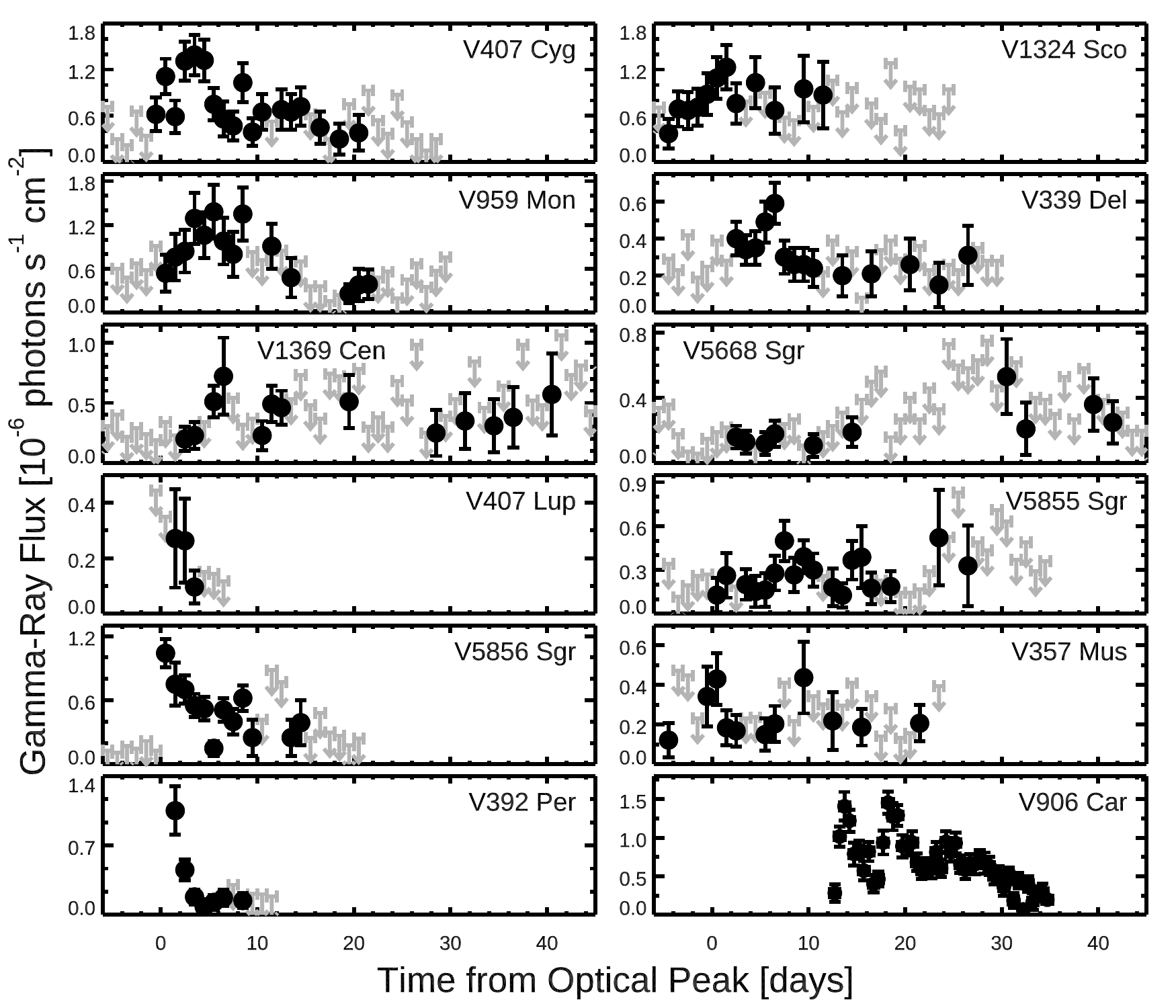}
\vspace{-0.25in}\caption{GeV gamma-ray light curves for 12 novae, as measured by \emph{Fermi}-LAT for photon energies $>$100 MeV. Dates with $> 2\sigma$ significance detections are marked as black dots, while 95\% upper limits are plotted as grey arrows for non-detections. Plotted times are relative to optical light curve maximum, except in the case of V959~Mon, which was discovered with \emph{Fermi} during solar conjuction; in this case $t = 0$ marks the first gamma-ray detection. Due to a solar panel issue, there are no \emph{Fermi}-LAT data available before the gamma-ray detections of V392~Per and before or after the gamma-ray detections of V906~Car. 
References for the gamma-ray analysis are listed in Supplemental Table 1.}
\label{fig:gammaLC}
\end{figure}

\emph{Fermi}-LAT has detected continuum gamma-ray emission in the photon energy range $\sim 0.1-10$ GeV 
from over a dozen Galactic novae \citep{Abdo+10,Ackermann+14,Cheung+16,Li+17,Aydi+20}, as summarized in Supplemental Table 1.  The gamma-rays are usually detected for a few weeks, starting around the time of optical light curve maximum  (Figure \ref{fig:gammaLC})\footnote{Unless otherwise noted, all quoted uncertainties and plotted error bars are 1$\sigma$ significance (i.e., 68\%).}. 

The gamma-rays are most naturally understood as non-thermal emission from relativistic particles
accelerated at shocks, likely through the process of diffusive shock acceleration (e.g., \citealt{Blandford&Ostriker78}).  
Typical detected gamma-ray luminosities are in the range of $L_{\gamma} \approx 10^{34}-10^{36}$ erg/s,
and the total energies emitted in gamma-rays are $E_{\gamma} \approx 10^{41}-10^{42}$ erg \citep{Ackermann+14,Cheung+16}, which are much smaller than the bolometric outputs of novae. However, only a small fraction of the power of the shock goes into relativistic particles, and only a fraction of the energy in relativistic particles is radiated as detectable gamma-rays, which means that the shock luminosities needed to produce the gamma-rays can be comparable to the bolometric output, $\sim 10^{38}$ erg/s.

\begin{figure}[h]
\includegraphics[width=4in]{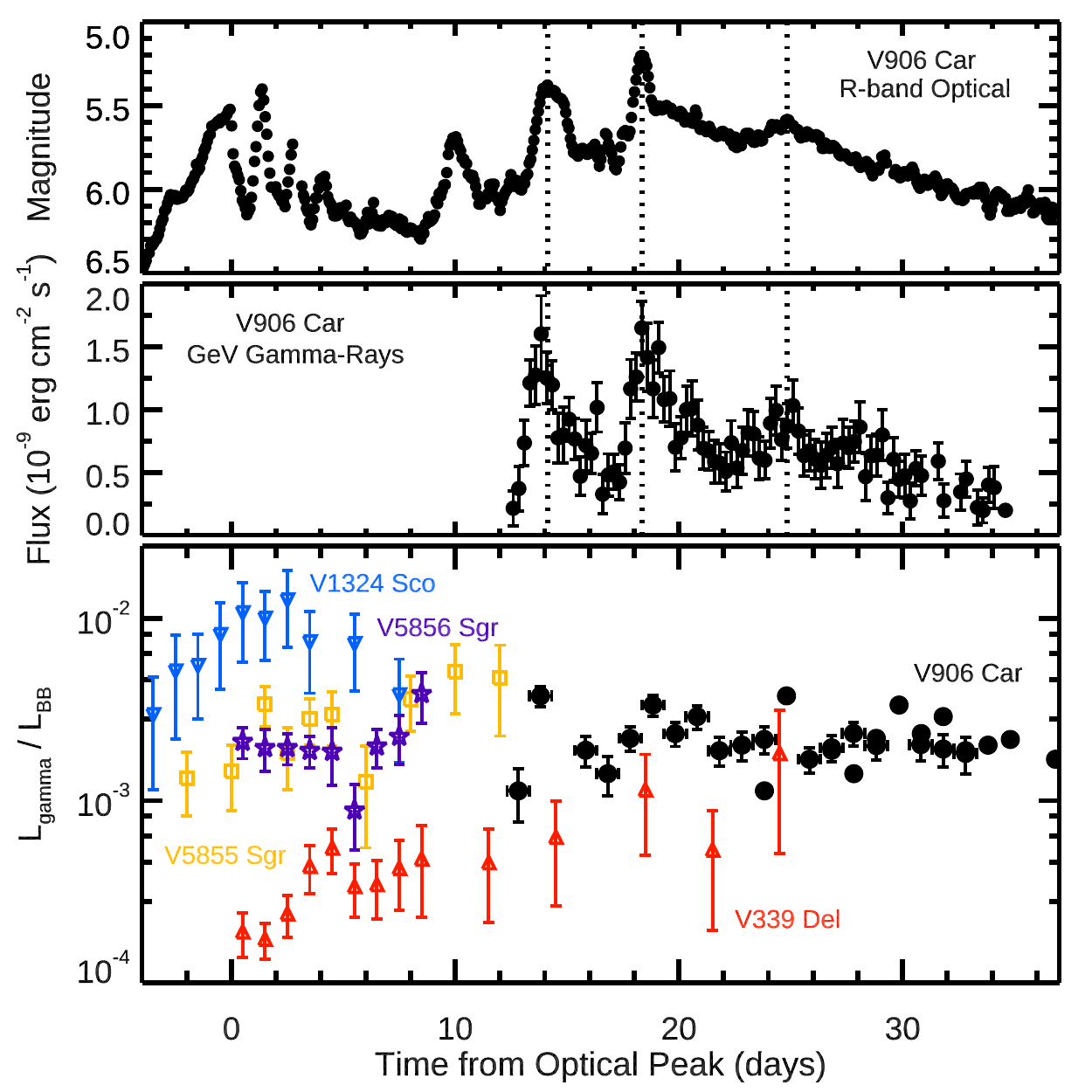}
\vspace{-0.25in}\caption{
The correlated flares in the optical and gamma-ray light curves of nova V906~Car. {\bf Top panel:} Optical $R$-band light curve of V906~Car, observed at high cadence and precision using the \emph{BRITE} constellation of nano-satellites. The three optical flares that occured during the \emph{Fermi} observations are marked with black dotted lines.
{\bf Middle panel:} \emph{Fermi}-LAT gamma-ray light curve ($>$100 MeV) of V906~Car. Unfortunately, \emph{Fermi}-LAT was not operational outside the times with plotted data.
{\bf Bottom panel:} The ratio of gamma-ray to blackbody luminosity as a function of time.
V906~Car (black circles) is compared with other \emph{Fermi}-detected novae: V5856 Sgr (purple stars), V5855 Sgr (gold squares), V339 Del (red triangles), and V1324 Sco (blue triangles).  Figure adapted from \citet{Aydi+20}, with $L_{\gamma}/L_{\rm BB}$ values from \citet{Metzger+15} and \citet{Li+17}. }
\label{fig:gamoptcorrelation}
\end{figure}

The high signal-to-noise gamma-ray light curves of V5856~Sgr and V906~Car reveal striking correlations between the optical and gamma-ray light curves \citep{Li+17, Aydi+20}. For example V906~Car showed maxima in gamma-rays alongside optical flares (top panels of Figure \ref{fig:gamoptcorrelation}). A correlation analysis finds that the brightenings occur nearly simultaneously, with the optical lagging the gamma-rays by $5.3\pm2.7$ hr \citep{Aydi+20}. Moderate ($2\sigma$) evidence also exists for correlated optical/gamma-ray light curves in V339 Del and V5855 Sgr, despite the poorer gamma-ray statistics \citep{Li+17}.  The bottom panel of Figure \ref{fig:gamoptcorrelation} shows the time evolution of the gamma-ray to blackbody luminosity ratio for a sample of gamma-ray detected novae, revealing a range of values $L_{\gamma}/L_{\rm BB} \approx 3\times 10^{-4}-10^{-2}$ (here $L_{\rm BB}$ is derived from a blackbody fit to the optical/IR spectral energy distribution, as a rough estimate of the bolometric luminosity).  This ratio is relatively constant for a given nova, although in V906~Car, $L_{\gamma}/L_{\rm BB}$ appears to peak during optical flares (Figure \ref{fig:gamoptcorrelation}). This correlation naturally results if a portion of the optical luminosity---particularly during strong flares---is reprocessed shock power (\S\ref{sec:shockoptical}).

\begin{textbox}[h]\section{BOX 5: Shocks in Embedded Novae}
Shocks have long been expected and observed in embedded novae, arising from the interaction of the nova ejecta with the giant companion's wind. The most famous example of these shocks is in RS~Oph, which showed bright radio synchrotron emission and hot X-ray emitting plasma during its 1985 and 2006 eruptions (e.g., \citealt{Hjellming+86, OBrien+06, Sokoloski+06}.
In the past decade, the sample of well-observed embedded novae has grown dramatically, establishing a class of novae with similar radio and X-ray signatures to RS~Oph \citep[e.g.,][]{Nelson+12, Linford+15,  Delgado&Hernanz19, Orio+20}.

These ``external" shocks with circumstellar material in symbiotic systems suffer significantly less absorption than shocks in classical novae that are presumably internal to the ejecta. For example, V407~Cyg is the only gamma-ray detected nova that showed 1--10 keV X-rays simultaneous with gamma-rays (Figure \ref{fig:xrays}). The gamma-ray luminosities of these embedded novae may be consistent with those of classical novae, but embedded novae are rarer and so tend to be discovered at larger distances (Supplemental Figure 1). V407 Cyg is the only one with a high-significance \emph{Fermi}-LAT detection, while three other embedded novae---V745 Sco, V1535~Sco, and V3890 Sgr---show hints of gamma-ray emission \citep{Cheung+14, Buson+19}, but at $< 3 \sigma$ significance (\citealt{Franckowiak+18}, K.\ Li, 2020, private communication).
\end{textbox}

\subsubsection{Do All Novae Emit Gamma-Rays?}

The first gamma-ray detection was from the 2010 nova in the symbiotic system V407 Cyg \citep{Abdo+10}, where the shocks were interpreted as the collision between the nova ejecta and the dense wind of the Mira giant companion \citep[e.g.,][]{Nelson+12,Martin&Dubus13}. However, essentially all of the other high-confidence gamma-ray detections have been classical novae with main sequence companions. The winds from main sequence stars are weak, and mass transfer onto the WD is quite conservative, so there is too little circumbinary material to generate powerful shocks. Therefore, the shocks must be internal to the nova ejecta. In fact, \citet{Martin+18} find that internal shocks could even give rise to the majority of the gamma-ray emission from V407 Cyg, without needing to invoke any interaction with the giant wind.

While it is plausible that all novae host internal shocks and emit gamma-rays, the rate of gamma-ray detected novae ($\sim 1$ yr$^{-1}$) is only a small fraction of the optical discovery rate ($\sim 5-10$ yr$^{-1}$).  The \emph{Fermi} detections tend to be of relatively nearby novae ($\sim 2-5$ kpc) and of marginal significance (Supplemental Table 1 and Supplemental Figure 1), suggesting that many novae go undetected due to the limited flux sensitivity of {\it Fermi}-LAT.

Nevertheless, the hypothesis that all novae possess the same gamma-ray luminosity can be firmly excluded, as \citet{Franckowiak+18} showed that the gamma-ray luminosities span at least two orders of magnitude.
However, to date no clear correlations have been found between gamma-ray luminosity and nova properties like $t_2$  (\citealt{Franckowiak+18}; Supplemental Figure 1). There is a suggestion of an anti-correlation between the total gamma-ray energy emitted and the duration of the gamma-ray emission \citep{Cheung+16}. In an internal shock model (\S\ref{sec:shockstheory}), we expect the gamma-ray luminosity to increase with ejecta mass and velocity, and future studies should test for such relationships.

\subsubsection{Leptonic or Hadronic Gamma-Rays?}
\label{sec:gammaspectra}

\begin{figure}[!b]
\includegraphics[width=6in]{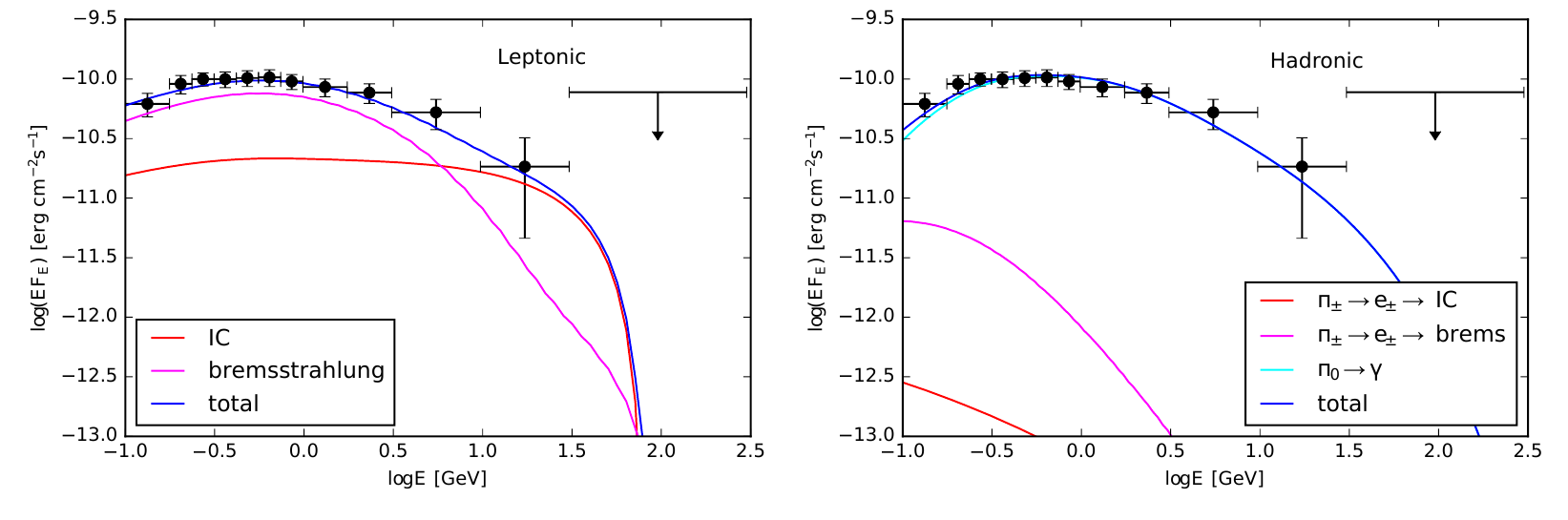}
\vspace{-0.2in}\caption{The {\it Fermi}-LAT GeV spectral energy distribution of V5856~Sgr (black points), compared with gamma-ray emission models (solid lines).  {\bf Left panel:} Leptonic model fit, which assumes electrons accelerated at the shock with a $q = 1.8$ momentum spectrum.  {\bf Right panel:}  Hadronic model fit, which assumes protons accelerated at the shock with $q = 2.4$.  Both models assume a maximum particle energy of 100 GeV.  Although both models describe the data well, the hadronic model is favored because it more robustly accounts for the break in the spectrum at low energies set by the pion creation threshold ($m_{\pi} c^{2} =$ 135 MeV).  The leptonic model also has several efficiency issues (see text for details).  Figure reproduced from \citet{Li+17}.}
\label{fig:Indrek}
\end{figure}

The time-integrated {\it Fermi}-LAT gamma-ray spectra of novae are typically modelled as a power-law with an exponential cut-off, 
$dN/dE \propto E^{-\Gamma}\exp\left(-E/E_{\rm c}\right)$,
where $E_{\rm c}$ is the cut-off energy at high photon energies and $\Gamma$ is the photon index.  The best-fit values for individual novae span the range $\Gamma \approx 1.3-2.3$ (Supplementary Table 1). 
While some novae are best fit with $E_{\rm c} \approx 2-8$ GeV, others are well described by a simple power law with no cutoff energy.
\citet{Franckowiak+18} show that all pre-2016 novae are consistent with a single universal spectrum with $\Gamma = 1.71\pm 0.08$ and $E_{\rm c} = 3.2 \pm 0.6$ GeV (excluding the embedded nova V407 Cyg from the sample, which exhibited a significantly harder photon index than the classical novae).   
Two more recent novae,  V5856 Sgr and V906 Car, were particularly bright in gamma-rays, enabling high-fidelity spectral fits and yielding values $\Gamma \approx 1.7-1.9$ and $E_{\rm c} \approx 3-8$ GeV, broadly consistent with the universal spectrum (Figure \ref{fig:Indrek}; \citealt{Li+17,Aydi+20}).  
Both spectra also show a deficit below a few hundred MeV compared to a simple power-law, the origin of which we return to below.  Although the evidence for a high-energy spectral cut-off is marginal based on the LAT data alone, a deep upper limit on the 0.1--10 TeV emission from V339 Del, obtained with MAGIC simultaneously with a LAT detection, necessitates a spectral cut-off or steepening in the 10--100 GeV range \citep{Ahnen+15}. 

Physical models for the gamma-ray emission divide into ``hadronic" and ``leptonic" scenarios, depending on whether the emitting particles are primarily relativistic ions or electrons. In leptonic scenarios, the accelerated electrons emit gamma-rays via bremsstrahlung emission due to interaction with ambient protons/electrons:
\begin{equation}
e^{\pm} + p \rightarrow e^{\pm} + p + \gamma; \,\,\,\,\,\,e^{\pm} + e^{-} \rightarrow e^{\pm} + e^{-} + \gamma
\label{eq:brems}
\end{equation}
Leptonic gamma-rays are also produced via inverse Compton (IC) scattering of optical photons:
\begin{equation}
e^{\pm} + \gamma_{\rm opt} \rightarrow e^{\pm} + \gamma
\label{eq:ic}
\end{equation}
In the hadronic scenario, ions collide with ambient ions such as protons, producing pions that decay into $\gamma$-rays as:
\begin{eqnarray}
p + p &\rightarrow& \pi^{0} \rightarrow \gamma + \gamma \nonumber \\
&\rightarrow& \pi^{\pm} \rightarrow \mu^{\pm} + \nu_\mu \rightarrow e^{\pm} + \nu_e + \nu_\mu,
\label{eq:pp}
\end{eqnarray}
where $\approx 1/3$ and  $\approx 2/3$ of the inelastic $p$-$p$ collisions go through the $\pi^{0}$ and $\pi^{\pm}$ channels, respectively \citep{Kelner&Aharonian08}.  Leptons produced in the $\pi^{\pm}$ channel can then produce gamma-rays through the bremsstrahlung and IC processes, although the $\pi^{0}$ channel is expected to dominate the gamma-ray luminosity (Figure \ref{fig:Indrek}). 
As a useful approximation, generating a photon of energy $E$ requires a proton of energy $10E$, so to produce gamma-rays up to $\sim 10$ GeV requires particle acceleration up to $\sim 100$ GeV \citep{Kelner&Aharonian08}.  In leptonic scenarios, the photon energy is comparable to that of the primary electron, so particle acceleration up to $\sim 10$ GeV is sufficient. 

Figure \ref{fig:Indrek} shows the {\it Fermi}-LAT gamma-ray spectrum of V5856~Sgr fit using the leptonic or hadronic models.  Relativistic particles (electrons or protons, respectively) are assumed to be accelerated at the shock with an energy spectrum $dN/dp \propto p^{-q}$, where $p = \beta\gamma_p$ is the particle momentum, while $\gamma_p$ and $\beta = v/c$ are the Lorentz factor and particle velocity, respectively.  As described in \cite{Vurm&Metzger18}, the thermodynamic evolution of the particles and their secondaries, including various emission and energy-loss processes (e.g., IC, bremsstrahlung, synchrotron, Coulomb scattering), can be followed downstream behind the shock to calculate the emerging radiation spectrum.  The best fit models require $q \lesssim 2$ in the leptonic scenario and $q \approx 2.4$ in the hadronic scenario.

The hadronic model naturally predicts a low-energy spectral turnover as observed, near the pion rest mass energy of $m_{\pi} c^{2} \approx 135$ MeV thaat sets the threshold energy for $p$-$p$ interactions (Eq.~\ref{eq:pp}). A similar break can appear in leptonic models due to Coulomb interactions competing for the energy of low-$\gamma_p$ electrons, but the location of this feature has to be fine-tuned because it depends on the optical radiation field.

Other considerations also favor the hadronic scenario.
 A strong magnetic field near the shock is required to confine and accelerate the particles 
(\S\ref{sec:Emax}; Eq.~\ref{eq:Emax}).  However, electrons embedded in such a magnetic field cool quickly via synchrotron emission, suppressing their contribution to the gamma-ray luminosity and hence making leptonic emission channels inefficient \citep{Li+17}.  The shallow spectral indices $q < 2$ required in leptonic models are also in tension with the values of $q > 2$ predicted by diffusive shock acceleration (\citealt{Blandford&Ostriker78}) or inferred from observations of other astrophysical shocks such as SN remnants \citep{Green19}.  Nevertheless, such low values of $q$ are required for the leptonic scenario to avoid an ``energy crisis": electron Lorentz factors $\gamma_p \gtrsim 10^{3}-10^{4}$ are needed to produce the gamma-rays, but for $q > 2$ the total energy is dominated by low-$\gamma_p$ electrons.  Leptonic models also predict an extension of the LAT gamma-ray emission down to the hard X-ray band \citep{Vurm&Metzger18} that is in tension with \emph{NuSTAR} observations of V906 Car \citep{Aydi+20}.  Naively, the correlation between gamma-ray and optical light curves (Figure \ref{fig:gamoptcorrelation}) could be interpreted within leptonic scenarios as time-variable IC emission driven by the fluctuating optical target photon background (Eq.~\ref{eq:ic}).  However, electrons with sufficient energy to generate the observed gamma-rays are already in a deeply fast-cooling regime \citep{Metzger+15}, and hence the non-thermal energy is converted to gamma-ray luminosity with maximal efficiency, regardless of optical luminosity.  Models that systematically fit nova gamma-ray spectra by following particle acceleration and emission within fully time-dependent one-dimensional hydrodynamical models also conclude that the hadronic channel dominates \citep{Martin+18}.

\begin{textbox}[h]\section{BOX 6: MeV Gamma-Ray Emission from Radioactive Nuclei}
Gamma-ray emission at $\sim$MeV photon energies  has long been predicted from novae due to nuclear decay processes \citep{Clayton&Hoyle74,Hernanz+02}.  This includes continuum and 511 keV line emission from $e^{-}$/$e^{+}$ annihilation, which are predicted to last a day or so following the TNR \citep{Gomez-Gomar+98}.  Also predicted are emission lines from decays of individual isotopes such as $^{7}$Be (478 keV) and $^{22}$Na (1275 keV) in CO and ONe novae, respectively.   Unfortunately, no MeV gamma-ray emission has yet been detected from novae, despite decades of upper limits measured using the \emph{Compton Gamma Ray Observatory} (e.g., \citealt{Hernanz+00}), the \emph{Wind} satellite (\citealt{Harris+91}), \emph{Swift}-BAT \citep{Senziani+08}, and \emph{INTEGRAL} \citep{Siegert+18}.  Detection of MeV gamma-ray emission, which would provide crucial diagnostics of the TNR, requires a new generation of gamma-ray satellites with greater sensitivity
(e.g., COSI, AMEGO; \citealt{Tomsick+19, McEnery+19}) or an extremely nearby event ($\lesssim$1 kpc).
 \end{textbox}

\subsection{X-rays from Hot Plasma}
\label{sec:Xrays}

In addition to the supersoft X-ray emission that arises from residual nuclear burning on the WD surface ($\S\ref{sec:SSS}$), novae often emit harder X-rays ($\gtrsim 1$ keV; e.g., \citealt{Lloyd+92,Krautter+96,Mukai&Ishida01,Orio+01,Mukai+08}). 
The harder component of the X-ray spectrum is distinct from the supersoft X-ray component and, is well modelled by free-free emission and atomic lines from an optically thin $10^7-10^8$ K thermally emitting plasma.
This X-ray emission also evolves differently from the supersoft X-ray component, and is visible both before and after the supersoft phase.
These observations imply a distinct physical origin for the harder X-ray component, and lead to an association with shocks.

A strong shock generated by the collision of two flows at velocity $v$ heats gas to a characteristic temperature \citep{Landau&Lifshitz59}
\be
kT_{\rm sh} \simeq \frac{3}{16}\mu m_p v^{2} \approx 1.2 \,{\rm keV} \left(\frac{v}{10^{3}{\rm\, km/s}}\right)^{2},
\label{eq:Tsh}
\ee
where we have taken $\mu = 0.62$ for the mean molecular weight of solar composition material.  
Thermal plasma emission models of nova X-ray spectra imply temperatures $kT \approx 1-10$ keV \citep{Mukai+08}, pointing to shock velocities $\approx 1,000-3,000$ km/s.
These velocities are broadly similar to the difference in velocity between the fast and slow components inferred from optical spectroscopy (e.g., Figure \ref{fig:v906spec}), suggesting that collisions between these two components may be the origin of the shocks.

\begin{marginnote}[]
\entry{Temperature in keV}{In X-ray astronomy, temperatures are often given in units of keV, calculated as $kT$, where $k$ is the Boltzmann constant ($1.38 \times 10^{-16}$ erg/K). 
1 keV = $1.16 \times 10^7$ K.}
\end{marginnote}

\begin{figure}[!t]
\includegraphics[width=5in]{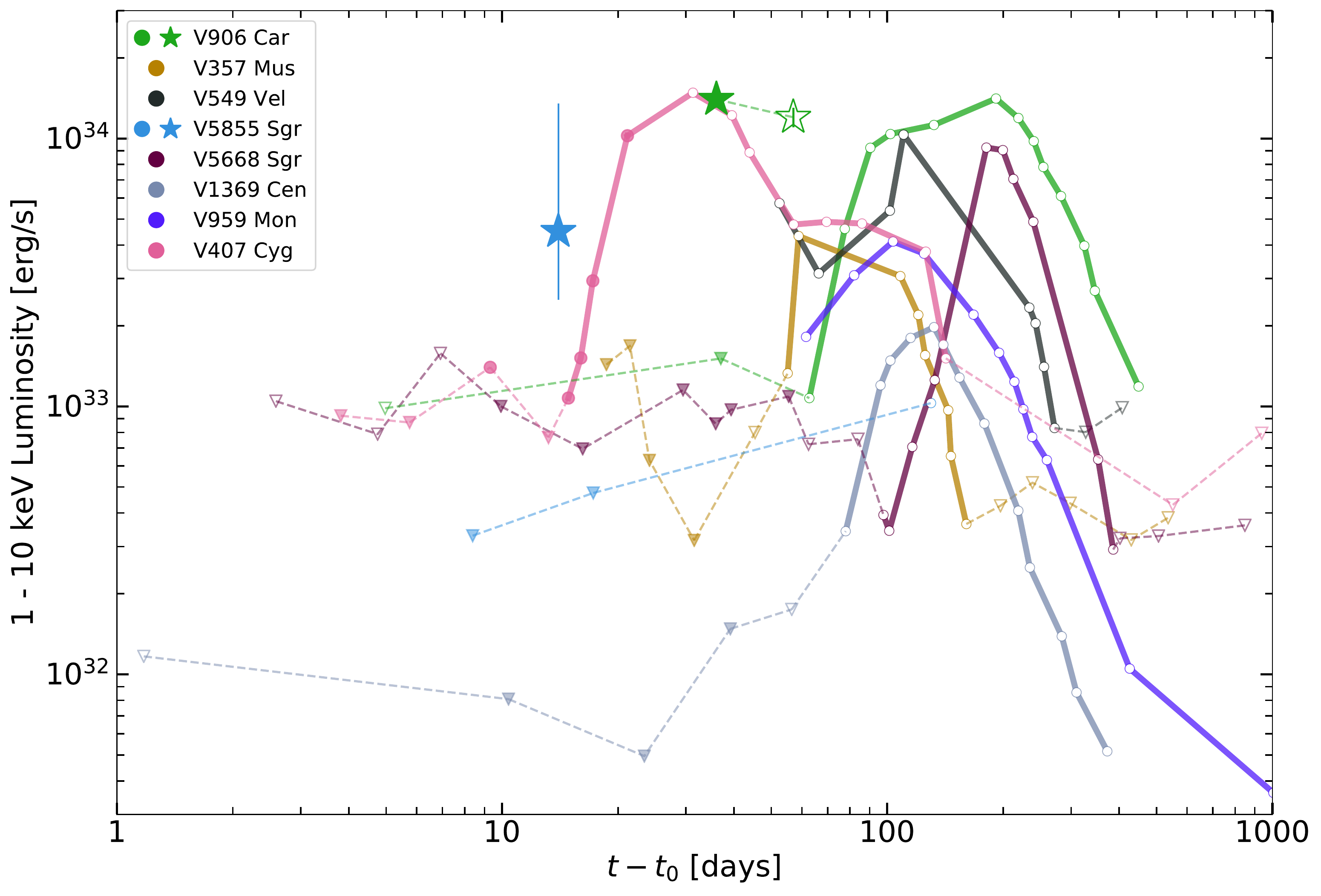}
\vspace{-0.2in}\caption{Hard X-ray (1--10 keV) light curves for eight gamma-ray detected novae. If an observation is concurrent with gamma-ray detection, the symbol is filled. Detections with \emph{Swift}-XRT are marked as circles and are not corrected for internal absorption by the nova ejecta, while \emph{Swift} upper limits are denoted with triangles. 
\emph{NuSTAR} detections are marked as stars, and are corrected for internal absorption \citep{Nelson+19, Sokolovsky+20}.
Assumed distances are listed in Supplementary Table 1.
Figure created by A.\ Gordon.}
\label{fig:xrays}
\end{figure}

Figure \ref{fig:xrays} shows 1--10 keV X-ray light curves for eight gamma-ray detected novae, primarily obtained with the X-ray Telescope (XRT) on {\it Swift}.
The shock-generated X-ray emission typically becomes detectable about a month after eruption and reaches a peak luminosity $L_{\rm X} \approx 10^{33}-10^{34}$ erg/s, which is $\lesssim 10^{-4}$ of the bolometric luminosity and $\lesssim$1\% of the gamma-ray luminosity.  
With the exception of embedded nova V407 Cyg (see {\bf Box 5}), no X-ray emission has been observed in the 1--10 keV band simultaneous with a {\it Fermi}-LAT detection \citep{Gordon+20}.

What is responsible for the delayed rise of the harder X-rays?  In some cases, shock formation itself may be delayed due to a lag between the TNR and expansion of the bulk ejecta (e.g., \citealt{Chomiuk+14b, Nelson+20}). 
However, the early onset of the gamma-ray emission in many novae indicate that some shocks occur well before the first X-ray detection.  In such cases, the lack of X-rays is attributed to absorption that delays detectable X-ray emission until the shocks emerge into an environment with $\tau_{\rm abs} \lesssim 1$ (Figure~\ref{fig:taueff}). 
This is supported by observations of the hard X-ray spectrum, which shows that the absorbing column density decreases with time after eruption \citep[e.g.,][]{Mukai&Ishida01}. 

The shocks can be detected earlier in harder X-rays ($\gtrsim 10$ keV), because of the decreasing ejecta opacity at higher photon energies (Figure~\ref{fig:taueff}).  {\it NuSTAR} has detected hard (3--79 keV) X-rays from three novae simultaneously with \emph{Fermi}-LAT (stars in Figure \ref{fig:xrays}; \citealt{Nelson+19,Sokolovsky+20,Sokolovsky+20_ret20}). 
The hard X-ray emission is best described with optically thin thermal plasma models with inferred temperatures, $kT \approx 6-11$ keV, and absorbing column densities $\Sigma \approx 10^{23}-10^{24}$ cm$^{-2}$ that will easily prevent detection by \emph{Swift}-XRT.  This X-ray emission is too soft to be a low-energy extension of the non-thermal {\it Fermi}-LAT emission \citep{Vurm&Metzger18}.

\begin{marginnote}[]
\entry{$\Sigma$}{Column density of gas between the source and the observer. For gas with number density $n(l)$ along a path length $l$, $\Sigma = \int n(l) dl$.}
\end{marginnote}

Even after correcting for absorption, the observed X-ray luminosities, $L_{\rm X} \approx 10^{33}-10^{34}$ erg/s, are $\sim$100 times lower than the gamma-ray luminosities detected concurrently, and orders of magnitude smaller than theoretically expected \citep{Nelson+19,Sokolovsky+20}.
One solution to this mystery is that the {\it NuSTAR}-detected shocks are distinct from those powering the gamma-rays (or arise from only a portion of the shock front), in which case the latter must be even more deeply embedded.  Alternatively, the X-ray emission from radiative shocks in novae are significantly suppressed relative to the naive expectation of one-dimensional models (\S\ref{sec:multiD}).  Regardless, this implies the bulk of the shock luminosity is reprocessed into UVOIR emission (\S\ref{sec:shockoptical}).

\subsection{Non-Thermal Radio Emission}
\label{sec:radiononthermal}

In addition to the thermal radio emission discussed in \S\ref{sec:radiothermal},
many novae also display non-thermal synchrotron radio emission that must also be associated with shocks \citep{Taylor+87,Krauss+11,Chomiuk+14, Weston+16b, Weston+16a, Finzell+18}.
The primary evidence for synchrotron emission is the high radio surface brightness sometimes observed in the first months of eruption.
Radio surface brightness is characterized by the brightness temperature:
\be
    \frac{T_B}{\rm K}  =  1200 \bigg(\frac{S_\nu}{\rm mJy}\bigg) \bigg(\frac{\nu}{\rm GHz}\bigg)^{-2} \bigg(\frac{\theta}{\rm arcsec}\bigg)^{-2} ,
	\label{eq:brightness_temperature}
\ee
where $S_{\nu}$ is the observed flux density at frequency $\nu$. The angular diameter of the source, $\theta$, can be estimated if the distance, expansion velocity, and expansion time are known, or by directly resolving the emission with e.g., very long baseline interferometry.
On timescales of $\lesssim$100 days, brightness temperatures inferred for novae are often $\gtrsim 10^{5}-10^{6}$ K, while the maximum $T_B$ of a photo-ionized thermal emitter is $\sim10^4$ K.
This implies an additional emission source, 
either from shock-heated thermal gas or shock-induced synchrotron emission. Pioneering work by \citet{Taylor+87} modelled the early radio emission from QU~Vul as shocks between a high-velocity outflow and slower earlier ejecta, similar to the scenario emphasized by this review.

X-ray observations of novae reveal that shocks can heat gas to temperatures $T_{\rm sh} \gtrsim 10^{7}$~K, exceeding the observed radio brightness temperatures (\S\ref{sec:Xrays}).  However, the column of hot gas available at any time is generally insufficient to explain the observed radio fluxes, as constrained by both X-ray observations  (e.g., \citealt{Weston+16a}) and modelling of nova shocks \citep{Metzger+14, Vlasov+16}.
The high radio luminosities and brightness temperatures instead favor synchrotron radiation from relativistic electrons gyrating in magnetic fields behind shocks.

\begin{figure}[!t]
\includegraphics[width=4in]{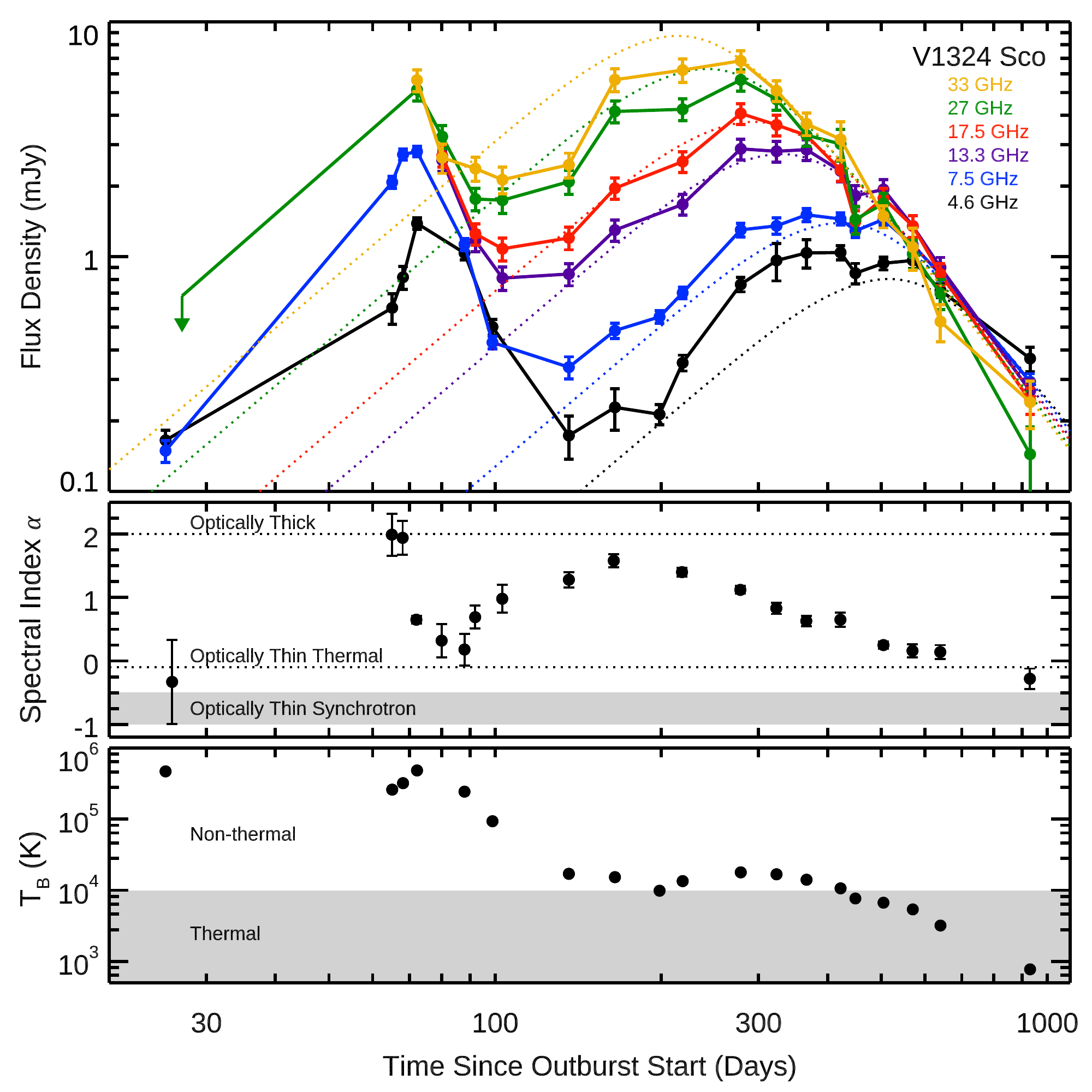}
\vspace{-0.25in}\caption{The radio light curve of gamma-ray detected nova V1324~Sco shows two distinct maxima. The first is likely associated with synchrotron emission, showing $T_B \approx 5 \times 10^5$ K and peaking on day $\sim$70. The second is powered by thermal emission with $T_B \approx 10^4$ K, and peaks $\sim$1 year after eruption. {\bf Top panel:} multi-frequency light curves observed at 4.6--33 GHz with the Karl G.\ Jansky Very Large Array (VLA). A simple ``Hubble flow" fit (dashed lines) demonstrates that the second maximum is reasonably well fit as expanding thermal ejecta \citep[e.g.,][]{Seaquist&Bode08}. {\bf Middle panel:} The spectral index of the radio emission as a function of time. 
{\bf Bottom panel:} The brightness temperature of the 4.6 GHz emission, assuming an expansion velocity of 1000 km/s starting on day 0.  Figure adapted from \citet[][$\copyright$AAS]{Finzell+18}, with permission.}
\label{fig:radiolc}
\end{figure}

Figure~\ref{fig:radiolc} shows an example of this high-$T_B$ early emission during the first $\sim$100 days of the gamma-ray detected nova V1324~Sco \citep{Finzell+18}. Several novae have shown similar distinct, early-time radio maxima that rival the later, $\sim10^4$ K thermal maxima in flux \citep{Taylor+87, Weston+16a}. Other novae show more subtle excesses at early times that nevertheless cannot be explained by $\sim10^4$ K expanding ejecta \citep{Chomiuk+14, Aydi+20}. 

Optically thin synchrotron radiation is characterized by a broad-band power-law spectrum, $F_{\nu} \propto \nu^{\alpha}$, where the spectral index $\alpha$ is determined by the energy spectrum of the emitting relativistic electrons ($dN/d\gamma_p \propto \gamma_p^{-p}$) as $\alpha = -(p+1)/2$. Typically $\alpha \approx -0.5$ to $-1$ is observed for synchrotron emission in relativistic shocks, implying $p= 2-3$ (e.g., SNe, SN remnants; \citealt{Weiler+02, Green19}). 
However, such a negative spectral index rarely describes the high brightness temperature emission in novae (Figure \ref{fig:radiolc}). 
Because of free-free absorption by the nova ejecta, as with the X-rays, early radio emission from the shock will be absorbed (Figure \ref{fig:taueff}).  The flux at lower frequencies will be preferentially suppressed, leading to a rising spectrum ($\alpha >0$).  For a single temperature absorbing medium with a fixed radial column, the low frequency spectral cut-off is exponential.  However, if absorption is important in the multi-temperature post-shock cooling layer where the radio emission is being produced, this causes the spectrum of radiation emerging from the shock (i.e., before absorption by the rest of the ejecta) to be flatter than that of optically thin synchrotron emission \citep{Metzger+14}.  

\begin{marginnote}[]
\entry{Spectral Index $\alpha$}{The spectral energy distribution of radio continuum emission is described as $S_{\nu} \propto \nu^{\alpha}$, where $S_{\nu}$ is the flux density at frequency $\nu$.}
\end{marginnote}

Radio emission (unlike gamma-rays or X-rays) can be imaged at very high angular resolutions ($\sim 10^{-3}$ arcsec), allowing long baseline interferometers to reveal where the shocks are taking place in the nova ejecta. In V959~Mon, synchrotron ``knots" were observed to hug the relatively slow thermal ejecta in the orbital plane, and hypothesized to originate in the collision of fast polar ejecta with this slower equatorial material (Figure \ref{fig:Chomiuk14}; \citealt{Chomiuk+14}). While the synchrotron emission over most of the shock surface is invisible due to free-free absorption (Figure \ref{fig:taueff}), portions of the shock front may peek out at low column densities and appear as synchrotron knots.

\section{THEORY OF INTERNAL RADIATIVE SHOCKS IN CLASSICAL NOVAE}
\label{sec:shockstheory}

The previous sections have described observational evidence in support of internal shocks as common---if not ubiquitous---features of nova outflows.  The most powerful shocks likely take place in the orbital plane of the binary, when an initial slow, equatorially focused outflow of velocity $v_{\rm s} \approx$ few hundred km/s is hit from behind by a fast, more spherically symmetric outflow of velocity $v_{\rm f} \approx$ few thousand km/s (\S\ref{sec:imaging}).
The slow flow may arise  from the outer Lagrange point as the nova envelope first encases the binary orbit (\S\ref{sec:RLOF}), while the fast flow may be a radiation-driven wind (\S\ref{sec:wind}). Whatever their origin, multiple episodes of such slow-fast transitions and collisions can take place in a single nova event, giving rise to distinct optical/gamma-ray flares (Figure~\ref{fig:gamoptcorrelation}).
  
\begin{figure}[!t]
\includegraphics[width=5in]{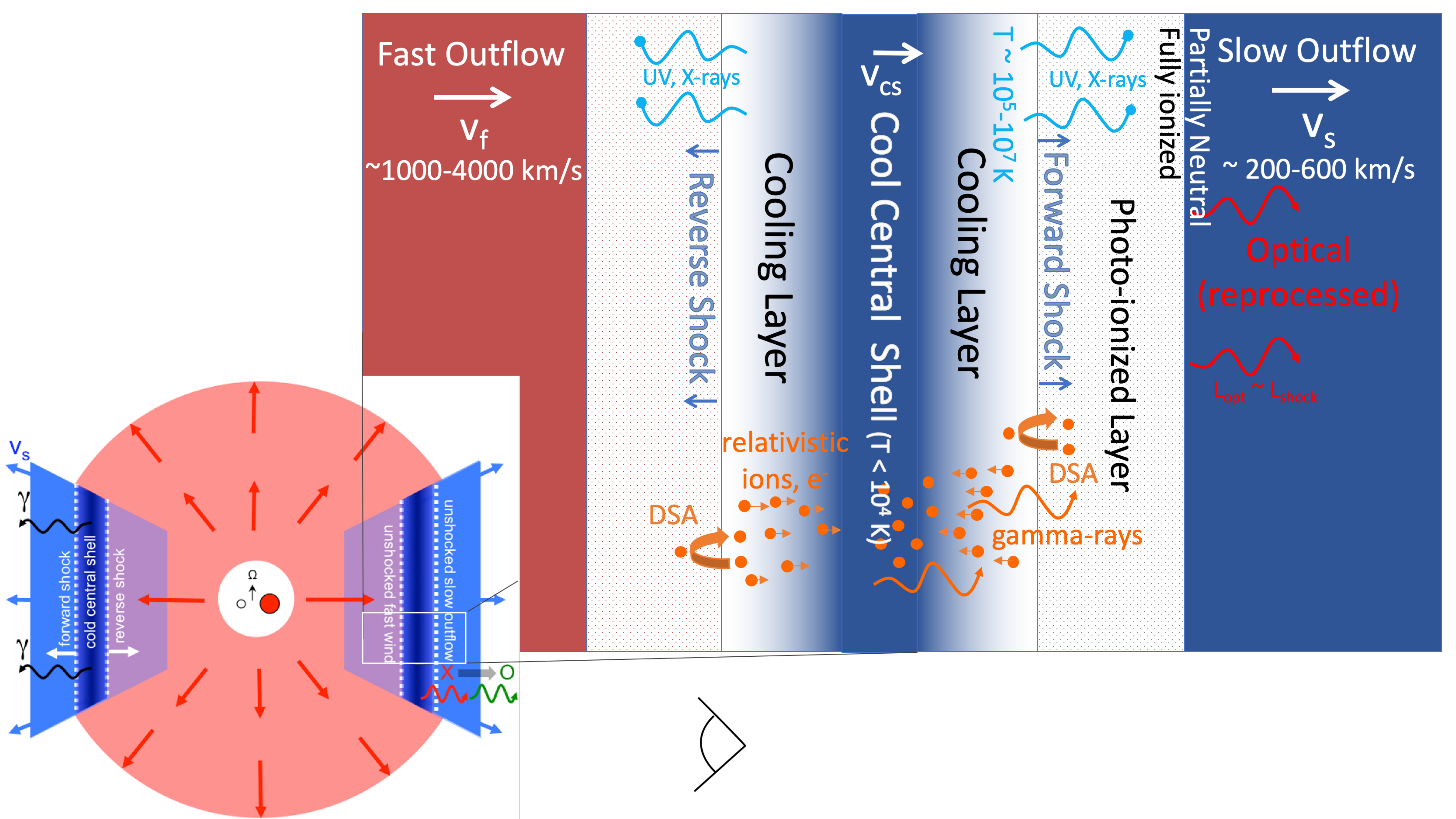}
\caption{Anatomy of a radiative internal shock, zoomed in on a small patch where a fast outflow of velocity $v_{\rm f}$ collides with a slower outflow of velocity $v_{\rm s}$.  A forward shock is driven forward into the slow outflow, while a reverse shock propagates backwards into the fast outflow.  Each shock heats the gas immediately behind it to temperatures $\gtrsim 10^{6}-10^{7}$ K, but the gas quickly cools via UV/X-ray emission in a narrow cooling layer. 
The cool gas collects into a thin central shell that will be corrugated on a scale of order its thickness due to thin-shell instabilities (not drawn; see Supplemental Animation 1). 
UV/X-rays from the shocks are absorbed by the partially neutral upstream flow or by the cool central shell, and their emission is reprocessed into the UVOIR and contributes to the optical light curve (\S\ref{sec:shockoptical}).  Most of the relativistic particles (electrons and ions; orange dots) accelerated at the shocks are advected into the cold central shell, where they emit gamma-rays.  Both optical and gamma-ray production occur with little delay, resulting in correlated optical and gamma-ray light curves.  Figure adapted from \citet{Metzger+15}.}
\label{fig:cartoon}
\end{figure}

As we demonstrate in \S\ref{sec:dynamics}, in the first weeks of many nova eruptions, the internal shocks are likely to be radiative---that is, they will radiate away their thermal energy faster than it can be reconverted back into kinetic energy via PdV expansion. This section presents the basic theory of internal radiative shocks  and how they can explain observations of novae.  We start by considering a one-dimensional toy model of the shock interaction (\S\ref{sec:dynamics}) and then outline particle acceleration (\S\ref{sec:Emax}) and radiative signatures (\S\ref{sec:shockoptical}).   In $\S\ref{sec:multiD}$ we discuss multi-dimensional effects not captured by a one-dimensional picture.  Figure \ref{fig:cartoon} shows a schematic diagram of the shock structure and its electromagnetic emissions.  

\subsection{Dynamics of Dual Radiative Shocks}
\label{sec:dynamics}

We model both the fast and slow flows as freely expanding winds with constant mass-loss rates of $\dot{M}_{\rm f}$ and $\dot{M}_{\rm s}$.  The fast outflow is assumed to be spherically symmetric, while the slow outflow is confined to a fractional solid angle $f_{\Omega} < 1$ in the orbital plane.  The number densities in the outflows at 
radius $r$ are:
\be
n_{\rm s} = \frac{\dot{M}_{\rm s}}{4\pi f_{\Omega} r^{2}  \mu m_p v_{\rm s}}\ \ \ \ \  {\rm and}\ \ \ \ \ n_{\rm f} = \frac{\dot{M}_{\rm f}}{4\pi r^{2} \mu m_p v_{\rm f}}.
\label{eq:nf}
\ee 
Given that novae typically eject $M_{\rm ej} \approx 10^{-5}-10^{-4}\,M_{\odot}$ over timescales of weeks (\S\ref{sec:massloss}), characteristic values are $\dot{M}_{\rm f}, \dot{M}_{\rm s} \approx  10^{-6}-10^{-5}\,M_{\odot}$/wk.  
This model is sufficient to explain the basic physics, even if the actual flows are more complex.

We define $t_{\rm sh} = 0$ as the time when the fast outflow turns on abruptly and collides with the earlier slow flow, driving a dual forward-reverse shock structure outwards in radius.  
As verified below, over at least the first few weeks of interaction, both shocks are likely to be radiative, and hence radiate all of the energy dissipated by the collision.  The cooled gas behind the shocks will collect in a thin, cold (and clumpy; \S\ref{sec:multiD}) shell of mass $M_{\rm cs}$ and velocity $v_{\rm cs}$.
Because the post-shock gas cools so quickly as to exert effectively no pressure,  the velocities of the forward and reverse shocks in the WD frame also equal $v_{\rm cs}$.  The mass and momentum of the cold shell thus grow, as  (e.g.,~\citealt{Metzger+14})
\be
\frac{dM_{\rm cs}}{dt_{\rm sh}} = f_{\Omega}\dot{M}_{\rm f}\left(\frac{v_{\rm f} - v_{\rm cs}}{v_{\rm f}}\right) + \dot{M}_{\rm s}\left(\frac{v_{\rm cs}-v_s}{v_s}\right),
\ee
\be \frac{d}{dt_{\rm sh}}\left(M_{\rm cs}v_{\rm cs}\right) = f_{\Omega}\dot{M_{\rm f}}(v_{\rm f}-v_{\rm cs}) + \dot{M}_{\rm s}(v_{\rm cs}-v_{\rm s}).
\ee
Numerical integration of these equations reveal an approximate steady state ($dv_{\rm cs}/dt_{\rm sh} \approx 0$) in cases when the shell gains most of its momentum from the fast wind, but most of its mass from the slow ejecta shell. In this case,
\be
v_{\rm cs} \approx \left(\frac{\dot{M}_{\rm f}}{\dot{M}_{\rm s}}\frac{v_{\rm f}f_{\Omega}}{v_{\rm s}}\right)^{1/2}v_{\rm s} \equiv \xi v_{\rm s},
\ee
yielding $\xi \sim$ few for typical parameters (e.g., $\dot{M}_{\rm f} \approx \dot{M}_{\rm s}; f_{\Omega} \approx 0.3; v_{\rm f} \approx 10v_{\rm s}$).  
The radius of the cold shell thus grows as 
\be
R_{\rm cs} \approx v_{\rm cs} t_{\rm sh} \approx 5.4\times 10^{13}\,{\rm cm}\, \left(\frac{\xi}{3}\right) \left(\frac{v_{\rm s}}{300\,{\rm km/s}}\right)\left(\frac{t_{\rm sh}}{1\rm \,wk}\right),
\label{eq:Rsh}
\ee 
and must lie outside the binary orbit in classical novae on the timescales of the observed gamma-rays (Eq.~\ref{eq:abin}).  

We can now confirm that the shocks are radiative, if we compare the expansion time, $t_{\rm exp} = R_{\rm cs}/v_{\rm cs} = t_{\rm sh}$, to the radiative cooling time of the shocked gas, $t_{\rm cool} = 3kT_{\rm sh}/(2n_{\rm sh}\Lambda[T = T_{\rm sh}])$. Here $T_{\rm sh}$ (Eq.~\ref{eq:Tsh} with $v = v_{\rm cs}$) and $n_{\rm sh}$ are the temperature and density of the shocked gas, respectively, 
and $\Lambda \approx 3\times 10^{-27}(T_{\rm sh}/{\rm K})^{1/2}\,{\rm erg\,cm^{3}\,s^{-1}}$ is the free-free cooling rate (although line cooling can dominate for low shock velocities). 

\begin{marginnote}[]
\entry{$t_{\rm exp}$}{The expansion time, or the timescale on which the radius of the central shell and shocks will double, $t_{\rm exp} = R_{\rm cs}/v_{\rm cs}$.}
\entry{$t_{\rm cool}$}{The radiative cooling timescale, where $\frac{d(ln T)}{dt}\,t_{\rm cool} \approx 1$. Here, $t_{\rm cool} = 3kT_{\rm sh}/(2n_{\rm sh}\Lambda[T = T_{\rm sh}])$.}
\end{marginnote}

Taking $n_{\rm sh} = 4n_{\rm f}[r = R_{\rm cs}]$ for the reverse shock ($n_{\rm sh} \approx 10^{10}$ cm$^{-3}$ for fiducial parameters at $t_{\rm sh} \sim 1$ week, where the factor of 4 comes from the shock jump conditions), we have:
\begin{eqnarray}
\frac{t_{\rm cool}}{t_{\rm exp}} \approx 0.1\left(\frac{\xi}{3}\right)^{2}\left(\frac{\dot{M}_{\rm f}}{10^{-5}\,M_{\odot}/{\rm wk}}\right)^{-1}\left(\frac{v_{\rm f}}{2000\,{\rm km/s}}\right)^{2}\left(\frac{v_{\rm s}}{300\,{\rm km/s}}\right)^{2}\left(\frac{t_{\rm sh}}{\rm 1\,wk}\right).  \label{eq:tcool}
\end{eqnarray}
The reverse shock is thus radiative ($t_{\rm cool} \ll t_{\rm exp}$) over the first few weeks for fiducial parameters. The forward shock can remain radiative for even longer, due to its lower velocity, and higher density of the slow flow.

\subsection{Relativistic Particle Acceleration}
\label{sec:Emax}

A small fraction of the charged particles that enter the shocks can be accelerated to relativistic velocities by the diffusive shock acceleration process, if 
a strong and turbulent magnetic field deflects particles back and forth across the shock front in the locally converging flow (e.g., \citealt{Blandford&Ostriker78}; see \citealt{Blasi19} for a review).  The magnetic field strength near the reverse shock can be estimated from equipartition arguments,
$B_{\rm sh} = v_{\rm cs}\sqrt{6\pi \epsilon_B \mu m_p n_{\rm f}}$, 
yielding $B_{\rm sh} \approx 3$\,G for fiducial parameters $\sim$1 week after eruption (\S\ref{sec:dynamics}) and an efficiency of magnetic field amplification $\epsilon_B = 0.01$.  A promising mechanism for amplifying the magnetic field in non-relativistic shocks is the cosmic ray current-driven instability (\citealt{Bell04}), where streaming relativistic particles amplify fluctuations in the magnetic field.
Numerical plasma simulations of this instability imply $\epsilon_{B} \gtrsim 10^{-4}-10^{-2}$, depending on the ion acceleration efficiency and the Alfv\'en Mach number (e.g.,~\citealt{Caprioli&Spitkovsky14b}).

Up to what energy can particles be accelerated?  In diffusive shock acceleration, as cosmic rays gain greater and greater energy $E$, they can diffuse back to the shock from a greater downstream distance, $z$, because of their larger gyroradii $r_{\rm g} = E/eB_{\rm sh}$.  The maximum energy to which particles are accelerated before escaping, $E_{\rm max}$, is found by equating the upstream diffusion time $t_{\rm diff} \sim D/v_{\rm cs}^{2}$ to the minimum of various particle loss timescales.  These include the downstream advection timescale $t_{\rm adv} \sim z_{\rm acc}/v_{\rm cs}$, where $z_{\rm acc}$ is the width of the acceleration zone, and (in hadronic scenarios) the pion creation timescale $t_{\pi} = (n_{\rm sh} \sigma_{\pi} c)^{-1}$, where  $\sigma_{\pi} \sim 2 \times 10^{-26}$ cm$^{2}$ is the inelastic cross section for p-p interactions \citep{Kamae+06}.  For example, equating $t_{\rm diff} = t_{\rm adv}$ and taking $D \approx r_{\rm g}c/3$ as the diffusion coefficient \citep{Caprioli&Spitkovsky14b}, one obtains
\begin{eqnarray}
&& E_{\rm max} \sim \frac{3 e B_{\rm sh} v_{\rm cs}z_{\rm acc}}{c} 
\approx  4\times 10^{5}{\rm GeV}\,\left(\frac{z_{\rm acc}}{R_{\rm cs}}\right)\left(\frac{\xi}{3}\right)\times \nonumber \\
&&\left(\frac{\epsilon_{B}}{10^{-2}}\right)^{1/2}\left(\frac{\dot{M}_{f}}{10^{-5}\,M_{\odot}/{\rm wk}}\right)^{1/2}\left(\frac{v_{\rm f}}{2000\,{\rm km/s}}\right)^{-1/2}\left(\frac{v_{\rm s}}{300\,{\rm km/s}}\right)
\label{eq:Emax}
\end{eqnarray}
if we normalize the radial width of the acceleration zone, $z_{\rm acc}$, by the shock radius, $R_{\rm cs}$ (Eq.~\ref{eq:Rsh}).  

\begin{marginnote}[]
\entry{$E_{\rm max}$}{The maximum energy to which relativistic particles are accelerated (e.g., by bouncing back and forth across the shock front).}
\end{marginnote}

In the case of fully ionized, non-radiative (adiabatic) shocks, it may be justified to take $z_{\rm acc} \sim R_{\rm cs}$, in which case particle acceleration up to $E_{\rm max} \gtrsim 10^{14}$ eV would appear possible.  However, in novae $z_{\rm acc}$ may be limited to much smaller values by several effects (e.g.~\citealt{Metzger+16}).  
For example, thermal cooling compresses the post-shock region to a characteristic width $z_{\rm cool} \sim v_{\rm cs}t_{\rm cool} \sim  (t_{\rm cool}/t_{\rm exp})R_{\rm cs} \lesssim (10^{-2}-10^{-1})R_{\rm cs}$ (Eq.~\ref{eq:tcool}), and would imply $E_{\rm max} \lesssim 10^{4}$ GeV if strong magnetic fields and particle acceleration are limited to this region (e.g.~due to damping of cosmic ray-driven instabilities by neutral-ion collisions; \citealt{Metzger+16}).  Energy loss due to pion creation may also limit the value of $E_{\rm max}$, particularly in the calorimetric limit that $t_{\pi} \ll t_{\rm adv}$ (see below).  Consistent with this, the gamma-ray spectra of some novae show evidence for high-energy cut-offs in their spectra above 10 GeV (Supplementary Table 1; \citealt{Ahnen+15})
which in hadronic models requires an intrinsic cut-off of the accelerated particle spectrum above $E_{\rm max} \sim 10E_{\gamma} \approx 100$ GeV, much less than suggested by equation (\ref{eq:Emax}) for $z_{\rm acc} \sim R_{\rm cs}$.


The nova ejecta can be sufficiently dense during the period of gamma-ray emission that the timescale for relativistic ions or electrons to radiate their energy in $\sim$ GeV gamma-rays is short compared to the ejecta expansion time (e.g., \citealt{Metzger+15}).  In other words, the shocks are radiative in terms of both their non-thermal and thermal particles.  This ``calorimetric" feature of nova shocks enables multi-wavelength observations to probe the fraction of the shock's kinetic luminosity placed into relativistic particles, $\epsilon_{\rm rel}$.  Since both forward and reverse shocks are radiative, the bolometric luminosity ($L_{\rm bol}$; dominated by the UVOIR bands around optical maximum) provides an upper limit on the total shock power (an upper limit because the remaining luminosity originates from the WD).  
Likewise, a fraction $\epsilon_{\gamma}$ of the relativistic particle energy is emitted as gamma-rays in the LAT bandpass ($\epsilon_{\gamma} \approx 0.3-0.4$ in the hadronic scenario; \citealt{Vurm&Metzger18}).  Thus, one can place a lower limit of $\epsilon_{\rm rel} \gtrsim \epsilon_{\gamma}^{-1}(L_{\gamma}/L_{\rm bol})$,
given the distant-independent luminosity ratio $L_{\gamma}/L_{\rm bol}$ \citep{Metzger+15}.  

\begin{marginnote}[]
\entry{$\epsilon_{\rm rel}$}{The fraction of the shock kinetic energy that is transferred to accelerating particles to relativistic speeds.}
\end{marginnote}

The bottom panel of Figure~\ref{fig:gamoptcorrelation} shows the time evolution of the ratio $L_{\gamma}/L_{\rm BB}$ for several gamma-ray novae, where $L_{\rm BB}$ is a best-fit blackbody luminosity that we take as an estimate of $L_{\rm bol}$.  The ratio varies from $3\times 10^{-4}$ in V339~Del to $\sim 10^{-2}$ in V1324~Sco, implying $\epsilon_{\rm rel} \gtrsim 10^{-3}-3\times 10^{-2}$.  In cases like V5856 Sgr and V906 Car, where the optical/gamma-ray correlation demonstrate that a large fraction of the bolometric luminosity is powered by the shocks (Figure~\ref{fig:gamoptcorrelation}), the lower limit on $\epsilon_{\rm rel}$ instead becomes an equality, enabling a measurement of $\epsilon_{\rm rel} \approx (2-5)\times 10^{-3}$ \citep{Li+17,Aydi+20}.

If the gamma-ray emission from novae is generated via the hadronic channel (\S\ref{sec:gammaspectra}), then an inescapable prediction is a luminosity of $\sim 0.1-10$ GeV neutrinos from $\pi^{\pm}$ decay that nearly mirrors the gamma-ray luminosity \citep{Razzaque+10,Metzger+16}.  
For an integration time of a few weeks (comparable to the duration of the observed gamma-ray emission), IceCube DeepCore \citep{Abbasi+12} can reach a sensitivity to neutrinos of $\approx 10^{-5}-10^{-4}$ GeV cm$^{-2}$ s$^{-1}$.
Unfortunately, this is $\gtrsim$ 4 orders of magnitude larger than the predicted neutrino fluxes $\approx 10^{-8}$ GeV cm$^{-2}$ s$^{-1}$ of the present sample of gamma-ray novae (Alex Pizzuto, private communication), making a detection unlikely with current generation experiments.  If the gamma-ray spectra in some novae extend to $\gtrsim$ TeV energies (\S\ref{sec:future}), then the prospects of a neutrino detection would be improved.

\subsection{Radiative Signatures of Radiative Shocks}
\label{sec:shockoptical}

When $v_{\rm f} \gg v_{\rm s}$, the reverse shock is more powerful than the forward shock and its luminosity
\be
L_{\rm sh} \approx \frac{9}{32}f_{\Omega}\dot{M}_{\rm f}v_{\rm f}^{2} \approx 1.0\times 10^{38}\,{\rm erg/s}\left(\frac{f_{\Omega}}{0.3}\right)\left(\frac{\dot{M}_{\rm f}}{10^{-5}\,M_{\odot}/{\rm wk}}\right)\left(\frac{v_{\rm f}}{2000\,{\rm km/s}}\right)^{2},
\label{eq:Lsh}
\ee
is comparable to the bolometric output of the nova near maximum.  Because the shocks are radiative, $L_{\rm sh}$ must escape as radiation.  However, the appearance of this radiation is strongly influenced by the medium ahead of the shocks. 

The column density of gas separating the forward shock ($r = R_{\rm cs}$) from an observer is 
\be
\Sigma = \int_{R_{\rm cs}}^{\infty}\mu n_{\rm s}dr \approx 6\times 10^{23}\,{\rm cm^{-2}} \left(\frac{\xi}{3}\right)^{-2}\left(\frac{\dot{M}_s}{10^{-5}\,M_{\odot}/{\rm wk}}\right)\left(\frac{v_{\rm s}}{300\,{\rm km/s}}\right)^{-2}\left(\frac{t_{\rm sh}}{1\,\,{\rm wk}}\right)^{-1},
\label{eq:tauT}
\ee
The column density to the reverse shock will be even higher, $\Sigma \approx 10^{24}-10^{26}$ cm$^{-2}$, because of the additional contribution from the central swept-up shell.  

Figure \ref{fig:taueff} shows the optical depth of the ejecta to absorption when $\Sigma \approx 10^{25}$ cm$^{-2}$, as expected roughly a week after the start of eruption for fiducial parameters.  At radio frequencies, the optical depth due to free-free absorption is extremely high at these early times, but it will decrease rapidly as $\propto t^{-3}$ or faster as the shocks propagate outwards, eventually enabling the escape of non-thermal synchrotron emission on timescales of months (\S\ref{sec:radiononthermal}).  Likewise, the far-UV/X-ray opacity is enormous when $\Sigma \approx 10^{25}$ cm$^{-2}$, but it decreases as $\Sigma \propto t^{-1}$ or faster as the shocks propagate outwards, 
eventually allowing the escape of X-rays (\S\ref{sec:Xrays}). In the first days--weeks of eruption, the substantial UV/X-ray luminosity of the shocks ($L_{\rm sh} \approx 10^{37}-10^{38}$ erg/s near peak; Eq.~\ref{eq:Lsh}) will be absorbed and reprocessed to wavelengths that can escape, contributing to the UVOIR (Figure \ref{fig:gamoptcorrelation}).
There is absorption in the OIR bands from a combination of free-free absorption and atomic absorption lines  
($\S\ref{sec:spectra}$), but it is significantly lower than at radio or X-ray wavelengths.

Shock emission combined with $\tau_{\rm abs} < 1$ in both the optical and {\it Fermi} bands is the key to producing correlated optical and gamma-ray light curves in novae (Figure~\ref{fig:gamoptcorrelation}). 
The correlation implies that a significant fraction of the nova optical luminosity around maximum is powered by shocks \citep{Metzger+14, Li+17,Aydi+20}, rather than by radiation from the nuclear-burning WD (\S\ref{sec:SSS}).  
Some of the puzzling variability observed in nova optical light curves (\S\ref{sec:lightcurves}; Figure~\ref{fig:rogues}) may be attributable to variability in the luminosity of the shocks (e.g., Eq.~\ref{eq:Lsh}), powered by fluctuations in the mass-loss rate and/or velocity of the fast or slow outflows.

A delay of days is observed between the start of the optical rise in novae and the onset of gamma-ray emission (Figure \ref{fig:gammaLC}).  This delay could be the timescale over which the fast wind is established, which then drives the shock interaction with the prior slow flow.  Alternatively, shock interaction could start even earlier than gamma-rays are observed, but at such large column densities, $\Sigma \gg 10^{26}$ cm$^{-2}$, that attenuation by photo-nuclear pair creation (orange line in Figure \ref{fig:taueff}) initially absorbs the gamma-rays \citep{Li+17,Franckowiak+18}.  
When the column densities are so high, the shocks are also mediated by radiation instead of collisionless plasma processes, leading to suppression of non-thermal particle acceleration (e.g.,~\citealt{Blandford&Payne81}) and hence  gamma-ray emission.

On the other hand, the faithful tracking of optical and gamma-ray light curves, with optical lagging the gamma-rays by $\lesssim$8 hours in V906~Car (Figure \ref{fig:gamoptcorrelation}), implies that particles are accelerated promptly, and  conversion of cosmic-ray to gamma-ray energy is effectively instantaneous.  The timescale required for the gamma-ray emitting particles to be accelerated to high energies at the shock is at most the advection time through the cooling layer (see derivation of Eq.~\ref{eq:Emax}), which is a small fraction of the expansion time.  Likewise, once accelerated, the timescale for relativistic ions to interact with the dense gas behind the shocks and produce pions and hence gamma-rays is extremely short ($\lesssim$ 1 day;  \citealt{Metzger+15,Vurm&Metzger18,Martin+18}).  A small lag in the arrival time of the optical radiation relative to the gamma-rays, such as the $5.3\pm 2.7$ hr lag observed in V906 Car \citep{Aydi+20}, can arise due to the finite diffusion time of optical photons through the ejecta, placing a constraint on the density profile ahead of the shocks.

\subsection{Multi-Dimensional Effects and Dust Formation}
\label{sec:multiD}

Although much insight can be gleaned from one-dimensional models, the structure of radiative internal shocks is intrinsically multi-dimensional.
On the largest scales, the interaction between a fast outflow and an equatorially concentrated slower flow will not only generate a dual shock structure in the equatorial plane (Figure \ref{fig:cartoon}), but 
may also redirect the fast flow into a bipolar/hourglass structure, consistent with the large-scale morphology of nova ejecta inferred from high-resolution imaging (Figures \ref{fig:opticalimages} and \ref{fig:Chomiuk14}).  This is qualitatively similar to the ``interacting winds'' scenario for the morphologies of planetary nebulae (\citealt{Soker&Livio89}).

On smaller scales, rapid cooling behind the shocks results in enormous increases in mean gas density by a factor up to $\mathcal{M}^{2} \sim 10^{4}(v/1000\,{\rm km/s})^{2}$, where $\mathcal{M} \equiv v/c_{\rm s} \approx 100$ is the Mach number of the shock of velocity $v$.
The compressed material collects into a thin shell with a thickness that is typically less than a few percent of its radius (e.g., \citealt{Steinberg&Metzger18}). 

Radiative shocks are prone to various instabilities that lead to complex dynamical and geometrical structures and play an important role in shaping their emission relative to adiabatic shocks.  These include thermal instabilities due to radiative cooling that can transform the temperature profile of the post-shock cooling region from a smooth gradient into a turbulent multi-phase plasma with distinct pockets of hot and cool gas (and little ``warm" gas in between; e.g., \citealt{Chevalier&Imamura82}).  The dense thin layer of gas behind dual radiative shocks is also subject to a ``non-linear thin-shell instability" (\citealt{Vishniac94}).  This occurs because lateral perturbations of the interaction front cause material to be diverted in such a way that the nominally planar shock front is given a corrugated shape \citep{Strickland&Blondin95}.  
Supplemental Animations 1 and 2 show the evolution of the temperature and density structure behind dual radiative shocks created by the head-on collision of two flows, illustrating the effects of these instabilities.

The high compression ratios and clumpy structure of the shocked ejecta may produce conditions conducive to grain nucleation and dust formation (\citealt{Derdzinski+17}; \S\ref{sec:dust}). The cold central shell achieves densities $\gtrsim10^{14}$ cm$^{-3}$, which can be maintained as the shell expands to large enough radii for the ambient temperature to drop sufficiently to enable dust nucleation. \cite{Derdzinski+17} find that dust grains readily grow to diameters $\sim 1\, \mu$m under these conditions, and yield dust masses sufficient to explain the observational signatures of dust in novae.
A prediction of this scenario is that dust extinction events will be preferentially observed for novae viewed within the binary plane; future imaging of nova dust emission with ALMA or IR interferometry can test this idea (\S\ref{sec:imaging}).

Another consequence of instabilities at radiative shocks is a reduction of the thermal X-ray luminosity as compared to the expectation from a one-dimensional analysis that most of the shock power will be radiated by gas of temperature $\sim T_{\rm sh}$ (Eq.~\ref{eq:Tsh}).  Firstly, the corrugated shape of the shock front imprinted by the thin-shell instability reduces $T_{\rm sh}$ from the case of a shock front that is strictly perpendicular to the upstream velocity.  Furthermore, the hot post-shock gas can drive weak shocks into under-pressurized cold regions in the thin shell, transferring energy from the hot to cool phase before it can be radiated and thus further reducing the effective temperature of the shock emission \citep{Steinberg&Metzger18}.  Both effects may contribute to the unusually weak thermal X-ray emission seen from novae, even at times when luminous gamma-ray emission points to powerful shocks (\S\ref{sec:Xrays}; Figure~\ref{fig:xrays}).  

Finally, the ion acceleration efficiencies $\epsilon_{\rm rel} \sim 0.3-1\%$ inferred for nova shocks (\S\ref{sec:Emax}; Figure \ref{fig:gamoptcorrelation}) are notably lower than the $\sim 10\%$ efficiency typically inferred for cosmic ray acceleration in SN remnants \citep{Ellison+07}.  However, the nature of the upstream magnetic field, and the radiative nature of the shocks in novae, are different from the SN case.  We expect that the magnetic field advected from the WD will be oriented perpendicular to the direction of the outflow velocity and thus parallel to the radially propagating shock front (similar to the ``Parker spiral" magnetic field topology of the solar wind).  Particle-in-cell plasma numerical simulations of non-relativistic magnetized shocks indicate that ion acceleration is substantially suppressed for this upstream magnetic field geometry \citep{Caprioli&Spitkovsky14}.  Nevertheless, due to the corrugated shape of radiative shock fronts (Supplemental Animations), local patches of the shock front may satisfy the conditions for efficient ion acceleration and a shock-front averaged efficiency of $\sim 1\%$ may be obtained \citep{Steinberg&Metzger18}.

\section{BRINGING IT TOGETHER: IMPLICATIONS AND OPEN QUESTIONS}\label{sec:implications}

In this final section, we discuss the implications of nova mass loss and shocks for the long-term evolution of WD binaries (\S \ref{sec:cvpop}) and more energetic astrophysical transients (\S\ref{sec:snia}, \S\ref{sec:transients}). We close by reviewing the most important open questions in nova studies, and promising avenues for addressing them (\S\ref{sec:future}).

\subsection{Long-Term Evolution of WD Binaries}
\label{sec:longterm}

Novae induce mass loss (and potentially, angular momentum loss) from accreting WD binaries, which has implications for how these binaries evolve and the end product of the WD itself.

\subsubsection{Implications for CV Populations}\label{sec:cvpop}
 
In standard models (e.g., \citealt{Rappaport+83}), mass transfer in CVs with orbital periods $P \approx 3-12$ h is driven by the loss of angular momentum due to braking by the secondary's magnetized wind.  At shorter orbital periods ($P \lesssim 2$ h), the secondary no longer maintains a strong or ordered magnetic field, and angular momentum loss is dominated by gravitational wave radiation.  As the period of a CV shrinks over time, the mean mass-transfer rate generally decreases (consistent with observations; \citealt{Patterson84,Pala+17}, although note the high accretion rate SW Sextantis stars; \citealt{Rodriguez+07}) and thus the interval between nova eruptions should increase.  By contrast, if the donor is a sub-giant or giant, mass transfer is instead governed by the companion's nuclear evolution, driven by the expansion of the secondary's envelope or its wind.

Does a nova eruption drive the WD and its companion closer together or further apart? If mass is lost from the binary (in the form of nova ejecta) but angular momentum is conserved, the binary separation will expand \citep{Frank+92}; this may lead to a decline in accretion rate, and is the origin of the hypothesis that CVs ``hibernate" after a nova eruption \citep{Prialnik&Shara86,Shara+86,Kovetz+88,Hillman+20}.
However, if angular momentum is removed from the binary (i.e., through common-envelope-like interaction with the ejecta; \S\ref{sec:RLOF}), the binary separation may decrease \citep{Livio+91}.
In addition to frictional angular momentum loss, other processes like the companion star's magnetic field ``braking" on the nova ejecta \citep{Martin+11}, asymmetric expulsion of nova ejecta \citep{Nelemans+16, Schaefer+19}, or angular momentum extracted by a cirumbinary disk \citep{Taam&Spruit01, Liu&Li16} may decrease the orbital separation during or following a nova eruption.

While the nova's impact on the orbital separation should be measurable by comparing the orbital period before and after nova eruption, such measurements have proved challenging---usually because we lack excellent time-series photometry for the pre-nova binary. The literature now includes $\sim$10 measurements of period changes across nova eruptions (Supplementary Table 2). They yield a mix of positive and negative values, and show no clear correlation with nova or binary properties \citep{Schaefer20b}. 
As a sample, however, they imply that---at least sometimes---angular momentum loss wins out over mass loss, and nova eruptions can drive their binaries to smaller separations. In the future, deep all-sky high-cadence surveys 
(e.g., the Legacy Survey of Space and Time with the Vera C.\ Rubin Observatory) will improve prospects for measuring changes in orbital period across nova eruptions.  
 
Understanding how much angular momentum is lost in nova eruptions---and how it scales with system parameters like WD mass---is a critical ingredient in modelling CV populations.
Although a wide range of observational evidence supports the standard model of CV evolution (e.g., \citealt{Townsley&Bildsten03,Knigge06,Schreiber+10}), several important discrepancies remain between observations and models.
The CV space density is observed to be $\sim$10-100 times lower than theoretically predicted \citep[e.g.,][]{Patterson98, Pala+20}, 
the minimum CV orbital period is observed to be longer than theoretically predicted (e.g., \citealt{Gansicke+09,Knigge+11}), and the masses of WDs in CVs are observed to be larger than those in their progenitor population \citep{Zorotovic+11}.
\citet{Schreiber+16} argue that all of these discrepant properties can be understood if CV binaries experience additional angular momentum loss beyond magnetic braking and gravitational waves during their evolution. As discussed above, nova eruptions provide several potential sinks of angular momentum. These are starting to be included in models of CV populations, but they must act preferentially on systems with low-mass WDs to explain the discrepancies between CV observations and theory \citep{Nelemans+16, Liu&Li16}.  Conveniently, nova angular momentum losses may be more severe for binaries with lower mass WDs, because lower-mass WDs typically generate more slowly expanding, higher mass ejecta (\S\ref{sec:TNR}), and may therefore have stronger ``common envelope" interactions (\S\ref{sec:RLOF}).
One speculative---and intriguing---mechanism for explaining the dearth of low-mass WDs in the CV population is merger of the WD with its companion brought on by slow nova eruptions.

\begin{textbox}[h]\section{BOX 7: Nova ``Feedback" on the Companion Star and Mass Transfer}
Although the standard picture of CV evolution predicts that the mass transfer rate $\dot{M}$ should mainly depend on the binary orbital period, CVs actually show a broad range of accretion rates at a given period, spanning more than two orders of magnitude \citep{Patterson84, Patterson11, Warner87}.  Novae may be the cause of this diversity, because, in addition to producing discrete changes in orbital separation (\S\ref{sec:cvpop}), the nova eruption ablates and irradiates the companion star \citep{Kovetz+88, Figueira+18}.  Irradiation of the secondary may lead to expansion of its atmosphere and a temporarily enhanced $\dot{M}$ for decades to millenia after the nova, until the WD has cooled down \citep[e.g.,][]{Mroz+16, Hillman+20}.

If the impact of the nova on the secondary star is sufficiently strong, a positive feedback process can be established, where the enhanced $\dot{M}$ results in more frequent novae and even greater heating of the companion \citep{Knigge+00}.  This effect would be stronger for smaller binary separations, and may explain the recurrent nova T Pyx, which has a very short orbital period $P = 1.83$\,h and a quiescent bolometric luminosity an order of magnitude greater than expected from CV theory \citep{Patterson+17, Godon+18}. Such positive feedback from nova cycles could lead to faster erosion of the companion star and the disappearance of short-period binaries from the CV population \citep{Patterson+17}.
\end{textbox}

\subsubsection{Progenitors of Thermonuclear Supernovae}\label{sec:snia}
A Type Ia SN results from the thermonuclear explosion of a CO WD.  One way to ignite runaway carbon burning in the center of a WD is by gradually increasing its mass up to a critical value near the Chandrasekar limit through accretion from a non-degenerate binary companion \citep{Whelan&Iben73,Nomoto82,Nomoto+84}.
Since the maximum initial mass of a CO WD is $\sim 1.1\, M_{\odot}$ \citep{Doherty+15}, the viability of this ``single degenerate" channel depends on whether the accreting WD
can accumulate and retain $\gtrsim$0.3$\, M_{\odot}$ of material.  Even if a CO WD can grow to $\sim M_{\rm Ch}$, it may not produce a detonation, but instead lead to a pure deflagration event (recently hypothesized to explain Type Iax SNe, a subtype of thermonuclear explosions that are dimmer and less energetic than ordinary Type Ia SNe; \citealt{Foley+13}).

The overabundance of C, O, and Ne observed in the ejecta of many novae is strong evidence for mixing between the underlying WD and the accreted envelope ($\S\ref{sec:TNR}$), so that the net effect of a nova is almost certainly to decrease rather than increase the WD mass due to the excavation and ejection of WD material. 
Only the most massive WDs and those with accretion rates just below $\dot{M}_{\rm stable}$ (Figure~\ref{fig:wolf13a}) should host gentle novae that do not lead to net mass loss from the WD \citep[e.g.,][]{Yaron+05}. These will also be the novae with the shortest recurrence times, and indeed, several recurrent nova systems are observed to host massive WDs and may have $M_{\rm acc} > M_{\rm ej}$ (\citealt{Thoroughgood+01, Osborne+11, Page+15}, see also {\bf Box 8}).

However, it is unclear if the compositions of the WDs in recurrent nova systems are primarily ONe or CO. An ONe WD cannot be a Type Ia SN progenitor, because the end result of its growth toward the Chandrasekhar limit is generally thought to be an ``accretion-induced collapse" to a neutron star instead of a thermonuclear explosion (\citealt{Nomoto&Kondo91,Schwab+15}; however see \citealt{Jones+19}). 
It also appears that there are not enough recurrent novae (or steady supersoft sources accreting at $\gtrsim \dot{M}_{\rm stable}$) in nearby galaxies to explain the Type Ia SN rate (e.g.,~\citealt{Gilfanov&Bogdan10}).
Furthermore, even if WDs manage to retain a significant fraction of the accreted mass, the accreted mass is typically too small to achieve $M_{\rm Ch}$, especially in old stellar populations \citep[e.g.,][]{Ruiter+09, Maoz&Mannucci12}.
  
Even if accreted material can be retained on the WD through nova eruptions (e.g., \citealt{Starrfield+20}), the sustained burning phase following a nova eruption necessarily leaves a layer of helium ash that accumulates over time (e.g.,~\citealt{Denissenkov+13}).  As in the hydrogen case, once this helium shell achieves a critical mass, it can undergo runaway helium burning, causing drastic expansion of the WD envelope and mass loss \citep{Cassisi+98}.  Such ``helium novae" can also occur in mass-transfer systems involving He-rich donors like V445 Puppis (see {\bf Box 1}).   

\begin{textbox}[h]\section{BOX 8: M31N 2008-12a: Type Ia Supernova Progenitor?}
A recurrent nova in the Andromeda Galaxy, M31N 2008-12a, exhibits a recurrence period of $\sim$1 year, the shortest known to date \citep{Darnley+15,Henze+15}. This is one of the best known candidates for a single-degenerate Type Ia SN progenitor, as the WD's very high effective temperature during the supersoft X-ray phase implies a mass close to the Chandrasekhar limit ($\sim 1.5 \times 10^6$ K; \citealt{Darnley+16}), and the very fast recurrence time is only achieved by nova models for the most massive WDs accreting at very high rates (Figure \ref{fig:wolf13a}).
M31N 2008-12a is also surrounded by a nova ``super-remnant" 130 pc in diameter, probably inflated by $\sim 10^5$ recurrent nova eruptions  \citep{Darnley+19}.
Estimates suggest that the M31N 2008-12a WD could reach the Chandrasekhar mass and thus undergo core burning in less than 20,000 years \citep{Darnley+17b}. However, whether the outcome will be a Type Ia SN, as opposed to a pure deflagration (e.g., Type Iax SN) or accretion-induced collapse event, depends on the core composition of the WD and the outcome of core ignition.
 \end{textbox}

\subsection{Implications for Extragalactic Shock-Powered Transients}\label{sec:transients}

As time-domain surveys uncover new classes of astrophysical transients, it has become clear that the classical mechanisms of nuclear fusion and radioactive decay are insufficient to explain many of their luminosities and timescales.  As an alternative, it has been proposed that many transients' light curves are instead powered by shock interaction, for classes as diverse as Type IIn SNe, Type Ia-CSM SNe, superluminous SNe, pulsational-pair instability SNe, stellar mergers, fast blue optical transients, and tidal disruption events \citep{Woosley+07,Smith+08, Silverman+13, Metzger&Pejcha17,Gal-Yam+19, Margutti+19}.

For example, interaction-powered SNe (e.g., Types IIn and Ia-CSM) reach peak luminosities $\sim$5--100 times brighter than SNe primarily powered by radioactive decay (i.e., Types Ia and Ib/c), 
and their optical light curves evolve more slowly \citep{Kiewe+12, Silverman+13}. 
The leading model is that these SNe are surrounded be large masses of circumstellar material, and the resulting shocks are deeply embedded and heavily absorbed, such that the shock emission is reprocessed to optical wavelengths before emerging \citep[e.g.,][]{Smith+08}. However, it had not been clearly demonstrated that this mechanism can indeed produce the bulk of the optical luminosity---until the recent observations of correlated optical/gamma-ray emission in much closer classical novae (Figure \ref{fig:gamoptcorrelation}).

With novae now demonstrating that reprocessed shock emission can dominate the luminosity of astrophysical transients, they can be used to study the physics of this reprocessing in detail.  Novae demonstrate that optical and gamma-ray wavelengths provide the most direct glimpses of the shock (Figure \ref{fig:taueff}).  They reveal how shock power is converted to gamma-ray luminosity, and by which mechanisms (Figure \ref{fig:Indrek}).  Novae show how other wavelength regimes like X-ray and radio can provide useful insights into the shocks---if studied while understanding that most astrophysical transients are aspherical (e.g., Figure \ref{fig:Chomiuk14}), that the emission from radiative shocks may not conform to expectations based on adiabatic shocks, and that optical depth effects play a key role in what we observe.  Novae also provide potential real-time probes of dust formation in shock-compressed media (\S\ref{sec:dust}, \S\ref{sec:multiD}), a phenomena shared with other astrophysical systems such as colliding wind binaries (e.g., \citealt{Tuthill+99}) and stellar mergers \citep{Munari+02,Metzger&Pejcha17}.

The ejecta morphology proposed throughout this review---a slow, denser equatorial component and a faster, isotropic flow (which becomes funneled out the poles)---is common amongst astrophysical transients, especially those with binary progenitors.  Similar ejecta morphologies occur in stellar mergers, SN\,1987A (and its circumstellar environment), and kilonovae \citep{Pejcha+16a, Blondin&Lundqvist93, Metzger19}.  Novae allow us to observe the consequences of the two flows interacting, grounding predictions for these other more distant and exotic explosions.

Shocks in novae tend to possess a combination of higher densities and lower velocities than many other astrophysical shocks (e.g., SNe and SN remnants, gamma-ray bursts), but nevertheless they sample a broad range of physical conditions.  As such, nova shocks provide an opportunity to study how characteristics of the particle acceleration process ($\epsilon_{\rm rel}$, $q$, E$_{\rm max}$; \S\ref{sec:gammaspectra}) vary with shock properties in the radiative regime.  In other shock environments, the cosmic ray acceleration efficiency $\epsilon_{\rm rel}$ and primary energy spectra are often difficult to measure, because of uncertainty in cosmic ray escape fraction and propagation effects \citep{Blasi13}.  However, this task is enormously simplified in novae because of their calorimetric nature (\S\ref{sec:Emax}).  As a recent example of ``lessons learned" from novae, applied to other astrophysical transients, \citet{Fang+20} convert the optical luminosities of diverse transients into upper limits on their shock luminosities, energies in relativistic particles, and neutrino fluxes.  Calibrating this conversion using particle acceleration properties measured from novae, \citet{Fang+20} found that known classes of shock-powered transients can only account for $<1-10$\% of the high-energy neutrino background measured by IceCube \citep{IceCube13}.

\subsection{Open Questions, Future Work}\label{sec:future}

\textbf{$\bullet$ How well does WD TNR theory match observed nova properties?}
The basic predictions of nova theory---ejecta masses and kinetic energies, and how these scale with WD mass and accretion rate---still lack precise tests over a broad parameter space. Observers should work to grow and diversify the sample of novae where these properties are all accurately measured. 
Many data are already in hand, thanks to sustained efforts by, e.g., the Swift Nova-CV collaboration \citep[e.g.,][]{Page+20}, the Stony Brook/SMARTS Atlas \citep{Walter+12}, and the eNova collaboration \citep[e.g.,][]{Roy+12}. 
Future progress can be made by synthesizing these large samples and multi-wavelength data sets.
A first step would be a meta-analysis of ejecta mass measurements, where discrepant estimates with different techniques are homogenized and the volume filling factor is properly accounted for.

On a related point, while several teams have developed codes for modelling nova eruptions over the last decades, no systematic comparison has ever been performed to assess  points of agreement or conflict.  Detailed code comparisons are a critical next step for establishing the biggest challenges and opportunities of the future.

\vspace{0.1in}
\noindent \textbf{$\bullet$ How does mixing between the accreted envelope and the underlying WD vary across novae, and depend on WD mass and accretion rate?}
Studies of diffusion-induced mixing and multi-dimensional simulations of shear-mixing driven by convection during the TNR have advanced our understanding of the mixing processes that are key for determining properties of the nova eruption.  
Future theoretical work should continue to examine these mechanisms throughout the entire parameter space of $M_{\rm WD}$ and $\dot{M}$, as well as investigate whether alternative processes, such as heating by convectively driven internal gravity waves, also contribute to mixing. Observers can probe this issue by systematically obtaining abundance measurements for large samples of novae using X-ray/UVOIR spectra, and test theoretical predictions for how mixing scales with nova properties by comparing the measured level of WD pollution with properties of the outburst and of the binary system (e.g., $M_{\rm WD}$ and $\dot{M}$).

\vspace{0.1in}
\noindent \textbf{$\bullet$ How is mass ejected in nova eruptions: impulsive ejection, radiation-driven wind, common envelope interaction, or a combination thereof?}
Basic features in nova optical light curves like  plateaus, jitters, and oscillations highlight that mass loss can persist over months, and fluctuate in rate and velocity. While the cadence and quality of observations are shedding new light on this complexity, theoretical work lags behind.
Current models are largely limited to one dimension (necessarily neglecting the effects of the binary) and 
utilize simple parameterizations of mass loss. 
A new generation of three-dimensional simulations, that holistically model the physics of the TNR, convection, mass loss, and interaction with the binary, is critically needed.
Such simulations are necessary for answering this---and many of the other open questions listed here---but will be computationally very challenging, as they will need to model enormous spatial and temporal dynamic ranges.

\vspace{0.1in}
\noindent \textbf{$\bullet$ What physical processes shape the 3D morphology of nova ejecta?}
While it is well established that nova ejecta are often  aspherical, work is needed to connect this complex morphology with the physics of mass loss in nova eruptions. As theorists pursue multi-dimensional simulations, observers should pursue high-resolution imaging at a range of wavelengths, using the VLA, ALMA, IR adaptive optics/interferometry, and \emph{HST}. Images should be obtained early and often during the eruption, to assess how the morphology of the ejecta changes as mass ejection persists and as the ejecta diffuse. Morphokinematic modelling software like {\tt SHAPE} will remain critical to interpreting these observations.  In the future, the Next Generation VLA will provide imaging of thermal emission at milliarcsecond scales \citep{Murphy+18}, enabling a large and exquisite survey of time-dependent nova morphologies.

\vspace{0.1in}
\noindent \textbf{$\bullet$ Where do shocks occur in the nova ejecta?} Throughout this review, we have suggested a picture of an equatorially focused slow flow impacted by a more isotropic faster flow. While there is observational support for this picture, it is far from clear that the scenario is universal.  Growing the sample of novae with high-resolution imaging will be helpful for testing the universality of this scenario---especially if combined with imaging of synchrotron emission. 
In addition, studies of how nova properties depend on orbital inclination will be illuminating. For example, if shocks primarily occur in the orbital plane,
we would expect radio, optical, and X-ray signatures of shocked gas to vary with orbital inclination.

A mystery persists as to why the X-rays observed concurrently with the GeV gamma-rays are so underluminous compared to naive expectations.  It is unclear if the corrugated and thermally unstable nature of radiative shocks can suppress the X-rays to the necessary degree, or if different shocks dominate the observed X-ray and gamma-ray emission.  Multi-dimensional simulations of shock interaction in novae, which capture the interaction of an isotropic fast flow with an equatorially concentrated slow flow at sufficient resolution to model relevant small-scale instabilities, would help address if the full range of observational signatures is consistent with a single dominant shock structure. 

\vspace{0.1in}
\noindent \textbf{$\bullet$ Where is dust produced in nova ejecta?} Future observations should test if dust is produced in the dense cool shell generated by radiative shocks in novae.  One scenario forwarded in this review predicts that deep dust dips in optical light curves should be correlated with shock signatures and occur in edge-on binaries.  Kinematics of the dust---measured from molecular line profiles---would test if the dust is associated with the slow flow.  Effort should also be invested in imaging the sites of dust formation with ALMA or near-IR interferometers.

\vspace{0.1in}
\noindent \textbf{$\bullet$ To what maximum energy do nova shocks accelerate particles?}
Additional observational constraints on $E_{\rm max}$---and if it varies across novae---would provide a valuable test of diffusive shock acceleration theories. Although there is evidence for a steepening in the gamma-ray spectra of novae above $\sim 10-100$ GeV from a single epoch in a single event (e.g.~\citealt{Ahnen+15}), additional observations of novae at $\sim$TeV gamma-ray energies (with VERITAS, MAGIC, HAWC, and eventually the Cherenkov Telescope Array; \citealt{CTA+19}) 
are needed to probe the highest particle energies accelerated at the shocks.  If novae produce a significant luminosity of $\sim$ TeV gamma-rays (implying a high $E_{\rm max}$) this would considerably improve their detection prospects as neutrino sources by IceCube and successor experiments.

\vspace{0.1in}
\noindent \textbf{$\bullet$ Why do the shock energies and gamma-ray luminosities span such a wide range?}    
While it is clear that the gamma-ray luminosities of novae span at least two orders of magnitude, we lack a clear explanation for this diversity.  Furthermore, no clear correlations have been identified between the gamma-ray emission and other nova properties.  Continued survey observations of GeV gamma-rays are necessary to grow the sample of gamma-ray detected novae to establish the full diversity of their properties.  The kinetic power of the shocks in novae should be determined by the masses and relative velocities of the slow and fast flows, and radio imaging has the potential to separately measure the properties of these distinct flows (Fig.~\ref{fig:Chomiuk14}).

\vspace{0.1in}
\noindent \textbf{$\bullet$ What powers super-Eddington luminosities in novae?} 
The apparently super-Eddington luminosities of some novae remain almost as much of a mystery as when they were first discovered, and the role of shocks (if any) in this behavior requires further elucidation.
Multi-wavelength observations, from the IR to the X-ray, should be combined to measure bolometric light curves for a substantial sample of novae.  With the availability of \emph{Gaia} parallaxes and three-dimensional Galactic dust maps, these will enable the construction of bolometric luminosity time evolution with unprecedented accuracy.
These can then be compared with shock signatures and the luminosity of the supersoft component to test the degree to which shocks contribute to the bolometric luminosities of novae. Comparisons between the bolometric luminosities of classical novae and embedded novae should also be carried out, as a test of the energetic contribution of a common-envelope phase in nova eruptions.

\vspace{0.1in}
\noindent \textbf{$\bullet$ Do CVs ever host dense circumbinary material?}
While the shocks in classical novae can be explained as internal shocks, and most observations of CVs are consistent with very low density circumbinary material, there are arguments to the contrary  to explain THEA lines, or to remove binary angular momentum. 
Novae themselves can serve as ``light bulbs" to illuminate material pre-existing the eruption. Observations of the early UV/X-ray flash and a better understanding of THEA lines have the potential to strongly constrain the presence of circumbinary material.

\vspace{0.1in}
\noindent \textbf{$\bullet$ Why are WDs in CVs so massive?}
Perhaps the most important mystery plaguing CV populations is why the WDs in CVs are substantially more massive than their progenitor population. Future observations should test if nova eruptions serve as a significant sink of angular momentum and cause CV binaries with low-mass WDs to merge, removing them from the population. Systematically comparing eruption properties for classical novae and embedded novae can shed light on how strongly the binary orbit couples to the nova ejecta.

Measurements of how binary orbital period changes across a nova eruption should be pursued at scale with  time-domain facilities like the Rubin Observatory, to quantify how much angular momentum is lost in nova eruptions.
Finally, we should test if all slow novae still contain a binary after eruption, or if some slow novae might result in the merger of the WD and its companion. Tests of this hypothesis will be supported by a deepening understanding of the observational signatures of stellar mergers (e.g., luminous red novae) and improved coverage of the time-domain sky.

\vspace{0.1in}
\noindent \textbf{$\bullet$ What can we learn from helium novae?} The basic physics of novae powered by hydrogen burning should apply to helium novae as well, whether they arise from direct helium accretion or from the accumulation of helium ash due to  hydrogen burning.  While much rarer than classical novae, helium novae can expand our understanding of all novae by extending the parameter space to higher ignition masses and ejecta velocities. The theoretical modelling discussed above should also be applied to helium novae, along with consideration of truly explosive phenomena (deflagrations or detonations) due to the removal of the $\beta$-decay constraint limiting the energetics of hydrogen novae.  Such modeling will help observational efforts to recognize the subset of helium novae among the much larger class of hydrogen novae and select them for dedicated follow-up.

\begin{summary}[SUMMARY POINTS]
\begin{enumerate}
\item The theory of nova eruptions as thermonuclear runaways (\S\ref{sec:TNR}) has successfully explained many observations (\S\ref{sec:observations}). However, there remain observations---as fundamental as optical light curves---that lack agreed-upon explanations. The challenge is to understand how mass loss proceeds in nova eruptions (\S\ref{sec:massejection}), including three-dimensional effects such as the interaction of the ejecta with the binary. 
\item The ejecta in novae are usually far from spherical. The general morphology may be a slower denser flow in the orbital plane and a faster flow that escapes in the polar directions, but it remains unclear if this can explain the morphology of all novae. 
\item Recent decades have shown dramatic improvements in our understanding of sustained nuclear burning (e.g., ``supersoft emission") on the WD following the nova (\S\ref{sec:SSS}). These observations solidified a belief in the 2000s that radiation from the hot WD determines the bolometric luminosity and evolution of novae.
\item Recent (post-2010) detections of GeV gamma-rays, and renewed attention to nova shocks, have shown that this paradigm is not complete. Models of nova shocks predicted---and observations have largely confirmed---that the internal shocks are radiative, with the kinetic luminosity promptly transformed to radiative luminosity. Observations of correlated optical and gamma-ray light curves imply that shocks can significantly contribute to the bolometric luminosity and eruption evolution.
\item Novae are therefore the nearest and most common ``interaction-powered" astrophysical transients, and the first to be detected and observed in detail from radio to gamma-ray wavelengths. They are valuable laboratories for understanding the physics---and observable signatures---of radiative shocks and relativistic particle acceleration.
\item To continue to make progress on nova mass loss and shocks, three-dimensional simulations that self-consistently include the full range of physics are urgently needed.  On the observational side, high-resolution imaging  will be critical for testing these simulations and further establishing the origin of nova shocks. 
\end{enumerate}
\end{summary}

\section*{DISCLOSURE STATEMENT}
The authors are not aware of any affiliations, memberships, funding, or financial holdings that might be perceived as affecting the objectivity of this review. 

\section*{ACKNOWLEDGMENTS}

We are deeply thankful to Elias Aydi for his work on figures and helpful comments.
We offer special thanks to Alexa Gordon, Kwan-Lok Li, Elad Steinberg, and Indrek Vurm and for supplying data to generate some of the figures, along with insights.  We thank Pavel Denissenkov, Jay Gallagher, Boris G\"{a}nsicke, Yael Hillman, Justin Linford, Koji Mukai, Kim Page, Alex Pizzuto, Josiah Schwab, Jeno Sokoloski, Kirill Sokolovsky, Elad Steinberg, Dean Townsley, and Bill Wolf for useful information and helpful feedback.  Special thanks to Chris Kochanek for thorough comments on an earlier draft of the text.  We acknowledge with thanks the variable star observations from the AAVSO International Database contributed by observers worldwide and used in this research. Figure \ref{fig:opticalimages} is based on observations made with the NASA/ESA Hubble Space Telescope, and obtained from the Hubble Legacy Archive, which is a collaboration between the Space Telescope Science Institute (STScI/NASA), the Space Telescope European Coordinating Facility (ST-ECF/ESA) and the Canadian Astronomy Data Centre (CADC/NRC/CSA).

LC acknowledges support from NSF awards NSF AST-1751874 \& AST-1907790, NASA awards Fermi-80NSSC18K1746 \& NuSTAR-80NSSC19K0522, and a Cottrell fellowship of the Research Corporation.  BDM acknowledges support from NSF award AST-1615084 and the Simons Foundation through the Simons Fellowship (grant 606260).  KJS acknowledges support from the NASA Astrophysics Theory Program (NNX17AG28G \& 80NSSC20K0544).

\newpage
\section*{Supplementary Material}

\setcounter{table}{0}
\setcounter{figure}{0}
\makeatletter 
\renewcommand{\thefigure}{S\@arabic\c@figure}
\renewcommand{\thetable}{S\arabic{table}}
\makeatother

\begin{table}[h]
\tabcolsep7.5pt
\caption{Gamma-Ray Detected Novae ($>3 \sigma$ Significance)}
\label{stab:gammaray}
\begin{center}
\begin{tabular}{@{}l|c|c|c|c|c|c@{}}
\hline
Nova & D& $t_{\gamma}^{\rm a}$ & $\Gamma^{\rm b}$ & $E_{\rm c}^{\rm c}$ & $F_{\gamma}^{\rm d}$ & Refs\\
& (kpc) & (days) & & (GeV) & ($10^{-7}$\,cm$^{-2}$\,s$^{-1}$)  & \\
\hline
V407 Cyg 2010 &$3.5\pm 0.3$ & 22 & $1.3\pm0.2$ & $2.0\pm0.5$ &$3.5\pm0.4$ & 1,14,20\\
V1324 Sco 2012 &$>6.5$ & 17 & $1.9\pm0.2$ & $7.7\pm4.7$ & $4.4\pm0.9$ & 2,14,21\\
V959 Mon 2012 & $1.4\pm0.4$ & 22 & $1.5\pm0.3$ & $1.3\pm0.5$ & $2.6\pm0.5$ & 2,14,22\\
V339 Del 2013 & $4.5\pm 0.6$ & 27 & $1.7\pm0.2$ & $3.0\pm1.8$ & $1.5\pm0.2$ & 2,14,23\\
V1369 Cen 2013 & $\sim$2 & 39 & $2.0\pm0.3$ & $2.0\pm1.0$ & $2.5\pm0.5$ & 3,14,24\\
V5668 Sgr 2015 & $2.8\pm0.5$ & 55 & $2.1\pm0.1$ & -  & $0.6\pm0.1$ & 3,14,5\\
V407 Lup 2016$^{\rm e}$ & $4.2\pm0.5$ & 3  & $2.2\pm0.3$ & -  & $1.6\pm0.7$ & 4,5\\ 
V5855 Sgr 2016 & $\sim$4.5  & 26 & $2.3\pm0.1$ & - & $3.0\pm0.8$ & 6\\
V5856 Sgr 2016$^{\rm f}$ & $2.5\pm0.5$ & 15 & $1.9\pm0.1$ & $5.9\pm2.6$ & $5.4\pm0.5$ & 7,5\\
V549 Vel 2017$^{\rm g}$ & $>$4 & 33 & $1.8\pm0.2$ & - & $0.4\pm0.2$ & 8,5\\ 
V357 Mus 2018 & $3.2\pm0.5$ & 27 & $2.2\pm0.1$ & - & $1.3\pm0.2$ & 5 \\ 
V906 Car 2018$^{\rm h}$ & $4.0\pm1.5$ & $>20^{\rm i}$ & $1.8\pm 0.1$ & $5.9 \pm 1.1$ & $12.2\pm0.4$ & 9\\
V392 Per 2018 & 4.1$^{+2.3}_{-0.4}$ & $\gtrsim$8$^{\rm j}$ & $2.0\pm0.1$ & - & $2.2\pm0.4$ & 10,25 \\ 
V1707 Sco 2019 & $>$6 & 5 & $2.1\pm0.2$ & - & $2.9\pm1.0$ & 11,12\\
YZ Ret & $2.7\pm0.4$ & 18 & $2.2\pm0.1$ & - & $2.6\pm0.2$ & 12,13,26\\
\hline
\end{tabular}
\end{center}
\begin{tabnote}
$^{\rm a}$Duration over which gamma-rays were detected at $>2 \sigma$ confidence, when light curves are binned at 1 day cadence.
$^{\rm b}$Gamma-ray photon index.
$^{\rm c}$Cut-off energy to a power-law gamma-ray spectrum (if an exponential cut-off yields a better fit than a simple power law).
$^{\rm d}$Average photon flux over the time period t$_{\gamma}$, for photon energies $>$100 MeV.
$^{\rm e}$ASASSN-16kt; $^{\rm f}$ASASSN-16ma; $^{\rm g}$ASASSN-17mt; $^{\rm h}$ASASSN-18fv;  
$^{\rm i}$This is a lower limit; LAT observations were not available for roughly 3 weeks before, and roughly 2 weeks after the gamma-ray detection interval. 
$^{\rm j}$This is a lower limit; LAT observations were not available for roughly 2 weeks before the gamma-ray detection interval. 
\\
References: 
(1) \citealt{Abdo+10};
(2) \citealt{Ackermann+14}; 
(3) \citealt{Cheung+16};
(4) \citealt{Cheung+16_16kt}; 
(5) Gordon, A.\ et al.\ 2020, in prep.; 
(6) \citealt{Nelson+19}; 
(7) \citealt{Li+17}; 
(8)  \citealt{Li+17_17mt}, 2020, in prep.; 
(9) \citealt{Aydi+20}; 
(10) \citealt{Li+18}, Blochwitz, C.\ et al.\ 2020, in prep.; 
(11) \citet{Li+19};
(12) K.\ Li, 2020, private communication;
(13) \citet{Li+20}
(14) \citet{Franckowiak+18}
(20) \citealt{Ozdonmez+16}; 
(21) \citealt{Finzell+15};  
(22) \citet{Linford+15};
(23) \citealt{Schaefer+14}; 
(24) \citealt{Mason+18}; 
(25) \citet{Schaefer18};
(26) \citet{Bailer-Jones+18}
\end{tabnote}
\end{table}

\begin{figure}[h]
\includegraphics[width=5in]{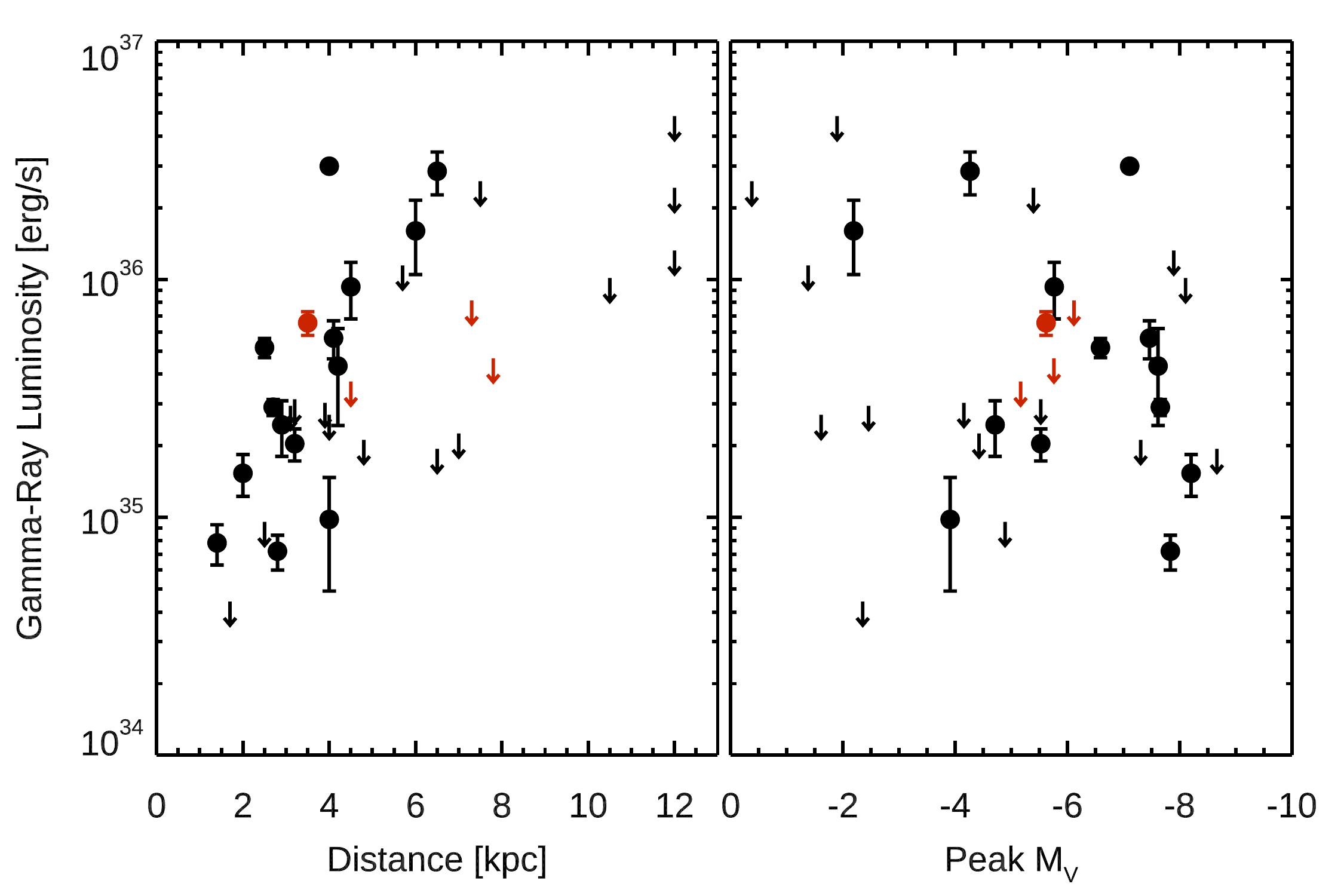}
\caption{Gamma-ray luminosity compared with nova distance (left panel) and peak absolute $V$-band magnitude. Classical novae are plotted in black, while embedded novae are shown as red symbols. For \emph{Fermi}-detected novae, gamma-ray fluxes and distances are as listed in Table \ref{stab:gammaray}. Upper limits for novae not detected by \emph{Fermi}-LAT come from \citealt{Franckowiak+18}, along with distances (also included is V3890~Sgr, as an upper limit, from \citealt{Buson+19}). 
Gamma-ray luminosities are calculated assuming the gamma-ray spectral shape of V906~Car \citep{Aydi+20}. Absolute $V$-band magnitudes are not corrected for dust extinction; peak apparent magnitudes originate from \citet{Franckowiak+18} and Gordon et al.\ 2020, in prep.}
\label{sfig:fermiprops}
\end{figure}


\begin{table}[h]
\tabcolsep7.5pt
\caption{Changes to Orbital Period over Nova Outbursts}
\label{stab:deltap}
\begin{center}
\begin{tabular}{@{}|l|c|c|c|c|c|c@{}}
\hline
Nova & P$_{\rm post}$ & $\Delta$P & Nova Year & Reference\\
& (day) & (ppm) & $\star$=recurrent &  \\
\hline
RR Pic & 0.1450238 & $-2004\pm1$ & 1925 & \cite{Schaefer20b}\\
HR Del & 0.214162 & $-472\pm5$ & 1967 & \cite{Schaefer20b}\\
QZ Aur & 0.3574970 & $-290\pm0.3$ & 1919 & \cite{Schaefer+19}\\
V1017 Sgr & 5.7860 & $-273\pm61$ & 1975 & \cite{Salazar+17}\\
DQ Her & 0.193620898 & $-4.46\pm0.03$ & 1934 & \cite{Schaefer20a}\\
CI Aql & 0.618361 & $-3.2\pm2.3$ & 2000$^\star$ & \cite{WH14} \\
U Sco & 1.230547 & $+3.5\pm5.6$ & 1999$^\star$& \cite{Schaefer11}\\
BT Mon & 0.3338149 & $+39.6\pm0.5$ & 1939 & \cite{Schaefer20a}\\
T Pyx & 0.0762336 & $+54\pm7$ & 2011$^\star$ & \cite{Patterson+17}\\
\hline
\end{tabular}
\end{center}
\begin{tabnote}
$^{a}$Binary orbital period measured after the nova outburst. $^{b}$Change in the orbital period, comparing before and after nova outburst in parts per million. Negative $\Delta$P implies the period is shorter after the nova outburst.
\end{tabnote}
\end{table}


\noindent  \href{https://drive.google.com/file/d/1bfEAxumdk9qho8DRZ1wQz1zKX0GbG1KB/view?usp=sharing}{\bf Supplemental Animation 1 (click for link)} Radiative shock fronts have corrugated and clumpy structure, due to thermal and thin-shell instabilities. This animation shows the complex density structure of the dual shocks created by a head-on collision of two flows (from the left and right, entering with 500 km/s corresponding to a Mach number $\mathcal{M} = 36$ for the adopted temperature floor of $10^{4}$ K). The colorscale tracks the log of density in units of g/cm$^3$. This animation is from the simulations of \citet{Steinberg&Metzger18}.\\
\vspace{0.1in}

\noindent \href{https://drive.google.com/file/d/1Yn8DXmHx7m2akPizaqIP3N0SdLW5jjVv/view?usp=sharing}{\bf Supplemental Animation 2 (click for link)} As in Supplemental Animation 1, but here the colorscale tracks the log of temperature rather than density. This animation is from the simulations of \citet{Steinberg&Metzger18}.


%
\bibliographystyle{ar-style2.bst}
\bibliography{nova}

\end{document}